\documentclass[11pt]{article}
\usepackage{amsfonts}
\usepackage{amssymb,latexsym}
\usepackage{amsmath,amscd}
\usepackage{theorem}
\usepackage{graphics}
\usepackage{subfigure}

\newtheorem{lemma} {Lemma} [section]
\newtheorem{proposition} [lemma] {Proposition}

\theorembodyfont{\rm}
\newtheorem{example}[lemma] {Example}
\newtheorem{remark}[lemma]{Remark}

\input{tcilatex}

\begin{document}

\title{\textbf{Base-controlled mechanical systems and geometric phases}}
\author{Alejandro Cabrera\thanks{\texttt{cabrera@mate.unlp.edu.ar}} \\
Departmento de Matematica, Universidad Nacional de La Plata\\
Calle 1 esq 115 (1900), La Plata\\
Argentina \\
}
\maketitle

\begin{abstract}
In this paper, we carry a detailed study of mechanical systems with
configuration space $Q\longrightarrow Q/G$ for which the base $Q/G$
variables are being controlled. The overall system%
\'{}%
s motion is considered to be induced from the base one due to the presence
of general non-holonomic constraints. It is shown that the solution can be
factorized into dynamical and geometrical parts. Moreover, under favorable
kinematical circumstances, the dynamical part admits a further factorization
since it can be reconstructed from an intermediate (body) momentum solution,
yielding a reconstruction phase formula. Finally, we apply this results to
the study of concrete mechanical systems.
\end{abstract}

\section{Introduction}

We shall describe a general formalism for studying classical mechanical
systems in which some of the configuration degrees of freedom are being 
\emph{controlled}, meaning that these are \emph{known functions of time}. We
will work under the (differential geometric-kinematical) assumption that the
controlled variables live in the base of a principal fiber bundle $%
Q\longrightarrow Q/G=B$. The remaining variables can be though of as living
in a Lie group $G$ and the equations for these \emph{fiber unknowns} are
derived by the hypothesis that overall motion respects some (general)\
non-holonomic constraints which are present in the system.

A special case is that in which the underlying momentum map give conserved
quantities, even when some of the variables are being acted by control
forces. In this case, it is clear that motion in the base variables must
induce motion in the remaining group variables in order for the momentum to
be constant during the resultant motion. A concrete example is given by a
self deforming body for which the shape evolution (base variables) is known
and global reorientation (group unknown) is induced by total angular
momentum conservation \cite{C def body}.

In this paper, we consider the more general situation in which fiber motion
is induced from the base one by the presence of (linear or affine)
non-holonomic constraints. These are represented by a distribution $D\subset
TQ\ $(\cite{BKMM,CMR1}) and we shall refer to them as $D-$\emph{constraints}%
. The information telling us how the base variables are moving is
represented by a base curve $\tilde{c}(t)\in B$ or, equivalently, by a curve 
$d_{0}(t)\in Q$ projecting onto $\tilde{c}$. The desired curve $%
c(t)=g(t)\cdot d_{0}(t)\in Q$ describing the full \emph{system%
\'{}%
s physical motion }is defined by the requirement that it projects onto $%
\tilde{c}$ on the base at each time (i.e. the base variables are the given
controlled ones)\ and that it satisfies the corresponding equations of \emph{%
motion} plus the $D-$\emph{constraints}. The base controlled hypothesis can
be seen as a set of \emph{time dependent constraints} and $g(t)\in G$ as the 
$d_{0}-$dependent (or \emph{gauge }dependent, see section \ref%
{subsubsec:Non-holonomic gauge}) fiber unknown.

The corresponding equations for $g(t)$ are derived by making \emph{dynamical}
assumptions, i.e. assumptions on the nature of the forces acting on the
system. By using variational techniques, we give explicitly the equations of
motion in section \ref{sec:Setting}. They correspond to the \emph{%
non-holonomic momentum equation} of \cite{BKMM} with time dependent
coefficients evaluated along $d_{0}(t)$. Using the kinematical structure of
the system, in sections \ref{subsubsec:Non-holonomic gauge} and \ref%
{subsubsec:Gauges and phases in Q}, we show how the solution $c(t)$ can be 
\emph{factorized} by considering specific gauges $d_{0}(t)$, yielding that
each factor has either a pure geometrical (kinematical) definition or it
obeys dynamical equations which are simpler than the overall fiber ones.

In section \ref{sec: special cases}, we shall carry out a detailed analysis
of systems with a special kinematical structure, focusing on the
geometric-dynamical factorization of the solution mentioned above. Moreover,
in section \ref{subsec: Reconstruction and phases}, we show that under
favorable kinematical circumstances (e.g. in the presence of \emph{%
horizontal symmetries} \cite{BKMM}), the dynamical factor $g(t)$ of the
solution $c(t)\in Q$ admits a further factorization. In fact, we can write
reconstruction \emph{phase formulas }\cite{MMR} for $g(t)$. The obtained
phase formulas relate the overall system%
\'{}%
s evolution to the geometry and dynamics of simpler intermediate solutions
which, in turn, live in smaller spaces (coadjoint orbits). Consequently,
these formulas generalize the ones obtained in \cite{Mont rig body} and \cite%
{C def body} for rigid bodies and self deforming bodies, respectivelly, to
the more general setting of $D-$constrained induced motion. Notice that
phase formulas become interesting and useful when the dynamical contribution
can be expressed in terms of the system%
\'{}%
s dynamical quantities like energy and/or characteristic times (see, e.g., 
\cite{Mont rig body}) . This is generically accomplished in section \ref%
{subsec: Reconstruction and phases} and exemplified in section \ref%
{subsec:Examples}.

The formalism presented in this work, for studying $D-$\emph{constrained,
base controlled} systems, applies to a larger class of mechanical systems
than the one encoded in \cite{C def body}. First, it applies to systems with
general configuration space $Q$ endowed with a principal bundle structure%
\footnote{%
Notice that in \cite{C def body}, $Q$ represented specifically the
configuration space of a deforming body.}. And, in the second place, it
allows for (linear or affine)\ $D-$constraints, and not only momentum
conservation, to rule the system%
\'{}%
s dynamics. Indeed, in examples \ref{Ex: 2 Balls} and \ref{subsubsec:Example
dipolar magnetic}, we are able to answer two natural questions which arise
from \cite{C def body}: what happens to the corresponding \emph{phase
formulas} when magnetic-type forces are acting upon a deforming body, and
thus, when the (angular)\ momentum is no longer conserved?; how does a self
deforming body move when there are additional (internal) non-holonomic
constraints between the (no longer controllable) shape variables?

To end this introduction, we would like to comment on the applications of
the present work to mechanical control theory. First, note that control
problems are, in a sense, \emph{orthogonal }to the one we described so far.
In this paper, we claim to \emph{know} the base variables dynamics and we
want to \emph{find} the induced fiber motion; while in control theory one 
\emph{starts} with a desired fiber dynamics and tries to \emph{find} which
base curve induces it (and, after that, how to implement this base motion
via control forces). Nevertheless, the spirit of this paper is to think of
the known base dynamics as coming from direct observation or measurement
(example \ref{Ex: 2 Balls} illustrates this point very clearly). Indeed, an
interesting feature of this kind of systems is the fact that the overall
motion $c(t)$ can be constructed from that in the base using only the \emph{%
kinematics} of $\tilde{c}(t)$ and without actually knowing the \emph{forces}
which are inducing such base motion. The results on the induced fiber
motion, obtained by the formalism we describe below, can be thus used for
theoretically correcting the a priori fiber dynamics prediction when the
observed base dynamics deviates from the control-theoretical desired one.
Also, analytical phase formulas provide interesting tools for directly
testing different control configurations and theoretical methods.

\noindent \textbf{Acknowledgements:}

I would like to thank Prof. J. Solomin and P. Balseiro for useful
discussions and suggestions. A.C. also thanks CONICET (Argentina) and
Universidad Nacional de La Plata (Argentina)\ for financial support.

\section{Controlled systems with additional Nonholonomic constraints}

\label{sec:Setting}

\subsection{The Kinematical Setting}

\label{subsec:Kinematical setting}In the remaining, we shall focus on
mechanical systems with general non-holonomic linear (or affine, see section %
\ref{subsec:affine constraints} below) constraints. More precisely, our
setting will consists of a mechanical system described by the data $%
(Q,L,G,D) $:

\begin{itemize}
\item Let $Q$ denote the \textbf{configuration space }and $G$ a symmetry Lie
group acting on $Q$ by the left such that $Q\overset{\pi }{\longrightarrow }%
Q/G$ is a \textbf{principal} $G-$\textbf{bundle}. We shall call, as usual, $%
B:=Q/G$ the \textbf{shape space }(see \cite{Mont gauge}). We denote the
action by $g\cdot q$ and the induced \emph{infinitesimal action} $\rho
_{g\ast }:TQ\longrightarrow TQ$.

\item Let $k_{q}(\cdot ,\cdot )$ denote a $G-$\textbf{invariant Riemannian
metric} on $Q$ and $k_{q}(\cdot ):T_{q}Q\longrightarrow T_{q}^{\ast }Q$ the
induced bundle isomorphism.

\item Let $L:TQ\longrightarrow \mathbb{R}$ denote the $G-$\textbf{invariant
Lagrangian} (with respect to the lifted $G-$action on $TQ$) given by the ($%
k_{q}-$)\emph{kinetic energy} minus $G-$invariant potentials (see also
appendix \ref{sec:App Kin En}).

\item Let $D\subseteq TQ$ be a \textbf{constraint distribution}.
\end{itemize}

We shall assume further:

\begin{description}
\item[$\left( H1\right) $] $D\ $is $G-$\textbf{invariant }and $%
D_{q}+V_{q}=T_{q}Q$, for all $q\in Q$ and $Ver_{q}=Ker(\pi _{\ast q})$
denoting the vertical subspace of $T_{q}Q$. This is referred to as the 
\textbf{principal case} in \cite{BKMM}.
\end{description}

Now, suppose that, for such a system, the \textbf{base variables are being
controlled in a certain known way}. This means, that

\begin{description}
\item[$\left( H2\right) $] we are given a curve $d_{0}(t)$ in $Q$ for $t\in
I:=[t_{1},t_{2}]$ or, equivalently, a map $\tilde{c}:[t_{1},t_{2}]%
\longrightarrow Q/G$ s.t. $\pi (d_{0}(t))=\tilde{c}(t)$. The \textbf{time
evolution of the controlled system} is then described by a curve $c(t)\in Q$
such that 
\begin{equation*}
\pi (c(t))=\tilde{c}(t)
\end{equation*}%
\textbf{for each} $t\in \lbrack t_{1},t_{2}]$.
\end{description}

The above means that%
\begin{equation}
c(t)=g(t)\cdot d_{0}(t)  \label{Eq constr def c(t)}
\end{equation}%
for an unknown ($d_{0}(t)-$dependent)\ curve $g(t)$ in $G$.

\begin{description}
\item[Definition:] We shall refer to the data $(Q,L,G,D,\tilde{c})$ as a 
\textbf{base-controlled }($D-$)\textbf{constrained dynamical system}.
\end{description}

The curve $g(t)$ is the \textbf{unknown} for our \emph{controlled mechanical
problem} as stated above. Note that, if the controlled problem has a unique
solution $c(t)$ for each initial value $c(t_{1})\in \pi ^{-1}(\tilde{c}%
(t_{1}))$, then for each curve $d_{0}(t)$ in $Q$ lying over $\tilde{c}(t)\in
Q/G$, there is a unique $g(t)$ satisfying $\left( \ref{Eq constr def c(t)}%
\right) $. In this case, the initial condition for the unknown in $G$ reads%
\begin{equation}
c(t_{1})=g(t_{1})\cdot d_{0}(t_{1}).  \label{Eq constr cond init g(t)}
\end{equation}

The curve $d_{0}(t)$ will be called \textbf{gauge choice }or, simply, 
\textbf{gauge. }This terminology is motivated by the analogy between the
freedom in choosing among such curves projecting to the same $\tilde{c}$ in
shape space and gauge freedom in classical gauge field theories (see \cite%
{Mont gauge}, \cite{SW}, references therein and also section \ref%
{subsubsec:Bundle formulation}).

\begin{remark}
\emph{(Restricted configuration space) }Note that $\left( H2\right) $
implies (but it is not equivalent to!) the following \textbf{holonomic
constraint:}%
\begin{equation*}
c(t)\in \pi ^{-1}(\tilde{c}(I)).
\end{equation*}%
For a specific problem in which $\tilde{c}$ is fixed, one can restrict the
analysis to $\tilde{Q}=\pi ^{-1}(\tilde{c}(I))$. Nevertheless, in what
follows, we continue with the study of generic $\tilde{c}$%
\'{}%
s and thus express the results in terms of the kinematical structure of the
whole $Q$. Notice that this is the more convenient procedure for studying
systems in which $\tilde{c}$ can be (dynamically)\ perturbed.
\end{remark}

\begin{remark}
\label{rmk: Vertical D}\emph{(Vertical }$D-$constraints\emph{) }Note that
the dimension assumption $(H1)$ states that the $D-$constraints are \emph{%
vertical}, in the sense that it ensures that the equations of motion
(locally) drop to the base $Q/G$ with no $D-$constraints remaining there. In
other words, the base curve $\tilde{c}(t)$ can be arbitrarily chosen within $%
Q/G$. For example, if the sum is direct, i.e., $D_{q}\oplus V_{q}=T_{q}Q$
then $D$ defines a principal connection and we are in the purely kinematic
case of \cite{BKMM}. In the case $D_{q}\cap V_{q}=0$ but $D_{q}\oplus
V_{q}\neq T_{q}Q$, then constraints are also to be considered in the motion
of the base variables and, thus, the base dynamics could not be
(arbitrarily) controllable.
\end{remark}

\subsubsection{Kinematical ingredients}

We now recall some known definitions and properties for mechanical systems
that we shall use through the paper.

First recall that for\emph{\ }simple mechanical systems with symmetry (\cite%
{AM,MR}) as described above, the lifted $G$ action on $TQ$ always has an
(equivariant) momentum map $J:TQ\longrightarrow \mathfrak{g}^{\ast }$ given
by 
\begin{equation*}
\left\langle J(v_{q}),X\right\rangle =\left\langle
k_{q}(v_{q}),X_{Q}(q)\right\rangle ,
\end{equation*}%
for $X\in \mathfrak{g}$. Let us also recall another ingredients (see ex. 
\cite{Mont gauge}):

\begin{itemize}
\item \textbf{Locked inertia tensor }$I_{q}:\mathfrak{g}\longrightarrow 
\mathfrak{g}^{\ast }$\textbf{\ ,} 
\begin{equation*}
I_{q}=\sigma _{q}^{\ast }\circ k_{q}\circ \sigma _{q}
\end{equation*}%
with $\sigma _{q}:\mathfrak{g}\longrightarrow T_{q}Q$ denoting the \emph{%
infinitesimal generator map}, $\sigma _{q}(X)=X_{Q}(q)$. Because the metric $%
k_{q}$ is $G-$invariant, $I$ satisfies the \emph{equivariance} property:%
\begin{equation*}
I_{g\cdot q}=Ad_{g}^{\ast }\circ I_{q}\circ Ad_{g^{-1}}.
\end{equation*}

\item The momentum map $J$ is $Ad^{\ast }-$equivariant, i.e.,%
\begin{equation*}
J(g\cdot m)=Ad_{g}^{\ast }J(m)
\end{equation*}%
with $Ad_{g}^{\ast }=(Ad_{g^{-1}})^{t}$ denoting the (left) \textbf{%
coadjoint representation} of $G$ on $\mathfrak{g}^{\ast }$ and $^{t}$ the
transpose. This follows from the identity%
\begin{equation*}
\sigma _{g\cdot q}(X)=\rho _{g_{\ast }q}(\sigma _{q}(Ad_{g^{-1}}X)).
\end{equation*}
\end{itemize}

Now, from $\left( \ref{Eq constr def c(t)}\right) $ we have that%
\begin{equation}
\frac{d}{dt}c(t)=\rho _{g(t)_{\ast }d_{0}(t)}(\sigma _{d_{0}(t)}(g^{-1}\frac{%
d}{dt}g(t)))+\rho _{g(t)_{\ast }d_{0}(t)}\frac{d}{dt}d_{0}(t)
\label{veloc c(t)}
\end{equation}%
and thus,%
\begin{equation}
J(\frac{d}{dt}c(t))=Ad_{g(t)}^{\ast }I_{d_{0}(t)}(g^{-1}\frac{d}{dt}%
g(t))+Ad_{g(t)}^{\ast }J(\frac{d}{dt}d_{0}(t)).  \label{J}
\end{equation}

We can think of $J_{0}(t):=J(\frac{d}{dt}d_{0}(t))$ as the \emph{apparent}
or \emph{internal} momentum along $d_{0}(t)$ and $I_{0}(t):=I_{d_{0}(t)}$ as
the locked inertia tensor changing with the gauge motion $d_{0}(t)$.

\subsection{Dynamical Hypothesis}

\label{subsec: D Dyn Hyp}The assumption $\left( H2\right) $ above can be
interpreted as giving a \emph{time dependent type of kinematical constraint }%
on the original system, in addition to the one represented by the
distribution $D\subseteq TQ$. To determine the motion of such a \emph{twice
kinematically constrained system}, i.e. to find\footnote{%
Equivalently, for a chosen gauge $d_{0}(t),$ to find the corresponding $g(t)$
in $G$.} $c(t)$ in $Q$, we need to add \emph{dynamical information}. This
information consists in assumptions about the nature of the \textbf{forces}
which are acting upon the system in order to satisfy the imposed kinematical
constraints.

For the set of constraints corresponding to the distribution $D$, we shall
assume

\begin{description}
\item[$\left( DH1\right) $] \textbf{D 
\'{}%
alambert 
\'{}%
s Principle:} The $D-$\emph{constraint forces} lie in the annihilator space
of the kinematical distribution $D$.
\end{description}

\bigskip Denoting $F^{D}:TQ\longrightarrow \mathbb{R}$ the $D-$forces (seen
as $1-$forms on $Q$) which act on the system enforcing the $D-$constraints, $%
(DH1)$ means that%
\begin{equation*}
F^{D}(v)=0
\end{equation*}%
for all "virtual displacement" $v\in D\subset TQ$. If $\left( DH1\right) $
is not satisfied by the system%
\'{}%
s forces, we must then know\footnote{%
Or to know some other information about them leading to the corresponding
equations of motion.} the $D-$constraint forces and add them to the
equations of motion (see \ref{rmk: external forces c(t)} below).

For the time dependent\emph{\ control like }constraints represented by the
shape space curve $\tilde{c}(t)\in Q/G$, the assumption takes a less usual
form:

\begin{description}
\item[$\left( DH2\right) $] The forces which are inducing the motion $c(t)$
to satisfy $\pi (c(t))=$ $\tilde{c}(t)$ are of a kind that we shall denote
as \textbf{good internal} ones. Good internal forces seen as $1-$forms $%
F_{int}^{c}:TQ\longrightarrow \mathbb{R}$ satisfy%
\begin{equation*}
F_{int}^{c}(\delta c)=0
\end{equation*}%
for all \textbf{vertical variations} $\delta c=\left. \frac{d}{ds}%
\right\vert _{s=0}(g(t,s)\cdot d_{0}(t))$ and any gauge $d_{0}(t)$ (see also
below).
\end{description}

In other words, good internal forces are such that they do not affect
dynamically (i.e. by adding extra terms) the \emph{vertical} part of the
equations corresponding to $(Q,L,G,D)$. This idea is already present in \cite%
{BKMM}, in terms of the validity of the nonholonomic momentum equations when
internal (shape space) control forces are present.

\begin{example}
\label{ex:Deforming Body internal forces}\emph{(}$\emph{Motion}$ \emph{of
self deforming bodies) }Let $Q=\mathbb{R}^{3N-3}$ be the configuration space
of an $N-$particles system modeling a deforming body. In this case, usual
internal forces between the particles of the system satisfying the \emph{%
strong action reaction principle} (\cite{Goldstein}) are \emph{good internal
forces} . For details, see \cite{C def body}.
\end{example}

\begin{remark}
\emph{(Non good internal forces)} If the constraint forces acting on the
controlled base variables are not of the \emph{good internal} type, then we
must add the extra piece of information missing, this is, how the equations
have to be modified by adding the non-vanishing terms $F^{c}(\delta c)$ (see %
\ref{rmk: external forces c(t)} below). In the case of the above example,
this means that if there are, say, electromagnetic forces acting on the self
deforming body which do not satisfy the \emph{strong} action-reaction
principle, then one must know the underlying magnetic field data and correct
the angular momentum conservation equations as usual (see e.g. \cite%
{Goldstein}, and also sections \ref{subsec: Phases for dipolar} and \ref%
{subsubsec:Example dipolar magnetic}).
\end{remark}

For control purposes, the equations of motion following from $\left(
DH1\right) $ and $\left( DH2\right) $ for the base variables $r\in B=Q/G$
can be locally written as (for details see \cite{BKMM})%
\begin{equation*}
M(r)\ddot{r}=-C(r,\dot{r})+N(r,\dot{r},J(g,\dot{g}))+F^{pot}+F_{int}^{c}
\end{equation*}%
where $g$ denotes the (local) vertical part of variables in $Q\simeq
Q/G\times G$, $J$ the (generalized, non holonomic) momentum, $F^{pot}$ the
potential forces acting on $B$ and $F_{int}^{c}$ the control\ forces
mentioned in $\left( DH2\right) $. Also, $M$ denotes the mass matrix of the
system, $C$ the Coriolis term (quadratic in $\dot{r}$) and $N$ a term being
quadratic in $\dot{r}$ and $\left\langle J,\xi \right\rangle $, where $\xi $
is a $q$ dependent element in $\mathfrak{g}=Lie(G)$.

In what follows, we shall assume that the system is being base-controlled,
so the control forces are inducing via the above equation the prescribed
motion $\tilde{c}(t)\equiv r(t)$. The problem is then to find the remaining
vertical part of the motion, which is induced by the one in the base $B$
because of the presence of the $D-$constraints.

\subsection{The variational principle}

\label{subsec: Variational principle}The equations of motion for the above
described base controlled $D-$constrained system, satisfying $\left(
H1,2\right) $ and $\left( DH1,2\right) $, can be deduced from an adapted 
\emph{variational principle}.

Explicitly, we shall assume that the solution curve $c(t)$ is an \textbf{%
extremal} of the action functional%
\begin{equation*}
S_{Q}=\dint\limits_{t_{1}}^{t_{2}}L(\overset{\cdot }{c})dt
\end{equation*}%
for deformations of the following specific kind:%
\begin{equation}
c(t,s)=g(s)\cdot c(t).  \label{Eq. variation c(t,s)}
\end{equation}%
These kind of deformations can be called \textbf{vertical} following the
ideas of \cite{CMR1}. Also following \cite{CMR1}, from $\left( DH1\right) $
we shall restrict the variations to the ones \textbf{satisfying the }$D-$%
\textbf{constraints}, i.e., $\delta c\in D_{c(t)}$.

Let $I=[t_{1},t_{2}]\ $and $\Omega (Q;\tilde{c}(t),q_{1},q_{2})$ denote the
space of smooth curves $I\longrightarrow Q$ with fixed end-points $%
q_{1},q_{2}\in Q$ such that $\pi (c(t))=\tilde{c}(t)$. Note that, for a
given (any) gauge choice $d_{0}(t)$ such that $\pi (d_{0}(t))=\tilde{c}(t)$,
any deformation can be written as%
\begin{equation}
c(t,s)=g(t,s)\cdot d_{0}(t).  \label{Eq deformations c(t,s) gauge}
\end{equation}%
and thus 
\begin{equation*}
\Omega (Q;\tilde{c}(t),q_{1},q_{2})\approx \Omega _{d_{0}}(G;g_{1},g_{2})
\end{equation*}%
with $\Omega _{d_{0}}(G;g_{1},g_{2})$ being the space of smooth curves $%
I\longrightarrow Q$ with fixed end points $g_{i}$, s.t. $g_{i}\cdot
d_{0}(t_{i})=q_{i}$ for $i=1,2$.

So, summing up, our problem is equivalent to the following (gauge invariant)
variational formulation:

\begin{description}
\item[P1] \emph{(Gauge invariant formulation)}\ Finding an extremal $c(t)$
of the action $S_{Q}$, i.e. $\delta S_{Q}=0$, among the curves in $\Omega (Q;%
\tilde{c}(t),q_{1},q_{2})$ for vertical deformations $\delta c(t)=\left. 
\frac{d}{ds}\right\vert _{s=0}c(t,s)$ induced by $\left( \ref{Eq. variation
c(t,s)}\right) $, vanishing at the end points, i.e. $\delta c(t_{i})=0$ for $%
i=1,2$, and with both $\overset{\cdot }{c}(t)$ and $\delta c$ satisfying the 
$D-$constraints.
\end{description}

Once the gauge is fixed, the action $S$ induces an equivalent \emph{non
autonomous Lagrangian system} on the $G$ which is, in turn, equivalent to
the following:

\begin{description}
\item[P2] \emph{(Gauge covariant formulation)\ }Finding extremal curve $g(t)$
in the set $\Omega _{d_{0}}(G;g_{1},g_{2})$ for the action%
\begin{equation*}
S_{G}[d_{0}]=\dint\limits_{t_{1}}^{t_{2}}L_{d_{0}}(g,\overset{\cdot }{g},t)dt
\end{equation*}%
i.e., for which $\delta S_{G}[d_{0}]=0$, satisfying the gauge induced $D-$%
constraints, i.e., 
\begin{equation}
\overset{\cdot }{d_{0}}(t)+(g^{-1}\overset{\cdot }{g})_{Q}(d_{0}(t))\in
D_{d_{0}(t)}  \label{Eq. D constraint for psi}
\end{equation}%
and for variations $\delta g(t)=\left. \frac{d}{ds}\right\vert _{s=0}g(t,s)$%
, $\delta g(t_{i})=0,$ $i=1,2$, satisfying the $D-$constraints:%
\begin{equation}
\left( g^{-1}\left. \frac{d}{ds}\right\vert _{s=0}g\right) _{Q}(d_{0}(t))\in
D_{d_{0}(t)}.  \label{Eq. constraint on delta g}
\end{equation}
\end{description}

Note that, although different gauge choices shall lead to different time
dependence of the $g(t)$ equation 
\'{}%
s coefficients, the full solution $c(t)$ is the same for all $d_{0}$%
\'{}%
s. In other words, though the equations for $g(t)$ (and thus $g(t)$ itself)
are not \emph{gauge invariant}, the solution $c(t)$ is. On the other hand, $%
g(t)$ can be seen as being \emph{gauge covariant\ }(see remark \ref{rmk:
gauge covariance} below).

In the above formulation $\left( P2\right) $, $L_{d_{0}}$ is $L(\overset{%
\cdot }{c})$ with $c(t)$ given by $\left( \ref{Eq constr def c(t)}\right) $.
It is easy to see that it takes the form of the (left) $G$-invariant non
autonomous Lagrangian given by:%
\begin{equation*}
L_{d_{0}}(g,\overset{\cdot }{g},t)=\frac{1}{2}k_{d_{0}}(\overset{\cdot }{%
d_{0}},\overset{\cdot }{d_{0}})+\frac{1}{2}\left\langle I(d_{0}(t))\xi ,\xi
\right\rangle +\left\langle J(\overset{\cdot }{d_{0}})(t),\xi \right\rangle
\end{equation*}%
with $\xi =g^{-1}\overset{\cdot }{g}$, $I(d_{0}(t)):\mathfrak{g}%
\longrightarrow \mathfrak{g}^{\ast }$ the \textbf{locked inertia tensor }map
and $J:TQ\longrightarrow \mathfrak{g}^{\ast }$ the \textbf{momentum map }%
(recall section \ref{subsec:Kinematical setting}).

Finally, note that variations $\delta \xi =\left. \frac{d}{ds}\right\vert
_{s=0}(g^{-1}\overset{\cdot }{g})$ induced by variations $\delta g=\left. 
\frac{d}{ds}\right\vert _{s=0}g(t,s)$ satisfy the following identity:%
\begin{equation}
\delta \xi -\frac{d}{dt}\left( g^{-1}\delta g\right) =[\xi ,g^{-1}\delta g]
\label{Eq. variations psi}
\end{equation}%
where $[,]$ denotes the Lie bracket on $\mathfrak{g}$.

\begin{remark}
\label{rmk: gauge covariance}\emph{(Gauge covariance) }Note that $\left( \ref%
{Eq. D constraint for psi}\right) $ is gauge dependent, meaning that it is
different for different choices of the gauge curve $d_{0}(t)$. Nevertheless,
since $D$ is $G-$invariant, it is \emph{gauge covariant}: if $\tilde{d}%
_{0}(t)=g_{cg}(t)\cdot d_{0}(t)$ is another gauge, then $g(t)$ in $\left( %
\ref{Eq constr def c(t)}\right) $ satisfies the $D-$constraint equation $%
\left( \ref{Eq. D constraint for psi}\right) $ for $d_{0}(t)$ iff $\tilde{g}%
(t):=g(t)g_{cg}^{-1}(t)$ satisfies the eq. analogous to eq. $\left( \ref{Eq.
D constraint for psi}\right) $ for the new gauge $\tilde{d}_{0}(t)$.
\end{remark}

\subsection{Equations of motion}

Note that, as a consequence of Newton 
\'{}%
s second law, the equations for the unknown $g(t)$ shall be \emph{second
order} ones. Also, by the time dependent control constraint, they shall also
be \emph{non autonomous }and \emph{gauge dependent}, i.e., its coefficients
will depend on time through the chosen $d_{0}(t)$.

We shall start with the \emph{gauge invariant formulation }$\left( P1\right) 
$. Taking into account that the variations are of the form $\left( \ref{Eq.
variation c(t,s)}\right) $,%
\begin{equation*}
\delta S_{Q}=0
\end{equation*}%
straightforwardly implies%
\begin{equation}
i_{c(t)}^{\ast }(\frac{d}{dt}J(\overset{\cdot }{c}))=0
\label{Eq. motion c(t)}
\end{equation}%
for $i_{c(t)}:\mathfrak{g}^{c(t)}\hookrightarrow \mathfrak{g}$ denoting the
inclusion and 
\begin{equation*}
\mathfrak{g}^{c(t)}:=\{X\in \mathfrak{g},\ X_{Q}(c(t))\in D_{c(t)}\}.
\end{equation*}%
The above equation is equivalent to the \emph{non-holonomic momentum equation%
} of \cite{BKMM}, evaluated on the controlled curve $c(t)$ of eq. $\left( %
\ref{Eq constr def c(t)}\right) $.

\begin{remark}
\label{rmk: No D+TG}\emph{(Non necessity of }$\left( H1\right) $ \emph{nor} $%
\left( H2\right) $\emph{) }Equation $\left( \ref{Eq. motion c(t)}\right) $
is one of the equations of motion of any system whose kinematics is as in
section \ref{subsec:Kinematical setting} without the need of $\left(
H1,2\right) $. The only dynamical hypothesis needed is $(DH1)$ plus the fact
that any other force acting on the system (seen as $1-$forms on $Q$) is such
that it vanishes when evaluated on vertical variations. What these last
kinematical hypothesis $\left( H1,2\right) \ $add is: that no $D-$%
constraints remain on the base variables and that these are being
controlled, so $\left( \ref{Eq. motion c(t)}\right) $ is the only \emph{%
equation of motion }(not of constraint) left to solve in the system.
\end{remark}

These are $k:=dim\mathfrak{g}^{c(t)}=dim\mathfrak{g}^{c(t_{1})}=const.$
equations coupled to the $(dim\mathfrak{g-}k)\ $number of $D-$constraint
equations:%
\begin{equation*}
\overset{\cdot }{c}(t)\in D_{c(t)}.
\end{equation*}

Notice that, since the shape space variables are being controlled, that is,
since $\left( H2\right) $ leave only $dim\mathfrak{g}$ degrees of freedom,
eq. $\left( \ref{Eq. motion c(t)}\right) $ and the above $D-$constraint
equations determine uniquely $c(t)$ because of $\left( H1\right) $.

Below, we shall give more explicit equations for the unknown $g(t)$ by
fixing a gauge choice $d_{0}(t)$ and working in the gauge covariant
formulation $\left( P2\right) $\footnote{%
We do need $\left( H1,2\right) $ for $\left( P2\right) $.}. To illustrate on
the underlying calculation, we shall derive the equations directly from $%
\left( P2\right) $, though they can be also derived from $\left( \ref{Eq.
motion c(t)}\right) $ using the decomposition $\left( \ref{Eq constr def
c(t)}\right) $. Let us, thus, evaluate%
\begin{equation*}
0=\delta S_{G}[d_{0}]=\dint\limits_{t_{1}}^{t_{2}}\left\langle
I(d_{0}(t))\xi +J(\overset{\cdot }{d_{0}})(t),\delta \xi \right\rangle .
\end{equation*}%
By eq. $\left( \ref{Eq. variations psi}\right) $ and integration by parts,
we have%
\begin{equation*}
=-\dint\limits_{t_{1}}^{t_{2}}\left\langle \frac{d}{dt}\left( I(d_{0}(t))\xi
+J(\overset{\cdot }{d_{0}})(t)\right) +ad_{\xi }^{\ast }\left(
I(d_{0}(t))\xi +J(\overset{\cdot }{d_{0}})(t)\right) ,g^{-1}\delta
g\right\rangle
\end{equation*}%
where $ad_{\xi }^{\ast }=-(ad_{\xi })^{t}$ denotes the (left) co-adjoint
action. Notice that $g^{-1}\delta g$ is arbitrary only among variations
satisfying the $D-$constraint $\left( \ref{Eq. constraint on delta g}\right) 
$, i.e. 
\begin{equation*}
g^{-1}\delta g\ (t)\in \mathfrak{g}^{d_{0}(t)}.
\end{equation*}%
Consequently, for $i_{d_{0}(t)}^{\ast }:\mathfrak{g}^{\ast }\hookrightarrow (%
\mathfrak{g}^{d_{0}(t)})^{\ast }$ denoting the canonical projection, 
\begin{equation}
i_{d_{0}(t)}^{\ast }\left[ \frac{d}{dt}\left( I(d_{0}(t))\xi +J(\overset{%
\cdot }{d_{0}})(t)\right) +ad_{\xi }^{\ast }\left( I(d_{0}(t))\xi +J(\overset%
{\cdot }{d_{0}})(t)\right) \right] =0  \label{Eq. motion for psi}
\end{equation}%
must hold. These are $k=dim\mathfrak{g}^{q}$ (constant $\forall q\in Q$)
equations of motion for the \emph{body velocity} $\xi (t)=g^{-1}\overset{%
\cdot }{g}(t)$ which are coupled to the $(dim\mathfrak{g}-k)$ \emph{%
nonholonomic constraint equations }eq. $\left( \ref{Eq. D constraint for psi}%
\right) \ $also for $\xi (t)$.

Before passing to the next section, we give some properties of the subspaces
which are involved in $\left( \ref{Eq. motion for psi}\right) $ and which
follow from the $G-$invariance of $D$.$\ $Recall that, in general, $%
\mathfrak{g}^{q}:=\{X\in \mathfrak{g},\ X_{Q}(q)\in D_{q}\}$ and $i_{q}:%
\mathfrak{g}^{q}\hookrightarrow \mathfrak{g}$ denotes the inclusion.

\begin{proposition}
\label{prop:properties of g^q}The following holds:

\begin{itemize}
\item $\mathfrak{g}^{g\cdot q}=Ad_{g}\mathfrak{g}^{q}$

\item $Ad_{g}\circ i_{q}=i_{g\cdot q}\circ Ad_{g}$
\end{itemize}
\end{proposition}

\begin{example}
\label{ex:Purely kinematical case} $\emph{(The}$ \emph{purely Kinematical
case of }\cite{BKMM}\emph{) }In this case, $D\cap TOrb_{G}(Q)$ is trivial
and thus $\mathfrak{g}^{q}=\{0\}$ for all $q\in Q$. Eq. of motion $\left( %
\ref{Eq. motion c(t)}\right) $ is trivial and the motion of the system is
only determined by the constraint equation $\dot{c}\in D$. See also section %
\ref{subsec: PK phases}.
\end{example}

\begin{example}
\label{ex:Horizontal symmetries copy} $\emph{(The}$ \emph{case of Full
Horizontal Symmetries }\cite{BKMM}\emph{) }In this case, there exists a
subgroup $H\subset G$ such that $\mathfrak{g}^{q}$ is constantly $\mathfrak{h%
}=Lie(H)$ for all $q\in Q$. Then, $i_{\mathfrak{h}}^{\ast }(J(\overset{\cdot 
}{c}))$ is a conserved quantity along the solution $c(t)$. See also section %
\ref{subsubsec:Phases Horizontal symmetries}.
\end{example}

\begin{example}
\label{ex:Eqs with No D Constraints} $\emph{(The}$ \emph{case }$D=TQ$ \emph{%
and momentum conservation) }When $D=TQ$ and so the $D-$constraints are
trivial, equations $\left( \ref{Eq. motion c(t)}\right) $ (equivalently, $%
\left( \ref{Eq. motion for psi}\right) $) imply the conservation of the
momentum $J$ along the solution $c(t)$. This is the case, for example, of a
self deforming body which freely rotates around its center of mass with
conserved angular momentum (\cite{C def body,SW}).
\end{example}

\subsubsection{Applied Forces}

\label{rmk: external forces c(t)}In the presence of arbitrary additional
external forces $F:TQ\longrightarrow \mathbb{R}$, the corresponding
equations of motion are%
\begin{equation*}
i_{c(t)}^{\ast }(\frac{d}{dt}J(\overset{\cdot }{c}))=i_{c(t)}^{\ast }\circ
\sigma _{c(t)}^{\ast }(F_{c(t)})\ 
\end{equation*}%
where $\sigma _{c(t)}:\mathfrak{g}\longrightarrow T_{c(t)}Q$ denotes the
infinitesimal $G-$action on $Q$ along $c(t)$.

Now, equation $\left( \ref{Eq. motion c(t)}\right) $ can be rewritten as%
\begin{equation*}
\frac{d}{dt}J(\overset{\cdot }{c})=\Gamma _{c}(t)\ 
\end{equation*}%
for a curve $\Gamma _{c}(t)\in Ker(i_{c(t)}^{\ast })$. This $\Gamma _{c}(t)$
is fixed by the $D-$constraint equations and can be interpreted as an 
\textbf{external (generalized)\ torque} caused by the forces implementing
the $D-$constraints (see example \ref{subsubsec:Example dipolar magnetic}).

Within the gauge covariant formulation, the corresponding equations of
motion are%
\begin{equation}
\frac{d}{dt}\left( I(d_{0}(t))\xi +J(\overset{\cdot }{d_{0}})(t)\right)
+ad_{\xi }^{\ast }\left( I(d_{0}(t))\xi +J(\overset{\cdot }{d_{0}}%
)(t)\right) =\sigma _{d_{0}(t)}^{\ast }\left( \rho _{g\ast d_{0}(t)}^{\ast
}F_{c(t)}\right) =:\Gamma _{d_{0}}(t).  \label{Eq:external forces d0}
\end{equation}

Assuming that there are no external forces and that $\left( DH1,2\right) $
hold, we arrive at eq. $\left( \ref{Eq:external forces d0}\right) $ with the
forces $F=F^{D}$ representing the $D-$constraint forces acting on the
system. Notice that, since $\left( DH1\right) $ holds, eq. $\left( \ref{Eq.
motion for psi}\right) $ above follows by projecting via $i_{d_{0}(t)}^{\ast
}:\mathfrak{g}^{\ast }\hookrightarrow (\mathfrak{g}^{d_{0}(t)})^{\ast }$. If
we choose a splitting $\mathfrak{g}=\mathfrak{g}^{d_{0}(t)}\oplus \mathfrak{O%
}^{d_{0}(t)}$ with $P^{\mathfrak{O}}:\mathfrak{g}\longrightarrow \mathfrak{g}%
^{d_{0}(t)}$ the corresponding projector, we get that the \emph{external
torque }$\Gamma _{d_{0}}(t)\in Ker$ $i_{d_{0}(t)}^{\ast }$ present in the
r.h.s. of eq. $\left( \ref{Eq:external forces d0}\right) $ can be also
written as\footnote{%
The above expression yields the same $\Gamma _{d_{0}}(t)$ for any choice of $%
P^{\mathfrak{O}}$ since $F_{c(t)}$ vanishes on $D\subset TQ$ by $\left(
DH1\right) $.}%
\begin{equation*}
\Gamma _{d_{0}}(t)=\left( 1-P^{\mathfrak{O}}\right) ^{\ast }\circ \sigma
_{d_{0}(t)}^{\ast }\left( \rho _{g\ast d_{0}(t)}^{\ast }F_{c(t)}^{D}\right) .
\end{equation*}%
Expression $\left( \ref{Eq:external forces d0}\right) $ gives $dim\mathfrak{g%
}$ equations coupled to the $(dim\mathfrak{g-}k)$ equations of $D-$%
constraints. Nevertheless, notice that in $\left( \ref{Eq:external forces d0}%
\right) $ we have $\left( dim\mathfrak{g-}k\right) $ new unknowns: the $D-$%
constraint forces $F^{D}$.

\subsubsection{Bundle Formulation}

\label{subsubsec:Bundle formulation}The gauge invariant formulation $\left(
P1\right) $ and the gauge fixed formulation $\left( P2\right) $ of the
problem, both have as underlying $G-$bundle $Q_{I}\longrightarrow I$ which
is related to $Q\longrightarrow Q/G$ by the pull-back diagram%
\begin{equation*}
\begin{array}{ccc}
Q_{I} & \longrightarrow & Q \\ 
\downarrow &  & \downarrow \\ 
I & \overset{\tilde{c}}{\longrightarrow } & Q/G.%
\end{array}%
\end{equation*}

Moreover, $Q_{I}$ is a trivial $G-$bundle and the corresponding global
sections are the \emph{gauge curves} $d_{0}(t)$ projecting to $\tilde{c}(t)$
on shape space. Choosing a section, so $Q_{I}\approx I\times G$, we arrive
at the non autonomous system on $G\ $as described by $\left( P2\right) $.

\begin{remark}
\label{rmk: 1d fields}\emph{(Relation to }$1-d$\emph{\ gauge field theories) 
}The setting above gives a description of our time-dependent problem in
terms of a $1-$dimensional \emph{gauge field theory}. Here, the \emph{fields 
}are the sections $I\longrightarrow Q_{I}\approx I\times G$ and $G$ is the
gauge group. See also \cite{Mont gauge} and references therein. Notice that,
in this context, the corresponding field theory is not gauge invariant
since, actually, the problem consists in finding the correct gauge taking $%
d_{0}$ into the desired solution $c=g\cdot d_{0}$.
\end{remark}

There is also another set of bundles which are relevant for this problem,
specially for the study of the equations of motion. These are the vector
bundles $\mathfrak{g}^{D}$, $\mathfrak{g}_{I}^{D}$. These are related by the
pull back diagram%
\begin{equation*}
\begin{array}{ccc}
\mathfrak{g}_{I}^{D} & \longrightarrow & \mathfrak{g}^{D} \\ 
\downarrow &  & \downarrow \\ 
I & \overset{\tilde{c}}{\longrightarrow } & Q%
\end{array}%
\end{equation*}%
with $\mathfrak{g}^{D}=\sqcup _{q\in Q}\mathfrak{g}^{q}$. The vector bundle $%
\mathfrak{g}^{D}$ can be also defined as $\sigma ^{-1}(D)$, for the vector
bundle morphism%
\begin{eqnarray*}
\sigma &:&Q\times \mathfrak{g}\longrightarrow TQ \\
&:&(q,X)\longmapsto X_{Q}(q)
\end{eqnarray*}%
with $D\subseteq TQ$ seen as a vector subbundle. Note that bundle $\mathfrak{%
g}_{I}^{D}$ is also trivial since $I$ is contractible. For a given choice of
gauge curve $d_{0}(t)$, there must exist a smooth curve $T(t)\in GL(%
\mathfrak{g})$ such that the set%
\begin{equation}
\{T(t)X_{i}\}_{i=1}^{dim\mathfrak{g}^{d(t_{1})}}  \label{Eq. sections of gD}
\end{equation}%
is a basis of $\mathfrak{g}^{d_{0}(t)}$ if $\{X_{i}\}_{i=1}^{dim\mathfrak{g}%
^{d(t_{1})}}$ is a basis of the vector space $\mathfrak{g}%
^{d_{0}(t_{1})}\subseteq \mathfrak{g}$.

This is the pull-back (to $\mathfrak{g}_{I}^{D}$) version of the moving
basis formulation of \cite{BKMM,CMR1}.

\begin{remark}
\emph{(Vector Bundle non triviality)\ }From example \ref{ex:Horizontal
symmetries copy} we see that the geometry of the bundle $\mathfrak{g}^{D}$
plays a crucial role in the form of the equations of motion. In other words,
the geometry of $\mathfrak{g}^{D}$ enters in the \emph{noncommutativity} of $%
\frac{d}{dt}$ and \ $i_{c(t)}^{\ast }$ in eq. $\left( \ref{Eq. motion c(t)}%
\right) $. Even though the bundle $\mathfrak{g}_{I}^{D}$ is always
trivializable, if it is not directly \emph{trivial}, the need of using time
dependent sections $T(t)$ enters non-trivially in the equations of motion.
See also sections \ref{subsec: Trivial Q case} and \ref{subsec: G abelian}
where this effect is isolated from others.
\end{remark}

\subsubsection{Non-Holonomic Gauges}

\label{subsubsec:Non-holonomic gauge}Recall the constraint equations $\left( %
\ref{Eq. D constraint for psi}\right) $ which are coupled to the motion ones 
$\left( \ref{Eq. motion for psi}\right) $. Being explicitly \emph{gauge
dependent}, a natural question that follows is: Is there a gauge, i.e. a
choice of $d_{0}(t)$, which simplifies these equations?

For eq. $\left( \ref{Eq. D constraint for psi}\right) $ we see that if $%
d_{0}(t)$ satisfies%
\begin{equation}
\frac{d}{dt}d_{0}(t)\in D_{d_{0}(t)},\ \forall t\in I  \label{Eq. d0 in D}
\end{equation}%
then, $\left( \ref{Eq. D constraint for psi}\right) $ is equivalent to the
simpler condition 
\begin{equation}
\xi (t)\in \mathfrak{g}^{d_{0}(t)}.  \label{Eq. D const for psi in NHgauge}
\end{equation}

We shall call a gauge $d_{0}$ satisfying $\left( \ref{Eq. d0 in D}\right) $
a \textbf{non-holonomic gauge }and denote it as $d_{0}^{NH}$.

Following \cite{BKMM}, given the base curve $\tilde{c}(t)\in Q/G$, a
geometrically defined candidate for non-holonomic gauge $d_{0}^{NH}$
fulfilling eq. $\left( \ref{Eq. d0 in D}\right) $ is given by the \textbf{%
horizontal lift }of $\tilde{c}(t)$ with respect to the \textbf{non-holonomic
connection}. The gauge $d_{0}^{NH}$ obtained in this way is defined by%
\begin{equation}
\begin{array}{c}
\left( i_{d_{0}^{NH}}^{\ast }I(d_{0}^{NH}(t))i_{d_{0}^{NH}}\right)
^{-1}\left( i_{d_{0}^{NH}}^{\ast }J(\dot{d}_{0}^{NH}(t))\right) =0 \\ 
\mathcal{A}_{d_{0}^{NH}}^{Kin}(\overset{\cdot }{d_{0}^{NH}}(t))=0%
\end{array}
\label{Eq. nonholonomic gauge relation}
\end{equation}%
where $i_{d_{0}^{NH}}^{\ast }I(d_{0}^{NH}(t))i_{d_{0}^{NH}}:\mathfrak{g}%
^{d_{0}^{NH}}\longrightarrow \left( \mathfrak{g}^{d_{0}^{NH}}\right) ^{\ast
}\ $and $\mathcal{A}^{Kin}:TQ\longrightarrow \mathfrak{U}$ denotes a $%
\mathfrak{U}-$valued 1-form that projects $\mathfrak{U}_{q}$ onto itself and
has $D_{q}$ as kernel. The subbundle $\mathfrak{U}\subset TQ$ can be defined
to be, at each $q\in Q$, the (kinetic energy metric) orthogonal complement
of $\left( \mathfrak{g}^{q}\right) _{Q}(q)$ within the subspace $T_{q}\left(
Orb_{G}(q)\right) $: $T_{q}\left( Orb_{G}(q)\right) =\left( \mathfrak{g}%
^{q}\right) _{Q}(q)\overset{\bot }{\oplus }\mathfrak{U}_{q}$ (see \cite{BKMM}
for details).

In this case, the gauge factor $d_{0}^{NH}(t)$ of the solution $c(t)$ can be 
\emph{kinematically determined} from the base-controlled dynamics%
\'{}
$\tilde{c}(t)$.

In an non-holonomic gauge, we also have

\begin{proposition}
\label{prop:nonholonomic gauge} Let $d_{0}^{NH}(t)$ be a \textbf{%
non-holonomic gauge} and define the \emph{non-holonomic body momentum} by 
\begin{equation}
\Pi (t):=I(d_{0}^{NH}(t))(g^{-1}\overset{\cdot }{g}(t))+J(\overset{\cdot }{%
d_{0}^{NH}})(t).  \label{Eq: def Pi non holonomic}
\end{equation}%
The following holds:
\end{proposition}

\begin{itemize}
\item $J(\overset{\cdot }{c})=Ad_{g(t)}^{\ast }\Pi (t)$, so $\Pi (t)$
represents the momentum as seen from the \emph{moving reference frame}
defined by $g(t)$ in $Q$,

\item the \textbf{constraints} read $g^{-1}\overset{\cdot }{g}(t)\in 
\mathfrak{g}^{d_{0}^{NH}(t)}$,

\item the \textbf{reconstruction} of $g(t)$ from $i_{d_{0}^{NH}}^{\ast }\Pi
(t)$ is:%
\begin{gather*}
g^{-1}\overset{\cdot }{g}(t)=\left( i_{d_{0}^{NH}}^{\ast }\circ
I(d_{0}^{NH}(t))\circ i_{d_{0}^{NH}}\right) ^{-1}\left( i_{d_{0}^{NH}}^{\ast
}\Pi (t)-i_{d_{0}^{NH}}^{\ast }J(\overset{\cdot }{d_{0}^{NH}})(t)\right) \\
\overset{Eq.\left( \ref{Eq. nonholonomic gauge relation}\right) }{=}\left(
i_{d_{0}^{NH}}^{\ast }\circ I(d_{0}^{NH}(t))\circ i_{d_{0}^{NH}}\right)
^{-1}i_{d_{0}^{NH}}^{\ast }\Pi (t),
\end{gather*}

\item the equation of motion for $\Pi (t)$ reads%
\begin{equation*}
i_{d_{0}^{NH}(t)}^{\ast }\left( \frac{d}{dt}\Pi (t)+ad_{\left(
i_{d_{0}^{NH}}^{\ast }\circ I(d_{0}^{NH}(t))\circ i_{d_{0}^{NH}}\right)
^{-1}\left( i_{d_{0}^{NH}}^{\ast }\Pi (t)\right) }^{\ast }\Pi (t)\right) =0,
\end{equation*}

\item which is coupled to the constraint equation for $\Pi (t)$:%
\begin{equation*}
\ I_{0}^{-1}(t)(\Pi (t)-J(\overset{\cdot }{d_{0}^{NH}}))\in \mathfrak{g}%
^{d_{0}^{NH}(t)}.
\end{equation*}
\end{itemize}

\begin{remark}
\emph{(No constraints and the Mechanical gauge)\ }Note that when $D=TQ$,
i.e., when there are no constraints, the non-holonomic connection coincides
with the \textbf{mechanical connection} (see for example \cite{Mont gauge}
and references therein)\ and, thus, the non-holonomic gauge reduces to the 
\textbf{mechanical gauge} $d_{0}^{Mech}$ defined by 
\begin{equation}
J(\overset{\cdot }{d_{0}^{Mech}}(t))=0.  \label{Eq: Mech gauge}
\end{equation}
\end{remark}

\section{Special Cases}

\label{sec: special cases}

\subsection{The conserved momentum case}

\label{subsubsec: Cons Mom Case}Here we describe base controlled systems
with no additional $D-$constraints, but whose motion is governed by \emph{%
momentum conservation}. This case encodes an important class of systems in
which the fiber motion is induced from the base in order to keep the
momentum constant, so we shall give a detailed description of the underlying
Hamiltonian structure. In section \ref{subsubsec:Reconstruction Cons Mom},
we shall use this description to the study of reconstruction phases for this
systems.

\label{subsec:setting}\label{subsec:Eqsmotion}There are two ways of encoding
this \emph{conserved momentum} case in the general $D-$constrained case
described above. One is to think that $D=TQ$ and the momentum $J$ as giving
conserved quantities due to \emph{horizontal symmetries} of the whole $G$
(see \cite{BKMM} and sec. \ref{subsubsec:Phases Horizontal symmetries}
below). Another, is to think 
\begin{equation*}
J(\dot{c})=\mu =const.
\end{equation*}%
as an \emph{affine constraint on the system }(see sec. \ref{subsec:affine
constraints} below). Both strategies lead to the same results that we shall
derive below in a (third possible)\ direct way, by analyzing the
corresponding equations of motion.

\label{subsubsec:dynam hyp}Since no $D-$constraints are present in the
system, we only need to assume $\left( H2\right) $ and $\left( DH2\right) $
from sections \ref{subsec:Kinematical setting} and \ref{subsec: D Dyn Hyp},
respectively. From these, using the variational techniques of \ref{subsec:
Variational principle}, it follows that the momentum map $J$ is conserved
along the physical motion of the system $c(t)\in Q$, i.e.%
\begin{equation}
\frac{d}{dt}J(\dot{c}(t))=0,\ \forall t.  \label{Eq conservac. J}
\end{equation}

\label{subsubsec:Eqs g(t)}The \textbf{non autonomous, second order equations}
\textbf{of motion} for $g(t)$, derived from $\left( \ref{Eq conservac. J}%
\right) $ read 
\begin{eqnarray}
0 &=&ad_{\xi (t)}^{\ast }(I_{0}(t)\xi (t)+J_{0}(t))+I_{0}(t)\frac{d}{dt}(\xi
(t))  \label{Eq g(t)} \\
&&+\frac{d}{dt}(I_{0}(t))\xi (t)+\frac{d}{dt}J_{0}(t)  \notag
\end{eqnarray}%
where we have denoted 
\begin{equation*}
\xi (t)=g^{-1}\frac{d}{dt}g(t)\in \mathfrak{g}
\end{equation*}%
and the initial values $(g(t_{1}),\overset{\cdot }{g}(t_{1}))$ must be such
that $g(t_{1})\cdot d_{0}(t_{1})=c(t_{1})$, $\frac{d}{dt}(g\cdot
d_{0})(t_{1})=\overset{\cdot }{c}(t_{1})$ are the \emph{initial values of
our mechanical problem}. We shall now focus on the Hamiltonian structure of
the equations of motion.

First, note that $I_{q}$ is a linear isomorphism for each $q\in Q$ and
defines a symmetric scalar product $(,)$ on $\mathfrak{g}$ by $\left(
X,Y\right) =\left\langle I_{q}X,Y\right\rangle $. Let $d_{0}$ denote any
gauge. If we call%
\begin{equation}
\Pi =I_{0}(t)\xi +J_{0}(t),  \label{eq. def Pi}
\end{equation}%
the map sending $\xi \mapsto \Pi $, which can be seen as a \emph{time
dependent Legendre transformation}, is invertible for all $t$. In fact, 
\begin{equation}
\xi (t)=I_{0}^{-1}(t)(\Pi -J_{0}(t)).  \label{Eq: Cons mom reconstr}
\end{equation}%
We also see that%
\begin{equation*}
J(\frac{d}{dt}c(t))=Ad_{g(t)}^{\ast }\Pi (t)
\end{equation*}%
and, hence, $\left( \ref{Eq conservac. J}\right) $ is equivalent to%
\begin{equation}
\frac{d}{dt}\Pi (t)=-ad_{I_{0}^{-1}(t)(\Pi (t)-J_{0}(t))}^{\ast }\Pi (t).
\label{Ec Pi}
\end{equation}

\label{subsec:Eq T*G}We will now transform eq. $\left( \ref{Eq g(t)}\right) $
to first order non autonomous equations on $T^{\ast }G$ making use of
underlying geometrical structures. Recall that $T^{\ast }G$ is isomorphic as
a vector bundle to $G\times \mathfrak{g}^{\ast }$ via \emph{left translations%
}, i.e., by taking \emph{body coordinates }(\cite{AM,MR}). Also recall the
two maps $G\times \mathfrak{g}^{\ast }\overset{L}{\underset{\pi }{%
\rightrightarrows }}\mathfrak{g}^{\ast }$ given by $L(g,\Pi )=Ad_{g}^{\ast
}\Pi $ and $\pi (g,\Pi )=\Pi $.

We can now state the following

\begin{proposition}
\label{prop:J-1}Let $g(t)$ be a curve in $G$ and $\Pi (t)=I_{0}(t)g^{-1}%
\frac{d}{dt}g(t)+J_{0}(t)$. The curve $g(t)$ is a solution of $\left( \ref%
{Eq g(t)}\right) $ iff the curve $(g(t),\Pi (t))$ is an integral curve of
the time dependent vector field 
\begin{equation*}
X(g,\Pi ,t)=(g(I_{0}^{-1}(t)(\Pi -J_{0}(t))),-ad_{I_{0}^{-1}(t)(\Pi
-J_{0}(t))}^{\ast }\Pi )
\end{equation*}%
on $G\times \mathfrak{g}^{\ast }$($\sim T^{\ast }G$). In this case, if $%
L(g(t_{1}),\Pi (t_{1}))=\mu $, then $(g(t),\Pi (t))\in L^{-1}(\mu )\approx G$
for all $t\in I$.
\end{proposition}

\begin{remark}
\emph{(Time dependent reduction)} Recall that we started with an, a priori, $%
2\times dimG$ dimensional problem, defined by the non-autonomous second
order equation $\left( \ref{Eq g(t)}\right) $ for $g(t)$. Now, due to the
conservation of the momentum $J$, we were also able to reduce the
dimensionality to $dimG=dim(L^{-1}(\mu ))$ because $\Pi (t)$ must be $%
Ad_{g^{-1}(t)}^{\ast }\mu $. See also the next subsection.
\end{remark}

Note that, from the above proposition (equiv. form eq. $\left( \ref{Ec Pi}%
\right) $) we have that%
\begin{equation*}
\Pi (t)\in O_{\mu }\subset \mathfrak{g}^{\ast }
\end{equation*}%
with $O_{\mu }$ denoting the $G-$\textbf{coadjoint orbit} through $\mu $ in $%
\mathfrak{g}^{\ast }$.

Finally, to solve for $g(t)\in G$,

\begin{enumerate}
\item we have to solve the \textbf{non autonomous first order differential
equation} $\left( \ref{Ec Pi}\right) $ on $O_{\mu }$ to obtain $\Pi (t)$ and

\item then \textbf{reconstruct }$g(t)$ from $\Pi (t)$ in the $G_{\mu }-$%
bundle $L^{-1}(\mu )\approx G\longrightarrow O_{\mu }$.
\end{enumerate}

This last step is studied in section \ref{subsubsec:Reconstruction Cons Mom}
below (see also Appendix \ref{sec:App Reconstr}).

\subsubsection{Hamiltonian structure for the time dependent system}

We shall now add \emph{time }and \emph{energy} variables to the above
non-autonomous equations on $T^{\ast }G$ in order to get a usual \emph{%
Hamiltonian }structure.

Let us then consider the \emph{extended phase space }$P_{E}=T^{\ast
}(G\times \mathbb{R})\simeq T^{\ast }G\times \mathbb{R\times R}^{\ast }$
with its standard \emph{symplectic structure} $\Omega $. By taking \emph{%
body coordinates }on $T^{\ast }G$, i.e., by trivializing via left
translations, 
\begin{equation*}
P_{E}\overset{L^{\ast }}{\simeq }G\times \mathfrak{g}^{\ast }\times \mathbb{%
R\times R}^{\ast },
\end{equation*}%
the form $\Omega $ becomes, for $(g,\Pi ,t,E)\in G\times \mathfrak{g}^{\ast
}\times \mathbb{R\times R}^{\ast }$,%
\begin{equation*}
\Omega _{(g,\Pi ,t,E)}^{L}=-\left\langle d\Pi \hat{,}g^{-1}dg\right\rangle
+\left\langle \Pi ,[g^{-1}dg\overset{\otimes }{,}g^{-1}dg]\right\rangle
+\left\langle dt\hat{,}dE\right\rangle .
\end{equation*}

\begin{remark}
\emph{(The new variable }$E$\emph{) }The last term in the above expression,
tells us that $E$ is the \emph{momentum conjugated to time }$t$. Adding this
momentum is the usual way of taking into Hamiltonian form time dependent
systems (see \cite{AM}).
\end{remark}

Next, we consider the \emph{Hamiltonian function }$H:P_{E}\longrightarrow 
\mathbb{R}$, given by%
\begin{equation*}
H_{E}(g,\Pi ,t,E)=\frac{1}{2}\left\langle \Pi ,I_{0}^{-1}(t)\Pi
\right\rangle -\left\langle \Pi ,I_{0}^{-1}(t)J_{0}(t)\right\rangle
+E+T_{0}(t)
\end{equation*}%
with 
\begin{equation*}
T_{0}(t)=K_{int}(t)-\frac{1}{2}\left\langle
I_{0}^{-1}(t)(J_{0}(t)),J_{0}(t)\right\rangle
\end{equation*}%
and where $K_{int}$ denotes the \emph{internal kinetic energy }defined in
Appendix \ref{sec:App Kin En}.

The equations of motion corresponding to the \textbf{Hamiltonian system }$%
(P_{E},\Omega ^{L},H_{E})$ for the desired solution curve $\gamma (s)=(g,\Pi
,t,E)(s)\in P_{E}$ are 
\begin{equation}
\left\{ 
\begin{array}{c}
g^{-1}\frac{d}{ds}g=I_{0}^{-1}(t)(\Pi -J_{0}(t)) \\ 
\frac{d}{ds}\Pi =-ad_{I_{0}^{-1}(t)(\Pi -J_{0}(t))}^{\ast }\Pi \\ 
\frac{dt}{ds}=1 \\ 
-\frac{dE}{ds}=\frac{1}{2}\left\langle \Pi ,\frac{d}{ds}(I_{0}^{-1}(t))\Pi
\right\rangle -\left\langle \Pi ,\frac{d}{ds}(I_{0}^{-1}(t)J_{0}(t))\right%
\rangle +\frac{d}{ds}T_{0}(t).%
\end{array}%
\right. \bigskip  \label{Eq Hamilt t E}
\end{equation}%
The above third equation, tells us that $\frac{d}{ds}=\frac{d}{dt}$ and that
if we choose as \emph{initial value} $t_{1}(s_{1})=s_{1},$ then $s=t$. Thus,
the first two eqs. above become $\left( \ref{Eq: Cons mom reconstr}\right) $
and $\left( \ref{Ec Pi}\right) $ respectively. Equivalently, they say that $%
(g,\Pi )(s=t)$ is an integral curve of the time dependent vector field on $%
T^{\ast }G$ of Proposition \ref{prop:J-1}. The last equation for $E$, endows
this \emph{momentum conjugated to time} with a physical interpretation in
terms of the \emph{kinetic energy }$K(\frac{d}{dt}c(t))$ of the mechanical
system on $Q$\emph{\ }(see Appendix \ref{sec:App Kin En}):%
\begin{equation*}
E=-K(\frac{d}{dt}c(t))+\left\langle \Pi
(t),I_{0}^{-1}(t)J_{0}(t)\right\rangle .
\end{equation*}

\begin{remark}
\emph{(Non conservation of Kinetic Energy)} Note that though the Hamiltonian 
$H$ is a conserved quantity along the solutions of $\left( \ref{Eq Hamilt t
E}\right) $ on $P_{E}$, it does \emph{not }represent the kinetic energy of
the original mechanical system on $Q$. In general, the time-dependent
control forces on the base variables do work on the system, implying that
the \emph{energy is not conserved}. Also notice that thus, in general, the
variable $E$ will not be a conserved quantity, but it will obey non trivial
dynamics. See also Appendix \ref{sec:App Kin En} for more details.
\end{remark}

\begin{remark}
\emph{(Mechanical gauge)} In the \textbf{mechanical gauge }$\left( \ref{Eq:
Mech gauge}\right) $, i.e., when $d_{0}(t)$ is horizontal with respect to
the mechanical connection on $Q$, we obtain%
\begin{equation*}
E=-K(\frac{d}{dt}c(t)).
\end{equation*}
\end{remark}

\begin{remark}
\emph{(Symmetries of }$(P_{E},\Omega ^{L},H_{E})$\emph{) }The $G$ action on $%
P_{E}$ given by%
\begin{equation*}
(g,\Pi ,t,E)\overset{h\cdot }{\longrightarrow }(hg,\Pi ,t,E)
\end{equation*}%
is \emph{Hamiltonian}. The corresponding conserved momentum map is 
\begin{eqnarray*}
L_{E} &:&P_{E}\longrightarrow \mathfrak{g}^{\ast } \\
&:&(g,\Pi ,t,E)\mapsto Ad_{g}^{\ast }\Pi .
\end{eqnarray*}%
The corresponding \emph{reduced space }(\cite{AM}) is%
\begin{equation*}
L_{E}^{-1}(\mu )/G_{\mu }=O_{\mu }\times \mathbb{R\times R}^{\ast }
\end{equation*}%
giving the Hamiltonian configuration for the \emph{non autonomous} equations 
$\left( \ref{Ec Pi}\right) $ on $O_{\mu }$ described above.\bigskip
\end{remark}

\subsection{Affine $D-$Constraints}

\label{subsec:affine constraints}In this subsection we shall follow \cite%
{BKMM} and \cite{CMR1} to show how to handle \emph{affine }$D-$\emph{%
constrained controlled }systems. By an affine $D-$constraint we mean one of
the type%
\begin{equation}
\mathfrak{A}_{q}^{D}(\dot{q})=\gamma (q,t)  \label{Eq: aff constr}
\end{equation}%
where $\mathfrak{A}^{D}:TQ\longrightarrow TQ$ is a linear fiber projector
defining an Eheresmann connection with $Ker\mathfrak{A}^{D}=D\subset TQ$. We
shall denote, as usual, the vertical subbundle by $V=Im\mathfrak{A}%
^{D}\subset TQ$. The field $\gamma (q,t)$ is then vertical valued, that is, $%
\gamma (q,t)\in V_{q}$ $\forall q,t$. Since our setting involves the
geometry of the principal $G-$bundle\ $Q\longrightarrow Q/G$, we assume the
following compatibility conditions to hold:

\begin{itemize}
\item[$\left( i\right) $] $\mathfrak{A}^{D}$ is $G-$invariant, that is $\rho
_{g\ast q}\circ \mathfrak{A}_{q}^{D}=\mathfrak{A}_{g\cdot q}^{D}\circ \rho
_{g\ast q}$,

\item[$\left( ii\right) $] $\gamma $ is $G-$invariant, that is $\gamma
(g\cdot q,t)=\rho _{g\ast q}\gamma (q,t).$
\end{itemize}

From the $G-$invariance of $\mathfrak{A}^{D}$ follows the $G-$invariance of $%
D$. We further assume the \emph{dimension condition }on $D$, namely, $(H1)$
of section \ref{subsec:Kinematical setting}. Now, we consider the \emph{%
affine} version of the Lagrange-D 
\'{}%
alambert principle present in \cite{CMR1}:

\begin{description}
\item[PAff] The curve $q(t)\ $is a solution to the above stated nonholonomic
affine constrained system iff $\dot{q}(t)$ satisfies the affine constraints $%
\left( \ref{Eq: aff constr}\right) $ and if for any variation $q(t,s)$ with
fixed end-points such that $\delta q\in D_{q}$, then%
\begin{equation*}
\delta \dint\limits_{t_{1}}^{t_{2}}L(q,\dot{q})dt=0.
\end{equation*}%
As in section \ref{subsec: Variational principle}, we adapt this variational
formulation to the \emph{base controlled case }by considering only \emph{%
vertical} variations $c(t,s)=g(t,s)\cdot d_{0}(t)$ for some gauge $d_{0}(t)$.
\end{description}

From this, it follows

\begin{proposition}
\label{prop:affine equations} The equations for $g(t)$ in order for $%
c(t)=g(t)\cdot d_{0}(t)$ to be a solution for the affine constrained and
controlled system satisfying $(i)$ and $(ii)$ described above are:\ the 
\emph{equation of motion}%
\begin{equation}
i_{d_{0}(t)}^{\ast }\left( \frac{d}{dt}\left( I(d_{0}(t))\xi +J(\overset{%
\cdot }{d_{0}})(t)\right) +ad_{\xi }^{\ast }\left( I(d_{0}(t))\xi +J(\overset%
{\cdot }{d_{0}})(t)\right) \right) =0  \label{Eq. motion affine}
\end{equation}%
with $\xi =g^{-1}\dot{g}$, coinciding with eq. $\left( \ref{Eq. motion for
psi}\right) $ for the linear (non affine) constraint case, and the \emph{%
constraint equation}%
\begin{equation}
\mathfrak{A}_{d_{0}(t)}^{D}\left[ \overset{\cdot }{d}_{0}(t)+(g^{-1}\dot{g}%
)_{Q}(d_{0}(t))\right] =\gamma (d_{0}(t),t).
\label{Eq: aff constr controlled}
\end{equation}
\end{proposition}

The fact that the equation of motion for $g^{-1}\dot{g}$ is the same for the
affine and linear cases is already commented, in terms of the \emph{%
nonholonomic momentum equation}, in \cite{BKMM} (see page 27).

As before, we can simplify the constraint equation by choosing suitable
gauges $d_{0}$. In a \emph{nonholonomic gauge }$d_{0}^{NH}$, eq. $\left( \ref%
{Eq: aff constr controlled}\right) $ become%
\begin{equation*}
\mathfrak{A}_{d_{0}^{NH}(t)}^{D}\left[ (g^{-1}\dot{g})_{Q}(d_{0}^{NH}(t))%
\right] =\gamma (d_{0}^{NH}(t),t)
\end{equation*}%
because $\mathfrak{A}_{d_{0}^{NH}(t)}^{D}(\dot{d}_{0}^{NH}(t))=0$. But, if
we define an \textbf{affine nonholonomic gauge} $d_{0}^{Aff}\ $to be one
satisfying%
\begin{equation}
\mathfrak{A}_{d_{0}^{Aff}(t)}^{D}(\dot{d}_{0}^{Aff}(t))=\gamma
(d_{0}^{Aff}(t),t)  \label{Eq: Aff gauge}
\end{equation}%
then, eq. $\left( \ref{Eq: aff constr controlled}\right) $ reads, 
\begin{equation*}
g^{-1}\dot{g}\in \mathfrak{g}^{d_{0}^{Aff}(t)}
\end{equation*}%
which is simpler to handle. Notice that eq. $\left( \ref{Eq: Aff gauge}%
\right) $ plus the requirement $\pi \left( d_{0}^{Aff}(t)\right) =\tilde{c}%
(t)$ do not determine $d_{0}^{Aff}(t)$ uniquely since $dimD$ can be grater
than $dimB$. On the other hand, when the field $\gamma =0$, a nonholonomic
gauge is an affine gauge.

In section \ref{Ex: ball on rotating table}, we apply this general
considerations to study the motion of a \emph{controlled} ball on a rotating
turntable.

\subsection{The case $G$ abelian}

\label{subsec: G abelian}We now illustrate on the structure of the equations
in the case $G$ is abelian. This allows us to isolate the contribution to
the motion coming from the non-trivial geometry of the vector bundle $%
\mathfrak{g}^{D}$ from the Lie algebraic part of the equations of motion
(i.e. terms involving $ad$).

When $G$ is abelian, $Ad_{g}$ is the identity for all $g\in G$, and thus

\begin{itemize}
\item $\mathfrak{g}^{g\cdot q}=\mathfrak{g}^{q}\ \forall g\in G$, ie, the
subspaces $\mathfrak{g}^{q}$are vertically constant in $Q$, thus $\mathfrak{g%
}^{d_{0}(t)}=\mathfrak{g}^{c(t)}$ and $i_{c(t)}^{\ast }=i_{d_{0}(t)}^{\ast }$%
,

\item $I(g\cdot q)=I(q)$, thus $I(c(t))=I(d_{0}(t))$,

\item $J(\overset{\cdot }{c})=I(d_{0}(t))g^{-1}\overset{\cdot }{g}(t)+J(%
\overset{\cdot }{d_{0}})=:\Pi (t)$,

\item the equation of motion reads%
\begin{equation}
i_{d_{0}(t)}^{\ast }(\frac{d}{dt}J(\overset{\cdot }{c}))=i_{d_{0}(t)}^{\ast }%
\left[ \frac{d}{dt}\left( I(d_{0}(t))g^{-1}\overset{\cdot }{g}(t)+J(\overset{%
\cdot }{d_{0}})\right) \right] =0,  \label{Eq. motion abelian}
\end{equation}

\item the constraint equation in a \emph{non-holonomic gauge} stays as%
\begin{equation*}
g^{-1}\overset{\cdot }{g}(t)\in \mathfrak{g}^{d_{0}^{NH}(t)}=\mathfrak{g}%
^{c(t)}.
\end{equation*}
\end{itemize}

By eq. $\left( \ref{Eq. nonholonomic gauge relation}\right) $, the
constraint equation in terms of $J(\overset{\cdot }{c})$ reads%
\begin{equation}
\ I^{-1}(d_{0}^{NH}(t))(J(\overset{\cdot }{c})-J(\overset{\cdot }{d_{0}^{NH}}%
))=\left( i_{d_{0}^{NH}}^{\ast }\circ I(d_{0}^{NH})\circ
i_{d_{0}^{NH}}\right) ^{-1}\left( i_{d_{0}^{NH}}^{\ast }J(\overset{\cdot }{c}%
)\right) \in \mathfrak{g}^{d_{0}^{NH}(t)}.  \label{Eq. constraints J abel}
\end{equation}

\begin{remark}
\label{rmk:gD bundle for G abel}\emph{(Base of the }$\mathfrak{g}^{D}\ $%
\emph{bundle)\ }Since $\mathfrak{g}^{g\cdot q}=\mathfrak{g}^{q}$ for abelian 
$G$, the vector bundle $\mathfrak{g}^{D}\longrightarrow Q$ descends to a
vector bundle over the shape space $\mathfrak{g}^{D}\longrightarrow Q/G$. In
this context,\ the objects $i_{c(t)}^{\ast }=i_{d_{0}(t)}^{\ast }=i_{\tilde{c%
}(t)}^{\ast }$ and $I(c(t))=I(d_{0}(t))=I(\tilde{c}(t))$ really depend on
the base curve $\tilde{c}(t)\in Q/G$.
\end{remark}

Now, we want to re-write the equation of motion for the momentum $J(\overset{%
\cdot }{c})$ in a usual first order differential form. As in section \ref%
{subsubsec:Bundle formulation}, consider a linear isomorphism $T_{t}:%
\mathfrak{g}^{\ast }\overset{\sim }{\longrightarrow }\mathfrak{g}^{\ast }$
taking the initial fiber $\left( \mathfrak{g}^{d_{0}^{NH}(t_{1})}\right)
^{\ast }$ to $\left( \mathfrak{g}^{d_{0}^{NH}(t)}\right) ^{\ast }$, 
\begin{equation*}
i_{d_{0}^{NH}(t)}^{\ast }\circ T_{t}=T_{t}\circ i_{d_{0}^{NH}(t_{1})}^{\ast
}.
\end{equation*}%
Eq. $\left( \ref{Eq. motion abelian}\right) $ becomes%
\begin{equation}
\frac{d}{dt}\left( i_{d_{0}^{NH}(t)}^{\ast }J(\overset{\cdot }{c})\right) =[%
\overset{\cdot }{T}T^{-1},i_{d_{0}^{NH}(t)}^{\ast }]\left( J(\overset{\cdot }%
{c})\right) ,  \label{Eq. abel motion T}
\end{equation}%
which is equivalent to the corresponding expressions in terms of moving
basis of \cite{BKMM}.

The above equation states how the non-triviality of the bundle $\left( 
\mathfrak{g}^{D}\right) ^{\ast }$ affects the evolution of the projected
momentum $i_{d_{0}^{NH}(t)}^{\ast }J(\overset{\cdot }{c})$. Note that even
when the bundle is trivializable, but not directly trivial, the
corresponding equation of motion also contains non-zero $\overset{\cdot }{T}%
T^{-1}$ term.

\begin{remark}
\label{rmk: trivial gi conservat}\emph{(Trivial }$\mathfrak{g}_{I}^{D}\ $%
\emph{bundle)\ }Recall from section \ref{subsubsec:Bundle formulation}, that
the pull back vector bundle $\mathfrak{g}_{I}^{D}$ is always trivializable.
But, when it is directly \emph{trivial}, the above equation read%
\begin{equation*}
\frac{d}{dt}\left( i_{d_{0}^{NH}(t)}^{\ast }J(\overset{\cdot }{c})\right) =0
\end{equation*}%
so it gives a \textbf{conservation law} related to the given base curve $%
\tilde{c}(t)$.
\end{remark}

More explicitly, let $\{e_{d_{0}^{NH}(t)}^{i}\}_{i=1}^{dim\mathfrak{g}%
^{d_{0}^{NH}(t)}}$ be a (moving) basis for the fiber $\mathfrak{g}%
^{d_{0}^{NH}(t)}$ along the gauge curve $d_{0}^{NH}(t)$. Then, constraints $%
\left( \ref{Eq. abel motion T}\right) $ for $J(\overset{\cdot }{c})$ imply
that%
\begin{equation}
J(\overset{\cdot }{c})=\dsum\limits_{i=1}^{dim\mathfrak{g}%
^{d_{0}^{NH}(t)}}\lambda _{i}(t)\ I(d_{0}^{NH}(t))e_{d_{0}^{NH}(t)}^{i}+J(%
\overset{\cdot }{d_{0}^{NH}})  \label{Eq: G abel constr moving basis}
\end{equation}%
for some time dependent coefficients $\lambda _{i}(t)\in \mathbb{R}$ to be
determined. From $\left( \ref{Eq. motion abelian}\right) $, we have that the 
$\lambda _{i}(t)$%
\'{}%
s must satisfy%
\begin{equation*}
A(t)\ \overset{\cdot }{\overset{\longrightarrow }{\lambda }}(t)=-B(t)\overset%
{\longrightarrow }{\lambda }(t)\ -\overset{\longrightarrow }{c}(t)
\end{equation*}%
where the time dependent real $(dim\mathfrak{g}^{d_{0}^{NH}(t)}\times dim%
\mathfrak{g}^{d_{0}^{NH}(t)})$ matrices $A(t)\ $and $B(t)$ are defined by%
\begin{eqnarray*}
A_{ij}(t) &=&\left\langle
I(d_{0}^{NH}(t))e_{d_{0}^{NH}(t)}^{i},e_{d_{0}^{NH}(t)}^{j}\right\rangle
=:I_{ij}^{\{e_{d_{0}^{NH}(t)}^{k}\}} \\
B_{ij}(t) &=&\left\langle \frac{d}{dt}\left(
I(d_{0}^{NH}(t))e_{d_{0}^{NH}(t)}^{j}\right)
,e_{d_{0}^{NH}(t)}^{i}\right\rangle
\end{eqnarray*}%
and the $dim\mathfrak{g}^{d_{0}^{NH}(t)}$ real vector $\overset{%
\longrightarrow }{c}(t)$ by 
\begin{equation*}
c_{j}(t)=\left\langle \frac{d}{dt}J(\overset{\cdot }{d_{0}^{NH}}%
),e_{d_{0}^{NH}(t)}^{j}\right\rangle .
\end{equation*}

Note that $A$ is symmetric and invertible. If we solved these equations for $%
J(\overset{\cdot }{c})(t)$, then the reconstruction of $g(t)$ from it is
straightforward because, since $G$ is abelian, we can make use of the
exponential map $exp:\mathfrak{g}\longrightarrow G$, yielding%
\begin{eqnarray}
g(t) &=&exp\left( \dint\limits_{t_{1}}^{t}ds\ I(d_{0}^{NH})^{-1}\left( J(%
\overset{\cdot }{c})(s)-J(\overset{\cdot }{d_{0}^{NH}})(s)\right) \right)
\label{Eq: G abel solution} \\
&=&exp\left( \dint\limits_{t_{1}}^{t}ds\ \left( i_{d_{0}^{NH}}^{\ast }\circ
I(d_{0}^{NH})\circ i_{d_{0}^{NH}}\right) _{(s)}^{-1}\left(
i_{d_{0}^{NH}(s)}^{\ast }J(\overset{\cdot }{c})(s)\right) \right) .  \notag
\end{eqnarray}

\begin{remark}
\label{rmk: G abel mech phase}\emph{(Mechanical connection phase formula)\ }%
As $G$ is abelian, the above expression yields%
\begin{equation*}
c(t)=exp\left( \dint\limits_{t_{1}}^{t}ds\ I(d_{0})_{(s)}^{-1}J(\overset{%
\cdot }{c})(s)\right) \cdot g_{Mech}(t)\cdot d_{0}(t_{1})
\end{equation*}%
with%
\begin{equation}
g_{Mech}(t)=exp\left( -\dint\limits_{t_{1}}^{t}ds\ I(d_{0}^{NH})^{-1}J(%
\overset{\cdot }{d_{0}^{NH}})(s)\right)  \notag
\end{equation}%
such that $g_{Mech}(t)\cdot d_{0}^{NH}(t)=Hor_{Mech}(\tilde{c})(t)$ gives
the horizontal lift of $\tilde{c}(t)\in B$ with respect to the \emph{%
mechanical connection }$\left( \ref{Eq: Mech gauge}\right) $ (see also sec. %
\ref{subsubsec:Gauges and phases in Q}). Notice that the equation of motion
for $J(\overset{\cdot }{c})$ (but not the constraint equation\footnote{%
For a general non \emph{non-holonomic }gauge, constraint equation becomes
the \emph{gauge covariant} eq. $\left( \ref{Eq. D constraint for psi}\right) 
$.}) is the same in any gauge $d_{0}(t)$.
\end{remark}

Finally, to better understand how the geometry of the bundle $\mathfrak{g}%
^{D}$ enters the equations of motion for $J(\overset{\cdot }{c})$, we
restrict ourselves to the interesting case in which the horizontal space
with respect to the \emph{nonholonomic connection} is (kinetic energy
metric) orthogonal to the whole vertical subspace $TOrb_{G}$ within $TQ$. In
this case, a mechanical gauge $d_{0}(t)$, for which $J(\overset{\cdot }{d_{0}%
})=0$,\ is also a non-holonomic one and \bigskip eq. $\left( \ref{Eq. motion
abelian}\right) $ yields the \emph{parallel transport equation:}%
\begin{equation}
D_{\overset{\cdot }{d}_{0}}\overset{\longrightarrow }{p}\equiv \frac{d}{dt}%
p^{i}-\dsum\limits_{j=1}^{dim\mathfrak{g}^{d_{0}(t)}}\gamma _{j}^{i}\
p^{j}=0,\ \ \ \ \ \forall \ 1\leq i\leq dim\mathfrak{g}^{d_{0}(t)}
\label{Eq: G abel paral transp}
\end{equation}%
for%
\begin{equation*}
p^{i}(t):=\left\langle J(\overset{\cdot }{c}),e_{d_{0}(t)}^{i}\right\rangle
=\dsum\limits_{j=1}^{dim\mathfrak{g}^{d_{0}(t)}}\lambda _{j}(t)\
\left\langle I(d_{0}(t))e_{d_{0}(t)}^{j},e_{d_{0}(t)}^{i}\right\rangle
\end{equation*}%
$\ $being the coordinates of $J(\overset{\cdot }{c})$ in a basis of $%
\mathfrak{g}^{\ast }$ dual to a basis $\{e_{d_{0}(t)}^{k}\}_{1}^{dim%
\mathfrak{g}}$ for which%
\begin{equation}
\left\langle I(d_{0}(t))e_{d_{0}(t)}^{i},e_{d_{0}(t)}^{i%
{\acute{}}%
}\right\rangle =0\ \forall \ 1\leq i\leq dim\mathfrak{g}^{d_{0}(t)},\ dim%
\mathfrak{g}^{d_{0}(t)}+1\leq i%
{\acute{}}%
\leq dim\mathfrak{g}.  \label{Eq. ortog basis abel}
\end{equation}%
Note that, above, for $dim\mathfrak{g}^{d_{0}(t)}+1\leq i%
{\acute{}}%
\leq dim\mathfrak{g\ }$then $p^{i%
{\acute{}}%
}=0$ by the orthogonality condition $\left( \ref{Eq. ortog basis abel}%
\right) $ and because (iff) the constraints $\left( \ref{Eq: G abel constr
moving basis}\right) $ are fulfilled. The \emph{linear connection
coefficients }$\gamma _{k}^{i}$ are defined by 
\begin{equation*}
D_{\left( \overset{\cdot }{d}_{0}\right) }e_{d_{0}(t)}^{i}:=\frac{d}{dt}%
e_{d_{0}(t)}^{i}=\dsum\limits_{k=1}^{dim\mathfrak{g}}\gamma _{k}^{i}\
e_{d_{0}(t)}^{k}.
\end{equation*}%
Consequently, for this case, the time evolution of $J(\overset{\cdot }{c})$
is \emph{geometrically determined}, because it moves \textbf{%
parallel-transported} along the base curve $\tilde{c}(t)\in Q/G$ in the
bundle $\mathfrak{g}^{D}\longrightarrow Q/G$ of remark \ref{rmk:gD bundle
for G abel} (see also \cite{BKMM}).

On the other hand, as noticed in remark \ref{rmk: trivial gi conservat},
when the involved geometry is trivial, i.e. $\mathfrak{g}^{D}=Q\times V$
with constant $V\subset \mathfrak{g}$, then $i_{V}^{\ast }J(\overset{\cdot }{%
c})$ is a \textbf{conserved quantity}. Indeed, since $\mathfrak{g}$ is
abelian, $V$ defines a subalgebra and we are in the case described in
section \ref{subsubsec:Phases Horizontal symmetries}.

In section \ref{Ex: vertical disk abel}, we apply these general
considerations to study the motion of a \emph{base controlled} vertical
rotating disk.

\subsection{The trivial bundle case $Q=G\times B$}

\label{subsec: Trivial Q case}To illustrate on how the controlled base
variables induce motion on the group variables, we now focus on the case in
which $Q=G\times B$, i.e., $Q\longrightarrow Q/G$ is a trivial principal $G-$%
bundle. Recall that we are considering the natural \emph{left} $G-$action on 
$G\times B$. In this case, 
\begin{equation*}
TQ=TG\oplus TB
\end{equation*}%
and thus, by hypothesis $(H1)$ of section \ref{subsec:Kinematical setting},%
\begin{equation*}
D_{(b,g)}=T_{b}B\oplus S_{(b,g)}
\end{equation*}%
with $S_{(b,g)}=T_{g}G\cap D_{(b,g)}$ as usual. Note that, since $D$ is $G-$%
invariant, for each $b\in B$, $S_{(b,g)}$ defines a $G-$invariant
distribution on $G$ which, in turn, is fixed by the subspace $%
S_{(b,e)}\subset T_{e}G=\mathfrak{g}$. So $D$ is characterized by a smooth
map $B\longrightarrow Gr_{dimS}(\mathfrak{g}):$= $\{$Grassmanian of\ $dimS\ $%
subspaces of $\mathfrak{g}\}$ or, equivalently, by a vector bundle 
\begin{equation}
V=\underset{b\in B}{\cup }S_{(b,e)}\longrightarrow B.
\label{Eq:vector bundle for trivial Q}
\end{equation}

Conversely, if $V\longrightarrow B$ is a vector bundle over the base $B$
with fibers $V_{b}\subset \mathfrak{g}$, it defines a $G-$invariant
distribution $D$ on $G\times B$ by setting $S_{(b,g)}=L_{g\ast e}V_{b}$. The
vector bundle $S\subset D$ thus corresponds to the map 
\begin{eqnarray*}
G\times B &\longrightarrow &Gr_{dimS}(\mathfrak{g}) \\
(b,g) &\longmapsto &L_{g\ast e}V_{b}.
\end{eqnarray*}

Now, the subspace (recall prop. \ref{prop:properties of g^q})$\ \mathfrak{g}%
^{(b,g)}:=\{X\in \mathfrak{g},\ X_{Q}(b,g)\in D_{(b,g)}\}$ is given by%
\begin{equation*}
\mathfrak{g}^{(b,g)}=\{X\in \mathfrak{g},\ \exists Y\in S_{(b,e)};\
X=Ad_{g}Y\}
\end{equation*}%
so, 
\begin{eqnarray*}
\mathfrak{g}^{(b,g)} &=&Ad_{g}\ \mathfrak{g}^{(b,e)} \\
\mathfrak{g}^{(b,e)} &=&S_{(b,e)}.
\end{eqnarray*}

At this point, we make an assumption on the metric on $Q=G\times B$:

\begin{description}
\item[$\left( HM\right) $] Suppose we have a smooth map 
\begin{eqnarray*}
B &\longrightarrow &\{Left\ invariant\ metrics\ on\ G\}\simeq \{metrics\ on\ 
\mathfrak{g}\} \\
b &\mapsto &\left( ,\right) _{b}.
\end{eqnarray*}%
The metric $k^{Q}(,)$ on $Q$ is assumed to be given by%
\begin{equation*}
k_{(b,g)}^{Q}(\left( \overset{\cdot }{b}_{1},\overset{\cdot }{g}_{1}\right)
,\left( \overset{\cdot }{b}_{2},\overset{\cdot }{g}_{2}\right) )=k_{b}^{B}(%
\overset{\cdot }{b}_{1},\overset{\cdot }{b}_{2})+(g_{1}^{-1}\overset{\cdot }{%
g}_{1},g_{2}^{-1}\overset{\cdot }{g}_{2})_{b}
\end{equation*}%
for $k^{B}(,)$ being a metric on $B$.
\end{description}

\begin{remark}
\emph{(Applicability)\ }This kind of metric on $Q=G\times B$ is the one
present on typical examples (see \cite{BKMM}). See also the examples of
section \ref{subsec:Examples}.
\end{remark}

Assuming $(HM)$, the momentum map $J:TQ\longrightarrow \mathfrak{g}^{\ast }$
corresponding to the left $G-$action on $Q$ is%
\begin{equation*}
J(\overset{\cdot }{b},\overset{\cdot }{g})=Ad_{g}^{\ast }\Psi _{b}(g^{-1}%
\overset{\cdot }{g})
\end{equation*}%
with $\Psi _{b}:\mathfrak{g}\longrightarrow \mathfrak{g}^{\ast }$ denoting
the isomorphism defined by the metric $\left( ,\right) _{b}$ on $\mathfrak{g}
$. The inertia tensor $I_{(b,g)}:\mathfrak{g}\longrightarrow \mathfrak{g}%
^{\ast }$ takes the form%
\begin{equation*}
I_{(b,g)}=Ad_{g}^{\ast }\circ \Psi _{b}\circ Ad_{g^{-1}}.
\end{equation*}

\bigskip Note that we have a natural lift $d_{0}^{NH}(t)=(\tilde{c}(t),e)\in
B\times G$ for a curve $\tilde{c}(t)\in B$. This gauge $d_{0}^{NH}(t)$
defines a non-holonomic gauge as defined in section \ref%
{subsubsec:Non-holonomic gauge}. In fact, this $d_{0}^{NH}(t)$ coincides
with the horizontal lift of $\tilde{c}(t)$ from $(\tilde{c}(t_{1}),e)$ with
respect to the non-holonomic connection of \cite{BKMM}. Moreover, it is also
a mechanical gauge $\left( \ref{Eq: Mech gauge}\right) $.

For this gauge choice, the inclusion%
\begin{equation*}
i_{d_{0}^{NH}(t)}:\mathfrak{g}^{d_{0}^{NH}(t)}=S_{(\tilde{c}%
(t),e)}\hookrightarrow \mathfrak{g}
\end{equation*}%
depends only on the base curve $\tilde{c}(t)\in B$ and coincides with the
inclusion%
\begin{equation*}
i_{\tilde{c}(t)}:V_{\tilde{c}(t)}\hookrightarrow \mathfrak{g}
\end{equation*}%
where $V_{\tilde{c}(t)}=S_{(\tilde{c}(t),e)}$ is the fibre of the vector
bundle $\left( \ref{Eq:vector bundle for trivial Q}\right) $.

The curve $c(t)\ $describing the motion on the constrained and controlled
system on $Q$ will thus be%
\begin{equation*}
c(t)=(\tilde{c}(t),g(t))=g(t)\cdot d_{0}^{NH}(t)
\end{equation*}%
and 
\begin{equation*}
J(\overset{\cdot }{c})=Ad_{g}^{\ast }\ I_{(\tilde{c},e)}(g^{-1}\overset{%
\cdot }{g})=Ad_{g}^{\ast }\ \Psi _{\tilde{c}(t)}(g^{-1}\overset{\cdot }{g}).
\end{equation*}%
In this case, equations of motion $\left( \ref{Eq. motion c(t)}\right) $ read%
\begin{equation*}
i_{c(t)}^{\ast }\left( \frac{d}{dt}J(\overset{\cdot }{c})\right) =0
\end{equation*}%
or, equivalently, 
\begin{equation}
i_{\tilde{c}(t)}^{\ast }\left( \frac{d}{dt}\left( \Psi _{\tilde{c}(t)}(g^{-1}%
\overset{\cdot }{g})\right) +ad_{g^{-1}\overset{\cdot }{g}}^{\ast }\Psi _{%
\tilde{c}(t)}(g^{-1}\overset{\cdot }{g})\right) =0.
\label{Eq: motion trivial case}
\end{equation}%
The constraints for $g(t)$ are%
\begin{equation}
g^{-1}\overset{\cdot }{g}(t)\in \mathfrak{g}^{(\tilde{c}(t),e)}=S_{(\tilde{c}%
(t),e)}=V_{\tilde{c}(t)}.  \label{Eq: constraint trivial case}
\end{equation}

Eqs. $\left( \ref{Eq: motion trivial case}\right) $ can be re-written using
a moving basis system on the vector bundle $V\longrightarrow B$ as done in
the previous section, yielding the local expression of the non-holonomic
momentum eqs. of \cite{BKMM} evaluated along $\tilde{c}(t)$.

Lets simplify the situation a bit more to try to isolate the Lie-algebraic
(vertical) contribution to the system%
\'{}%
s motion from the $\mathfrak{g}^{D}-$geometric (horizontal) contribution
studied in the previous section.

In case the bundle $V\longrightarrow B$ is \emph{trivial}, that is $%
S_{(b,e)}=S_{0}\subset \mathfrak{g}$ for all $b\in B$, then 
\begin{equation*}
i_{\tilde{c}(t)}^{\ast }=i_{0}^{\ast }\ \forall t
\end{equation*}%
and so equation $\left( \ref{Eq: motion trivial case}\right) $ reads%
\begin{equation*}
\frac{d}{dt}\left( i_{0}^{\ast }\Psi _{\tilde{c}(t)}(g^{-1}\overset{\cdot }{g%
})\right) =-i_{0}^{\ast }\left( ad_{g^{-1}\overset{\cdot }{g}}^{\ast }\Psi _{%
\tilde{c}(t)}(g^{-1}\overset{\cdot }{g})\right)
\end{equation*}%
which is an eq. for $\Psi _{\tilde{c}(t)}(g^{-1}\overset{\cdot }{g})$,
coupled to the constraint equation $\left( \ref{Eq: constraint trivial case}%
\right) $ for $g^{-1}\overset{\cdot }{g}$. Its algebraic structure is still
hard to handle in general. If we wanted to solve the above (general)
equation by using usual Lie-algebraic properties of $\mathfrak{g}$, then we
would need to assume some additional condition on how the subspace $S_{0}$
changes when moving vertically along the fiber $\left( \tilde{c}(t),e\right)
\rightsquigarrow \left( \tilde{c}(t),g\right) $.

Suppose, then, that $S_{0}$ is $Ad_{G}$ invariant. It follows that $%
\mathfrak{g}^{c(t)}=\mathfrak{g}^{\tilde{c}(t)}=S_{0}$ and that $%
S_{0}\subset \mathfrak{g}$ is a Lie subalgebra. By the constraints $g^{-1}%
\overset{\cdot }{g}\in S_{0}$ and eq. of motion $\left( \ref{Eq. motion c(t)}%
\right) $ becomes the \emph{conservation law} (as in remark \ref{rmk:
trivial gi conservat})%
\begin{eqnarray*}
\frac{d}{dt}\left( i_{0}^{\ast }J\left( \dot{c}\right) \right) &=&0 \\
\frac{d}{dt}\left( i_{0}^{\ast }\Psi _{\tilde{c}(t)}(g^{-1}\overset{\cdot }{g%
})\right) &=&-\left( ad_{g^{-1}\overset{\cdot }{g}}^{\ast }i_{0}^{\ast }\Psi
_{\tilde{c}(t)}(g^{-1}\overset{\cdot }{g})\right) .
\end{eqnarray*}%
Although being integrable, this equation is still hard to solve explicitly
in general (see \cite{MR} for the rigid body $\mathfrak{g}=\mathfrak{so}(3)$
case). Nevertheless, in this situation, the dynamical factor $g(t)$ of $c(t)$
can be \emph{reconstructed} from a solution to the above equation in $S_{0}$%
\emph{\ }yielding corresponding \emph{phase formulas}, as described in
section \ref{subsec: Reconstruction and phases} and Appendix \ref{sec:App
Reconstr}.

From the analysis of this section, we see that even under very favorable
hypothesis on the geometry of $Q$ and $D$, the equations of motion can be
very complicated and we cannot continue with the general study of $c(t)$.
Nevertheless, if we require deeper compatibilities (as above)\ between $D$
and the $G-$action, e.g. horizontal symmetries, in secs. \ref%
{subsubsec:Phases Horizontal symmetries} and \ref{subsec: Phases for dipolar}
we shall show that further \emph{phase formulas} can be given for
characterizing the solution $c(t)$.

\section{Reconstruction and Phases}

\label{subsec: Reconstruction and phases}In the following, we focus on
reconstruction phases (\cite{MMR}) for both the full solution $c(t)$ and
vertical (gauge dependent) unknown $g(t)$. The interested reader can find
various types of reconstruction phases in \cite{MRS}.

\subsection{Gauges and phases in $Q\longrightarrow Q/G$ for $D-$constrained
systems}

\label{subsubsec:Gauges and phases in Q}

Suppose that the base curve $\tilde{c}(t)\in Q/G$ is closed, $\tilde{c}%
(t_{1})=\tilde{c}(t_{2})$. Choice $\left( \ref{Eq. nonholonomic gauge
relation}\right) $ for the \emph{non-holonomic gauge} $d_{0}^{NH}(t)$
provides us with a geometric phase in the motion of the system in $Q$ as
follows. Being defined as a horizontal lift, $d_{0}^{NH}(t_{2})$ coincides
with the \emph{holonomy} of the associated to the base curve $\tilde{c}(t)$
measured from the initial condition $d_{0}^{NH}(t_{1})=c(t_{1})$ and with
respect to the \emph{non-holonomic connection}. Thus, the corresponding
phase formula is 
\begin{equation*}
c(t_{2})=g_{Dyn}(t_{2})\cdot g_{NH}\cdot d_{0}^{NH}(t_{1})
\end{equation*}%
with $g_{NH}$ uniquely defined by $d_{0}^{NH}(t_{2})=g_{NH}\cdot
d_{0}^{NH}(t_{1})$ and where $g_{Dyn}(t)$ is the solution of equations $%
\left( \ref{Eq. motion for psi}\right) $ and $\left( \ref{Eq. D const for
psi in NHgauge}\right) $, with $\xi (t)=g_{Dyn}^{-1}\overset{\cdot }{g_{Dyn}}
$ and time dependent coefficients evaluated along this gauge $d_{0}^{NH}(t)$.

Another geometric phase\emph{\ }$g_{MP}$ appears when using the \emph{%
mechanical gauge}. Let the gauge $d_{0}^{NH}(t)$ be as above and $%
g_{Mech}(t) $ be defined by requiring $\tilde{d}_{0}(t):=g_{Mech}(t)\cdot
d_{0}^{NH}(t)$ to be the horizontal lift with respect to the \emph{mechnical
connection }$\left( \ref{Eq: Mech gauge}\right) $ (see \cite{Mont gauge})%
\emph{\ }on $Q$ with $g_{Mech}(t_{1})=e$. If we write $g_{Dyn}(t)=g_{\tilde{D%
}}(t)\cdot g_{Mech}(t)$, the corresponding equations of motion for the
remaining dynamic contribution $g_{\tilde{D}}(t)$ are%
\begin{equation}
i_{\tilde{d}_{0}(t)}^{\ast }\left( \frac{d}{dt}\left( I(\tilde{d}_{0}(t))g_{%
\tilde{D}}^{-1}\overset{\cdot }{g}_{\tilde{D}}\right) +ad_{g_{\tilde{D}}^{-1}%
\overset{\cdot }{g}_{\tilde{D}}}^{\ast }I(\tilde{d}_{0}(t))g_{\tilde{D}}^{-1}%
\overset{\cdot }{g}_{\tilde{D}}\right) =0
\label{Eq. motion nonhol gauge + mech}
\end{equation}%
which are simpler from the original ones $\left( \ref{Eq. motion for psi}%
\right) $ because the $J(\frac{d}{dt}(\tilde{d}_{0}))$ term vanishes by $%
\left( \ref{Eq: Mech gauge}\right) $. But the constraint equations $\left( %
\ref{Eq. D const for psi in NHgauge}\right) $ in terms of $g_{\tilde{D}}$
read%
\begin{equation}
g_{\tilde{D}}^{-1}\overset{\cdot }{g}_{\tilde{D}}+\overset{\cdot }{g}%
_{Mech}g_{Mech}^{-1}\in \mathfrak{g}^{\tilde{d}_{0}(t)}
\label{Eq. constraint nonhol gauge + mech}
\end{equation}%
which are more complicated than the original ones for $g_{Dyn}$.

The relation between the above different gauge phases read%
\begin{eqnarray*}
c(t_{2}) &=&g_{Dyn}(t_{2})\cdot g_{NH}\cdot c(t_{1}) \\
&=&g_{\tilde{D}}(t_{2})\cdot g_{Mech}(t_{2})\cdot g_{NH}\cdot c(t_{1}) \\
&=&g_{\tilde{D}}(t_{2})\cdot g_{MP}\cdot c(t_{1}).
\end{eqnarray*}%
with the second \emph{geometric phase }being $g_{MP}=g_{Mech}(t_{2})\cdot
g_{NH}$.

\begin{remark}
\emph{(Sympifications from different gauges) }In the non-holonomic gauge,
the constraint equations are simpler and, in turn, in the mechanical gauge
the equations of motion become simpler. One would like to have both
simplifications to hold, but this cannot be achieved in general since the
horizontal lift with respect to the mechanical connection is not horizontal
with respect to the non-holonomic connection for general $D$. Finally, we
would like to observe that, in some situations, we have additional
information about the $D-$constraints and the non-holonomic gauge becomes
preferable (see, for example, the next sections).
\end{remark}

\subsection{Reconstruction Phases for systems with Conserved Momentum}

\label{subsubsec:Reconstruction Cons Mom}Now, we shall elaborate on the
reconstruction of $g(t)$ for a solution $\Pi (t)$ in $O_{\mu }\subset 
\mathfrak{g}^{\ast }$, as described in sec. \ref{subsubsec: Cons Mom Case}
in case there are no $D-$constraints. A concrete example of the \emph{phase
formulas} we obtain below can be found in \cite{C def body} for the motion
of a self deforming body.

Suppose that we have a solution $\Pi (t)=Ad_{g^{-1}(t)}^{\ast }J(\dot{c})\in
O_{\mu }$ for eq. $\left( \ref{Ec Pi}\right) $ with $\mu =J(\dot{c}%
)=const\neq 0$ and that we chose a linear projector $P:\mathfrak{g}%
\twoheadrightarrow \mathfrak{g}_{\mu }$ satisfying%
\begin{equation}
Ad_{h}\circ P=P\circ Ad_{h}.
\end{equation}%
From Appendix \ref{sec:App Reconstr}, we know that we can then write%
\begin{equation*}
g(t)=h_{D}(t)\cdot g_{G}(t)
\end{equation*}%
with the \emph{geometric phase} $g_{G}$ being the \emph{horizontal lift} of $%
\Pi (t)$ with respect to connection defined by $P$ in the $G_{\mu }-$bundle $%
G\longrightarrow O_{\mu }$ and the \emph{dynamic phase }$h_{D}\in G_{\mu }$
defined by%
\begin{equation}
\frac{d}{dt}h_{D}h_{D}^{-1}(t)=P\left( I_{c(t)}^{-1}(J(\dot{c})-Ad_{g}^{\ast
}J_{0}(t))\right)  \label{eq. dyn phase 1}
\end{equation}%
with $h_{D}(t_{1})=e$. The last step follows from eq. $\left( \ref{Eq: Cons
mom reconstr}\right) $ for $g(t)$ where $I_{c(t)}$ denotes the inertia
tensor evaluated along the physical motion $c(t)$.

Suppose now that $\mathfrak{g}$ has an $Ad-$invariant scalar product $(,)$,
as considered in Appendix \ref{sec:App Reconstr}. Let $u_{1}=\frac{\Psi (\mu
)}{\left\Vert \Psi (\mu )\right\Vert }$ and $\{u_{i}\}_{i=1}^{dim\mathfrak{g}%
_{\mu }}$ denote an orthonormal basis with respect to $(,)$ of the vector
subspace $\mathfrak{g}_{\mu }\subset \mathfrak{g}$. In this case, equation $%
\left( \ref{eq. dyn phase 1}\right) $ becomes%
\begin{eqnarray}
\frac{d}{dt}h_{D}h_{D}^{-1}(t) &=&\frac{1}{\left\Vert \Psi (\mu )\right\Vert
^{2}}\left( 2K(\dot{c}(t))-2K_{int}(t)+\left\langle
J_{0}(t),I_{0}^{-1}(t)J_{0}(t)\right\rangle -\left\langle
J_{0}(t),I_{0}^{-1}(t)\Pi (t)\right\rangle \right) \Psi (\mu )+  \notag \\
&&+\Sigma _{i=2}^{dim\mathfrak{g}_{\mu }}\left[ (u_{i},I_{c(t)}^{-1}\mu
)-(u_{i},I_{c(t)}^{-1}Ad_{g}^{\ast }J_{0}(t))\right] \ u_{i}
\end{eqnarray}%
where $K$ represents the \emph{kinetic energy }of the controlled system in $%
Q $ (see Appendix \ref{sec:App Kin En}).

When $d_{0}(t)=d_{0}^{Mec}(t)$ is the \emph{mechanical gauge }$\left( \ref%
{Eq: Mech gauge}\right) $,%
\begin{equation}
\frac{d}{dt}h_{D}h_{D}^{-1}(t)=\frac{1}{\left\Vert \Psi (\mu )\right\Vert
^{2}}(2K(\dot{c}(t))-2K_{int}(t))\Psi (\mu )+\Sigma _{i=2}^{dim\mathfrak{g}%
_{\mu }}(u_{i},I_{c(t)}^{-1}\mu )\ u_{i}.
\label{Eq: dyn phase cons mom metric}
\end{equation}

\begin{remark}
\emph{(Locked inertia tensor and physical information in }$h_{D}$\emph{)}
The above reconstruction phase formula, in the mechanical gauge, relates the
dynamical phase $h_{D}$ to the data of the locked inertia tensor $I_{c(t)}$
and the kinetic energy $K$, both along the physical solution curve $c(t)$ in 
$Q$, and to the gauge kinetic energy $K_{int}(\dot{d}_{0}^{Mec})$.
\end{remark}

\begin{remark}
\emph{(The case }$J(\dot{c})=0$\emph{) }In this case, the system%
\'{}%
s motion $c(t)$ coincides with the \emph{mechanical gauge }$d_{0}^{Mech}(t)$
motion because of $\left( \ref{Eq: Mech gauge}\right) $. We thus say that
the induced motion $c(t)$ is \emph{geometrical} with respect to the base one 
$\tilde{c}(t)$ (see also example \ref{ex:Def body zero ang momentum} below).
\end{remark}

\begin{remark}
\emph{(The case }$G_{\mu }$ \emph{abelian) }In this case, $%
I_{c(t)}=I_{d_{0}^{Mec}(t)}$ and, thus, the only dynamical (i.e.
non-kinematical) information needed to evaluate formula $\left( \ref{Eq: dyn
phase cons mom metric}\right) $ is the system%
\'{}%
s kinetic energy evolution $K(\dot{c}(t))$. In this case, $h_{D}(t)$ can
also be easily integrated by means of the corresponding exponential map $exp:%
\mathfrak{g}_{\mu }\longrightarrow G_{\mu }$.
\end{remark}

\subsection{Phases for $D-$constrained, Purely Kinematical systems}

\label{subsec: PK phases}We recall from \cite{BKMM},

\begin{description}
\item[Definition] A constrained system $(Q,L,G,D)\ $is said to have \textbf{%
purely kinematical (PK) }constraints if $TQ=Ver\oplus D$.
\end{description}

Since $D$ is $G-$invariant, it defines a principal connection on $%
Q\longrightarrow Q/G$. Let $A^{D}$ denote the corresponding $\mathfrak{g}-$%
valued $1-$form on $Q$. The constraint equation for $c(t)\ $then reads%
\begin{equation*}
A^{D}(\overset{\cdot }{c})=0
\end{equation*}%
and the vertical eqs. of motion $\left( \ref{Eq. motion for psi}\right) $
are trivial since $\mathfrak{g}^{q}=0$ for all $q$. So we have that

\begin{proposition}
\label{thm:Phases PK} The motion for a base-controlled system $(Q,L,G,D,%
\tilde{c})$ for which $D$ defines \emph{purely kinematical }PK constraints,
is of \emph{geometric nature }with respect to $\tilde{c}$. In other words,
the solution $c(t)$ is given by the \emph{horizontal lift} of the base
controlled curve $\tilde{c}(t)$ with respect to the principal connection on $%
Q\longrightarrow Q/G$ defined by the constraint distribution $D$.
\end{proposition}

\begin{description}
\item[Corollary] If $\tilde{c}$ is closed in $[t_{1},t_{2}]$, we then have a 
\emph{geometric phase} $g_{G}$ in the system%
\'{}%
s dynamics associated to the initial value $c(t_{1})$ and defined by $%
g_{G}=Hol(\tilde{c})$:%
\begin{equation*}
c(t_{2})=g_{G}\cdot c(t_{1}).
\end{equation*}
\end{description}

\begin{example}
\label{ex:Def body zero ang momentum} $\emph{(Deforming}$ \emph{bodies with
zero angular momentum) }If we regard $J(\dot{c})=0$ as a $D-$constraint for
the motion of a self deforming body, with $J$ being the angular momentum
map, then $D$ coincides with the mechanical connection%
\'{}%
s horizontal space. From the above proposition, we recover the known fact (%
\cite{SW})\ that global reorientation $g(t)\in SO(3)\ $of such a body is 
\emph{geometrical }with respect to the deformation $\tilde{c}(t)$.
\end{example}

\subsection{Phases for $D-$constrained systems with Horizontal Symmetries}

\label{subsubsec:Phases Horizontal symmetries}We now analyze a
geometric-kinematical favorable case leading to phase formulas for the
dynamical factor $g(t)$ of $c(t)$.

\begin{description}
\item[Definition] (\cite{BKMM}) A constrained system $(Q,L,G,D)\ $is said to
have (full) \textbf{horizontal symmetries (HS) }if there exists a subgroup $%
H\subset G$ such that
\end{description}

\begin{enumerate}
\item $\xi _{Q}(q)\in D_{q}$ $\forall q\in Q\ $when $\xi \in \mathfrak{h}%
:=Lie(H)\subset \mathfrak{g}$,

\item (\emph{Full} condition) $S_{q}:=D_{q}\cap T_{q}\left(
Orb_{G}(q)\right) =T_{q}\left( Orb_{H}(q)\right) \ \forall q\in Q.$
\end{enumerate}

Condition $(2)$ above states that horizontal symmetries exhaust the whole
vertical kinematics. The analysis we give below can be extended to the
non-full case, i.e. by assuming only $(1)$, but we keep hypothesis $(2)$ for
simplicity. Example \ref{Ex: 2 Balls} below illustrates the non-full case.

For an \textbf{HS }system, the bundle $\mathfrak{g}^{D}$ is the trivial one $%
Q\times \mathfrak{h}$. Since the inclusion map $i_{q}=i_{\mathfrak{h}}:%
\mathfrak{h}=\mathfrak{g}^{q}\hookrightarrow \mathfrak{g}$ becomes
independent of the point $q$, eq. $\left( \ref{Eq. motion c(t)}\right) $
reads%
\begin{equation}
i_{\mathfrak{h}}^{\ast }\left( \frac{d}{dt}(J(\overset{\cdot }{c}))\right) =%
\frac{d}{dt}(i_{\mathfrak{h}}^{\ast }J(\overset{\cdot }{c}))=0.
\label{Eq. motion J HS}
\end{equation}%
Consequently, $i_{\mathfrak{h}}^{\ast }J(\overset{\cdot }{c})$ gives a \emph{%
conserved quantity} during the motion of the system as at the end of section %
\ref{subsec: Trivial Q case}. This projection $i_{\mathfrak{h}}^{\ast }J(%
\overset{\cdot }{c})$ can be interpreted as the part of the total momentum
map which is compatible with the constraints (see also \cite{BKMM}).

Next, we shall enunciate a few results which follow from the definition of a
system with full \textbf{HS}.

\begin{proposition}
\label{prop:properties of HS}The following holds:

\begin{itemize}
\item $H\subset G\ $is a \emph{normal }subgroup of $G$, thus$\ \mathfrak{h}$
is $G-$invariant $Ad_{g}\mathfrak{h}=\mathfrak{h}$,

\item $i_{\mathfrak{h}}\ Ad_{g}=Ad_{g}\ i_{\mathfrak{h}},$

\item For each $q\in Q$, let $I_{q}^{\mathfrak{h}}=i_{\mathfrak{h}}^{\ast
}\circ I_{q}\circ i_{\mathfrak{h}}:\mathfrak{h}\longrightarrow \mathfrak{h}%
^{\ast }$ be the \textbf{restricted inertia tensor}, then%
\begin{equation*}
I_{g\cdot q}^{\mathfrak{h}}=Ad_{g}^{\ast }I_{q}^{\mathfrak{h}}Ad_{g^{-1}},\
\forall g\in H.
\end{equation*}
\end{itemize}
\end{proposition}

We shall now describe the appearance of phase formulas for the dynamical
factor $g(t)\ $of the motion $c(t)\ $of an \textbf{HS }system. First, recall
that in a \textbf{non-holonomic gauge} $d_{0}^{NH}(t)$ the constraint
equation for the body velocity $\xi =g^{-1}\overset{\cdot }{g}$ becomes eq. $%
\left( \ref{Eq. D const for psi in NHgauge}\right) $ which, for an HS
system, reduces to 
\begin{equation}
g^{-1}\overset{\cdot }{g}(t)\ \in \mathfrak{h}
\label{Eq. D const for psi NH and HS}
\end{equation}%
for all $t$. From the other side, if we consider the non-holonomic \emph{%
body momentum} $\Pi (t)\in \mathfrak{g}^{\ast }$ of eq. $\left( \ref{Eq: def
Pi non holonomic}\right) $, because of the constraint $\left( \ref{Eq. D
const for psi NH and HS}\right) ,$ we have that%
\begin{equation}
g^{-1}\overset{\cdot }{g}(t)=\xi (t)=(I_{0}^{\mathfrak{h}})_{(t)}^{-1}\left(
i_{\mathfrak{h}}^{\ast }\Pi (t)-i_{\mathfrak{h}}^{\ast }J(\dot{d}%
_{0}^{NH}(t))\right) .  \label{eq: reconstruc g from pi HS}
\end{equation}%
Thus, eq.$\left( \ref{Eq. motion for psi}\right) $, equiv. eq. $\left( \ref%
{Eq. motion J HS}\right) $, become%
\begin{equation}
\frac{d}{dt}(i_{\mathfrak{h}}^{\ast }\Pi (t))=-ad_{g^{-1}\overset{\cdot }{g}%
(t)}^{\ast }(i_{\mathfrak{h}}^{\ast }\Pi (t))=-ad_{(I_{0}^{\mathfrak{h}%
})_{(t)}^{-1}(i_{\mathfrak{h}}^{\ast }\Pi (t)-i_{\mathfrak{h}}^{\ast }J(\dot{%
d}_{0}^{NH}(t)))}^{\ast }i_{\mathfrak{h}}^{\ast }\Pi (t).
\label{Eq. total Pi HS}
\end{equation}%
The above expressions are equivalent to%
\begin{equation}
i_{\mathfrak{h}}^{\ast }\Pi (t)=Ad_{g^{-1}}^{\ast }(i_{\mathfrak{h}}^{\ast
}J(\overset{\cdot }{c}))  \label{Eq. HS for Pi}
\end{equation}%
because%
\begin{equation*}
i_{\mathfrak{h}}^{\ast }J(\overset{\cdot }{c})=i_{\mathfrak{h}}^{\ast
}Ad_{g}^{\ast }(\Pi (t))=Ad_{g}^{\ast }(i_{\mathfrak{h}}^{\ast }\Pi (t)),
\end{equation*}%
by proposition \ref{prop:properties of HS}. The constraint equation $\left( %
\ref{Eq. D const for psi NH and HS}\right) $ can be also put in terms of $%
\Pi (t)$ as follows%
\begin{equation}
I_{0}^{-1}(t)(\Pi (t)-J(\overset{\cdot }{d_{0}^{NH}(t)}))=(I_{0}^{\mathfrak{h%
}})_{(t)}^{-1}\left( i_{\mathfrak{h}}^{\ast }\Pi (t)-i_{\mathfrak{h}}^{\ast
}J(\dot{d}_{0}^{NH}(t))\right) \in \mathfrak{h}.
\label{Eq. constraint Pi HS}
\end{equation}%
Eqs. $\left( \ref{Eq. total Pi HS}\right) $ and $\left( \ref{Eq. constraint
Pi HS}\right) $, both determine the dynamics of $\Pi (t)\in \mathfrak{g}%
^{\ast }$ from the initial value $\Pi (t_{1})=J(\overset{\cdot }{c})=\mu $.

Now, from $\left( \ref{Eq. D const for psi NH and HS}\right) $ and $%
g(t_{1})=e$ it follows that $g(t)\in H$ for all $t\in \lbrack t_{1},t_{2}]$.
Thus, from $\left( \ref{Eq. HS for Pi}\right) $, we can deduce that 
\begin{equation*}
i_{\mathfrak{h}}^{\ast }\Pi (t)\in O_{i_{\mathfrak{h}}^{\ast }J(\overset{%
\cdot }{c})}^{H}
\end{equation*}%
where $O_{i_{\mathfrak{h}}^{\ast }J(\overset{\cdot }{c})}^{H}$ denotes the $%
H-$\emph{coadjoint orbit} in $\mathfrak{h}^{\ast }$ through the constant
element $i_{\mathfrak{h}}^{\ast }J(\overset{\cdot }{c})$. The following
(commutative) diagram summarizes the relevant geometric situation%
\begin{equation*}
\begin{array}{ccccc}
(g,i_{\mathfrak{h}}^{\ast }\Pi (t))\in L^{-1}(i_{\mathfrak{h}}^{\ast }J(%
\overset{\cdot }{c}))\simeq H & \hookrightarrow &  & H\times \mathfrak{h}%
^{\ast } &  \\ 
\downarrow &  & \pi \swarrow &  & \searrow L \\ 
i_{\mathfrak{h}}^{\ast }\Pi (t)\in O_{i_{\mathfrak{h}}^{\ast }J(\overset{%
\cdot }{c})}^{H} & \hookrightarrow & \mathfrak{h}^{\ast } &  & \mathfrak{h}%
^{\ast }\ni i_{\mathfrak{h}}^{\ast }J(\overset{\cdot }{c})%
\end{array}%
\end{equation*}%
for the maps $L(g,\alpha )=Ad_{g}^{\ast }\alpha $ and $\pi (g,\alpha
)=\alpha $, $(g,\alpha )\in H\times \mathfrak{h}^{\ast }$. Recall that $L$
is the momentum map corresponding to the left $H$ symplectic action on $%
H\times \mathfrak{h}^{\ast }\simeq T^{\ast }H$ and that $H\simeq L^{-1}(i_{%
\mathfrak{h}}^{\ast }J(\overset{\cdot }{c}))\overset{\pi }{\longrightarrow }%
O_{i_{\mathfrak{h}}^{\ast }J(\overset{\cdot }{c})}^{H}$ defines a principal $%
H_{i_{\mathfrak{h}}^{\ast }J(\overset{\cdot }{c})}-$bundle over the reduced
space $O_{i_{\mathfrak{h}}^{\ast }J(\overset{\cdot }{c})}^{H}$, as described
in Appendix \ref{sec:App Reconstr}.

\begin{remark}
\emph{(Initial conditions) }When the initial conditions are $g(0)\neq e\ $in 
$G$, so $c(0)=g(0)\cdot d_{0}^{NH}(0)$, then eq. $\left( \ref{Eq. D const
for psi NH and HS}\right) $ implies that $g(t)=g(0)\cdot g_{H}(t)$ where $%
g_{H}(t)\in H$ is the solution corresponding to the initial condition $%
g_{H}(0)=e$. Thus, below we shall focus on the $g(0)=e$ case.\bigskip
\end{remark}

We are now in position to apply the usual \textbf{reconstruction procedure}
of \cite{MMR} for the group unknown $g(t)\in H$ from a solution $i_{%
\mathfrak{h}}^{\ast }\Pi (t)\in O_{i_{\mathfrak{h}}^{\ast }J(\overset{\cdot }%
{c})}^{H}$. Let $P:\mathfrak{h}\longrightarrow \mathfrak{h}_{i_{\mathfrak{h}%
}^{\ast }J(\overset{\cdot }{c})}=Lie\left( H_{i_{\mathfrak{h}}^{\ast }J(%
\overset{\cdot }{c})}\right) $ be a linear projector s.t.%
\begin{equation*}
P\circ Ad_{g}=Ad_{g}\circ P
\end{equation*}%
for all $g\in H$. As described in Appendix \ref{sec:App Reconstr}, $P$
defines a \emph{principal connection }$A_{P}:TH\longrightarrow \mathfrak{h}%
_{i_{\mathfrak{h}}^{\ast }J(\overset{\cdot }{c})}$ and so:

\begin{proposition}
\label{prop:Phases HS copy(2)} Keeping the notations introduced above, let $%
\Pi (t)\in \mathfrak{g}^{\ast }$ be a solution of eqs. $\left( \ref{Eq.
total Pi HS}\right) $, $\left( \ref{Eq. constraint Pi HS}\right) $ and $i_{%
\mathfrak{h}}^{\ast }\Pi (t)\ $its projection onto $\mathfrak{h}^{\ast }$.
Then, the corresponding solution $g(t)$ of the reconstruction eq. $\left( %
\ref{eq: reconstruc g from pi HS}\right) $ which satisfies the constraints $%
\left( \ref{Eq. D const for psi NH and HS}\right) $ with $g(0)=e$ is such
that $g(t)\in H$ $\forall t\in I$ and%
\begin{equation*}
g(t)=h_{D}(t)\ g_{G}(t).
\end{equation*}%
Above, the \emph{geometric phase }$g_{G}(t)$ is the \emph{horizontal lift}
of $i_{\mathfrak{h}}^{\ast }\Pi (t)\in O_{i_{\mathfrak{h}}^{\ast }J(\overset{%
\cdot }{c})}^{H}$ from $g_{G}(0)=e$ with respect to the principal connection 
$A_{P}$ on the principal $H_{i_{\mathfrak{h}}^{\ast }J(\overset{\cdot }{c}%
)}- $bundle $H\overset{\pi }{\longrightarrow }O_{i_{\mathfrak{h}}^{\ast }J(%
\overset{\cdot }{c})}^{H}$ and the \emph{dynamic phase }$h_{D}(t)\in H_{i_{%
\mathfrak{h}}^{\ast }J(\overset{\cdot }{c})}$ is defined by the equation
\end{proposition}

\begin{eqnarray}
\frac{d}{dt}h_{D}h_{D}^{-1}(t) &=&A_{P}(\frac{d}{dt}g)_{g}=P\left(
Ad_{g(t)}(I_{0}^{\mathfrak{h}})_{(t)}^{-1}(i_{\mathfrak{h}}^{\ast }\Pi
(t)-i_{\mathfrak{h}}^{\ast }J(\dot{d}_{0}^{NH}(t)))\right) \\
&=&\ P\left( (I_{c(t)}^{\mathfrak{h}})^{-1}\left( i_{\mathfrak{h}}^{\ast }J(%
\dot{c})-Ad_{g(t)}^{\ast }i_{\mathfrak{h}}^{\ast }J(\dot{d}%
_{0}^{NH}(t))\right) \right) \\
h_{D}(0) &=&e.
\end{eqnarray}

\begin{remark}
\emph{(Physical content of }$h_{D}$\emph{) }The above dynamical phase $h_{D}$
depends on the (restricted)\ inertia tensor $I_{c(t)}^{\mathfrak{h}}$ and on
the gauge internal momentum $Ad_{g(t)}^{\ast }i_{\mathfrak{h}}^{\ast }J(\dot{%
d}_{0}^{NH}(t))$, both as seen from the reference system which is moving
along the physical evolution $c(t)\in Q$. Moreover, if the non-holonomic
gauge choice is the horizontal one $\left( \ref{Eq. nonholonomic gauge
relation}\right) $, then 
\begin{equation*}
\frac{d}{dt}h_{D}h_{D}^{-1}(t)=P\left( (I_{c(t)}^{\mathfrak{h}})^{-1}i_{%
\mathfrak{h}}^{\ast }J(\dot{c})\right)
\end{equation*}%
only depends on $I_{c(t)}^{\mathfrak{h}}$.
\end{remark}

\begin{remark}
\label{rmk: HS zero momentum}\emph{(The case }$i_{\mathfrak{h}}^{\ast }J(%
\overset{\cdot }{c})=0$\emph{) }In this case, $g(t)$ coincides with the
dynamical phase and is given by%
\begin{equation*}
g^{-1}\overset{\cdot }{g}(t)=-(I_{d_{0}^{NH}(t)}^{\mathfrak{h}})^{-1}i_{%
\mathfrak{h}}^{\ast }J(\dot{d}_{0}^{NH}(t))
\end{equation*}%
since $i_{\mathfrak{h}}^{\ast }\Pi (t)=0$ by $\left( \ref{Eq. HS for Pi}%
\right) $. Nevertheless, the full motion $c(t)\ $is \emph{geometric }with
respect to the base one $\tilde{c}(t)$. The reason is that $c(t)\ $coincides
with the horizontal lift $d_{0}^{NH}$of $\tilde{c}$ with respect to the 
\emph{non-holonomic connection} (\cite{BKMM}) because of equation $\left( %
\ref{Eq. nonholonomic gauge relation}\right) $. Notice that this is true for 
\emph{full} horizontal symmetries, i.e., when the conservation of $i_{%
\mathfrak{h}}^{\ast }J=0$ exhausts the whole vertical eqs. of motion (see
also \cite{BKMM}). This result generalizes the one of \cite{SW} (see ex. \ref%
{ex:Def body zero ang momentum}) on the geometric nature of base-induced
motion for zero momentum systems to the context of $D-$constrained\emph{\ }%
systems with full horizontal symmetries.
\end{remark}

When $\mathfrak{h}$ admits an $Ad-$invariant inner product, the dynamic
phase equation can be also related to other mechanical magnitudes.

\begin{proposition}
\label{prop:Phases HS} Keeping the notations introduced above, suppose that $%
\mathfrak{h}$ is endowed with an $Ad-$invariant inner product $(,)\ $%
inducing the isomorphism $\Psi :\mathfrak{h}^{\ast }\longrightarrow 
\mathfrak{h}$ and let $P:\mathfrak{h}\longrightarrow \mathfrak{h}_{i_{%
\mathfrak{h}}^{\ast }J(\overset{\cdot }{c})}$ be the orthogonal projector
onto $\mathfrak{h}_{i_{\mathfrak{h}}^{\ast }J(\overset{\cdot }{c})}$. Let$\
\{u_{i}\}$ be an orthonormal basis for $\mathfrak{h}_{i_{\mathfrak{h}}^{\ast
}J(\overset{\cdot }{c})}$ with $u_{1}=\frac{\Psi (i_{\mathfrak{h}}^{\ast }J(%
\overset{\cdot }{c}))}{\left\Vert \Psi (i_{\mathfrak{h}}^{\ast }J(\overset{%
\cdot }{c}))\right\Vert }$. Also, let the non-holonomic gauge $d_{0}^{NH}$
be defined by the horizontal lift $\left( \ref{Eq. nonholonomic gauge
relation}\right) $. Then, the corresponding dynamic phase equation becomes%
\begin{eqnarray*}
\frac{d}{dt}h_{D}h_{D}^{-1}(t) &=&\left( 2K(\frac{d}{dt}c(t))-2K_{int}(t)%
\right) \frac{\Psi (i_{\mathfrak{h}}^{\ast }J(\overset{\cdot }{c}))}{%
\left\Vert \Psi (i_{\mathfrak{h}}^{\ast }J(\overset{\cdot }{c}))\right\Vert
^{2}}+ \\
&&+\Sigma _{i=2}^{dim\mathfrak{h}_{i_{\mathfrak{h}}^{\ast }J(\overset{\cdot }%
{c})}}(u_{i},(I_{c(t)}^{\mathfrak{h}})^{-1}\left( i_{\mathfrak{h}}^{\ast }J(%
\dot{c})\right) )\ u_{i} \\
h_{D}(t_{1}) &=&e.
\end{eqnarray*}
\end{proposition}

In the above expression for the dynamic phase, ($K_{int}$) $K$ denotes the
(gauge-internal)\ kinetic energy of the controlled system in $Q$ (see
Appendix \ref{sec:App Kin En}). As before, $I_{c(t)}^{\mathfrak{h}}$
represents the (restricted)\ inertia tensor as seen from the reference
system which is moving along the physical evolution $c(t)\in Q$ . The above
formula relates this physical quantities, which are directly involved in the
system%
\'{}%
s dynamics, to the phases appearing during the full $H-$horizontally
symmetric motion (see Corollary bellow).

\begin{description}
\item[Corollary:] \label{cor:Phases in Q system HS}Finally, if the solution $%
\Pi (t)\in \mathfrak{g}^{\ast }$ is such that $i_{\mathfrak{h}}^{\ast }\Pi
(t_{1})=i_{\mathfrak{h}}^{\ast }\Pi (t_{2})$ then:
\end{description}

\begin{itemize}
\item $g_{G}(t_{2})$ is the \emph{holonomy} of the base path $i_{\mathfrak{h}%
}^{\ast }\Pi (t)$ in the $H_{i_{\mathfrak{h}}^{\ast }J(\overset{\cdot }{c}%
)}- $bundle $H\overset{\pi }{\longrightarrow }O_{i_{\mathfrak{h}}^{\ast }J(%
\overset{\cdot }{c})}^{H}$ with respect to the connection defined by $P$
measured from $g_{G}(t_{1})=e$.

\item the solution for the constrained and controlled system $c(t)\in Q$
satisfies the following \emph{phase relation }at time $t_{2}$:%
\begin{equation*}
\fbox{$c(t_{2})=h_{D}(t_{2})g_{G}(t_{2})\cdot d_{0}^{NH}(t_{2})$}
\end{equation*}%
where $d_{0}^{NH}(t_{2})$ is the horizontal lift of $\tilde{c}(t)$ with
respect to the non-holonomic connection (\cite{BKMM}), starting from $%
d_{0}(t_{1})=c(t_{1})$.

\item when, in addition, the base curve $\tilde{c}(t)\in Q/G$ is closed for $%
t\in \lbrack t_{1},t_{2}]$, so $\tilde{c}(t_{1})=\tilde{c}(t_{2})$, then $%
d_{0}(t_{2})=g_{G}^{NH}\cdot d_{0}(t_{1})$ where $g_{G}^{NH}$ is the \emph{%
holonomy} of the base path $\tilde{c}$ with respect to the non-holonomic
connection in the bundle $Q\longrightarrow Q/G$ measured from the initial
condition $d_{0}(t_{1})=c(t_{1})$. So, in this case,%
\begin{equation*}
\fbox{$c(t_{2})=h_{D}(t_{2})g_{G}(t_{2})\cdot g_{G}^{NH}\cdot c(t_{1})$}.
\end{equation*}
\end{itemize}

\subsection{Phases for systems with dipolar-magnetic-torque type of Affine
Constraints}

\label{subsec: Phases for dipolar}An interesting special case of affine
constrained systems which do not satisfy hypothesis $\left( ii\right) $ of
sec. \ref{subsec:affine constraints} but present \emph{reconstruction phase
formulas} is the following.

\begin{description}
\item[$\left( ii%
{\acute{}}%
\right) $] The affine constraints are of \textbf{external
dipolar-magnetic-torque} form, this is,%
\begin{equation*}
\mathfrak{A}_{q(t)}^{Mech}(\dot{q}(t))=I_{q(t)}^{-1}Ad_{h_{M}(t)}^{\ast }%
\hat{L}_{0}
\end{equation*}%
for $\mathfrak{A}^{Mech}$ denoting the \emph{mechanical connection} (see 
\cite{Mont gauge}). Equivalently, the affine constraint can be put in the
form%
\begin{equation*}
J(\dot{q}(t))=Ad_{h_{M}(t)}^{\ast }\hat{L}_{0}
\end{equation*}%
for some given curve $h_{M}(t)\in $ $G$, with $h_{M}(t_{1})=id$ and the
initial momentum value $\hat{L}_{0}\neq 0\in \mathfrak{g}^{\ast }$.
\end{description}

The time derivative of the above equation is equivalent to the following 
\emph{non-conservation of momentum equation}%
\begin{equation*}
\frac{d}{dt}J(\dot{q}(t))=ad_{\dot{h}_{M}h_{M}^{-1}}^{\ast }L(\dot{q}(t))
\end{equation*}%
where the right hand side represents a \emph{generalized torque} of a very
special kind.

In the section \ref{subsubsec:Example dipolar magnetic}, we shall study the
motion of a body with dipolar magnetic moment in an external magnetic field
which can be described as a system with affine constraints of type $(ii%
{\acute{}}%
)$ above. This justifies our terminology.

So we now assume $(ii%
{\acute{}}%
)$ to hold and that we have a base controlled curve $\tilde{c}(t)$. Next, we
choose the \emph{mechanical gauge }$d_{0}^{Mec}(t)$ $\left( \ref{Eq: Mech
gauge}\right) $ because $D$ for the above connection form $\mathfrak{A}%
_{q(t)}^{Mech}$ is exactly the horizontal space with respect to the
mechanical connection. Since constraints represent $dimG$ equations, they
fully characterizes the dynamics of the group unknown $g(t)$ in $%
c(t)=g(t)\cdot d_{0}^{Mec}(t)$. Indeed, $D$ defines a principal connection,
thus $\mathfrak{g}^{D}=0$ the zero bundle and so eq. of motion $\left( \ref%
{Eq. motion affine}\right) $ are trivial, i.e., $0=0$. These constraint
equations in $(ii%
{\acute{}}%
)$ can be written as%
\begin{equation*}
Ad_{h_{M}^{-1}(t)}^{\ast }Ad_{g(t)}^{\ast }I_{d_{0}^{Mec}(t)}\left( g^{-1}%
\dot{g}\right) =\hat{L}_{0}=const.
\end{equation*}%
From this, we see that if we call $R_{M}(t):=h_{M}^{-1}(t)g(t)\in G$ and $%
\Pi (t):=I_{d_{0}^{Mec}(t)}\left( g^{-1}\dot{g}\right) $, then%
\begin{equation}
Ad_{R_{M}(t)}^{\ast }\Pi (t)=\hat{L}_{0}  \label{Eq: conserv aff dip rec}
\end{equation}%
so $\Pi (t)\in O_{L_{0}}\subset \mathfrak{g}^{\ast }$, the coadjoint orbit
through $\hat{L}_{0}$, for all $t$. The corresponding equation giving the
dynamics of $\Pi (t)$ is%
\begin{equation*}
\frac{d}{dt}\Pi (t)=-ad_{R_{M}^{-1}\overset{\cdot }{R}_{M}}^{\ast }\Pi
(t)=-ad_{\left( I_{d_{0}^{Mec}(t)}^{-1}\Pi (t)-Ad_{g^{-1}}\dot{h}%
_{M}h_{M}^{-1}\right) }^{\ast }\Pi (t).
\end{equation*}%
Note that this equation is coupled to the one that defines $\Pi (t)$ from $%
g(t)$. Nevertheless, recall the map%
\begin{eqnarray*}
L &:&G\times \mathfrak{g}^{\ast }\longrightarrow \mathfrak{g}^{\ast } \\
L(R_{M},\Pi ) &=&Ad_{R_{M}}^{\ast }\Pi .
\end{eqnarray*}

Equation $\left( \ref{Eq: conserv aff dip rec}\right) $ implies that we are
in the situation described in Appendix \ref{sec:App Reconstr} and we can
thus apply the reconstruction procedure (\cite{MMR}) on the principal $G_{%
\hat{L}_{0}}-$bundle $G\simeq L^{-1}(\hat{L}_{0})\longrightarrow $ $O_{\hat{L%
}_{0}}$ to obtain $R_{M}(t)$ from a solution $\Pi (t)\in O_{\hat{L}_{0}}$.
This yields the phase formula $R_{M}(t)=R_{M}^{Dyn}(t)R_{M}^{Geom}(t)$ where
the dynamic phase $R_{M}^{Dyn}(t)\ $lies in $G_{\hat{L}_{0}}$ and $%
R_{M}^{Geom}(t)$ is a horizontal lift of $\Pi (t)$ with respect to some
chosen $P-$connection $A_{P}$ in the $G_{\hat{L}_{0}}-$bundle $L^{-1}(\hat{L}%
_{0})\simeq G\longrightarrow O_{\hat{L}_{0}}$ (see Appendix \ref{sec:App
Reconstr}). In this case, the dynamic phase equation, when put in terms of
the original $g(t)$, reads%
\begin{eqnarray}
\frac{d}{dt}R_{M}^{Dyn}R_{M}^{Dyn\ -1}(t) &=&A_{P}(\frac{d}{dt}R_{M}(t))_{g}
\\
&=&P\left( Ad_{h_{M}^{-1}}\left( I_{c(t)}^{-1}J(\dot{c})_{(t)}-\dot{h}%
_{M}h_{M}^{-1}\right) \right) \\
R_{M}^{Dyn}(t_{1}) &=&id.
\end{eqnarray}

In section \ref{subsubsec:Example dipolar magnetic}, we shall work out the
details of the above reconstruction formula in the magnetic dipole example.

Finally, if $\Pi (t_{1})=\Pi (t_{2})$ then we have a \emph{phase formula}
which fully characterizes the motion of the system $c(t)\in Q\ $at time $%
t_{2}$:%
\begin{equation*}
c(t_{2})=h_{M}(t_{2})\cdot R_{M}^{Dyn}(t_{2})\cdot Hol_{\Pi
(t_{1,2})}^{P}\cdot g_{MP}\cdot c(t_{1})
\end{equation*}%
where $g_{MP}\ $is the mechanical-gauge geometric phase (sec. \ref%
{subsubsec:Gauges and phases in Q}) and $Hol_{\Pi (t_{1,2})}^{P}$ is the
holonomy of the curve $\Pi (t)$ with respect to the $P-$connection in the
bundle $L^{-1}(\hat{L}_{0})\simeq G\longrightarrow O_{\hat{L}_{0}}$ measured
from the initial value $e\in G$.

\section{Examples}

\label{subsec:Examples}Here we illustrate our general considerations on
simple examples of \emph{base controlled}, $D-$constrained\emph{\ }systems.
Examples of shape-controlled self deforming bodies with conserved angular
momentum can be found in \cite{C def body}.

\subsection{Vertical Rotating disk}

\label{Ex: vertical disk abel}\emph{\ }We consider the vertical rotating
disk example from \cite{BKMM}. This gives an example of the systems
considered in section \ref{subsec: G abelian}. In this case, $Q=\mathbb{R}%
^{2}\times S^{1}\times S^{1}\ni q=(x,y,\theta ,\varphi )$ and we consider $G=%
\mathbb{R}^{2}\times S^{1}\ni g=(x,y,\theta )\ $(left)\ acting on itself.
The Lagrangian reads 
\begin{equation*}
L(\dot{x},\dot{y},\dot{\theta},\dot{\varphi})=\frac{1}{2}m(\overset{\cdot }{x%
}^{2}+\overset{\cdot }{y}^{2})+\frac{1}{2}I\overset{\cdot }{\theta }^{2}+%
\frac{1}{2}J\overset{\cdot }{\varphi }^{2}
\end{equation*}%
and the nonholonomic constraints (non sliding) are given by%
\begin{eqnarray*}
\overset{\cdot }{x} &=&Rcos\varphi \overset{\cdot }{\theta } \\
\overset{\cdot }{y} &=&Rsin\varphi \overset{\cdot }{\theta }
\end{eqnarray*}%
where $R$ is the radius of the disk. In this case, the base controlled curve
is $\tilde{c}(t)=\varphi (t)$ and 
\begin{equation*}
d_{0}^{NH}(t)=(x_{0},y_{0},\theta _{0},\varphi (t))
\end{equation*}%
is a nonholonomic gauge (which, in this example, also coincides with the
mechanical gauge). Also,%
\begin{equation*}
\mathfrak{g}^{q}=span\{(Rcos\varphi ,Rsin\varphi ,1)\in Lie(G)=\mathbb{R}%
^{2}\oplus \mathfrak{s(1)}\}
\end{equation*}%
and 
\begin{equation*}
J(\overset{\cdot }{c})=(m\overset{\cdot }{x},m\overset{\cdot }{y},I\overset{%
\cdot }{\theta }).
\end{equation*}%
From section \ref{subsec: G abelian}, the constraint equation in terms of $J(%
\overset{\cdot }{c})$ for this nonholonomic gauge reads $I_{0}^{-1}J(\overset%
{\cdot }{c})\in \mathfrak{g}^{q}$, or,%
\begin{equation*}
J(\overset{\cdot }{c})=\lambda (t)(mRcos\varphi (t),mRsin\varphi (t),I)
\end{equation*}%
for some $\lambda (t)\in \mathbb{R}$ to be determined by the corresponding
equation of motion $\left( \ref{Eq. motion abelian}\right) $ for $J(\overset{%
\cdot }{c})$:%
\begin{eqnarray*}
\left\langle i_{d_{0}(t)}^{\ast }(\frac{d}{dt}J(\overset{\cdot }{c}%
)),(Rcos\varphi ,Rsin\varphi ,1)\right\rangle &=&0 \\
\overset{\cdot }{\lambda }(mR^{2}+I)+\lambda \left[ \frac{d}{dt}%
(mRcos\varphi ,mRsin\varphi ,I)\right] \cdot (Rcos\varphi ,Rsin\varphi ,1)
&=&0.
\end{eqnarray*}%
Note that the second term in the last equation is zero because the two
vectors are orthogonal. Then, since $\lambda (t)=\overset{\cdot }{\theta }$
by the definition of the momentum $J(\overset{\cdot }{c})$, we have 
\begin{equation*}
\overset{\cdot }{\lambda }(mR^{2}+I)=\overset{\cdot \cdot }{\theta }%
(mR^{2}+I)=0
\end{equation*}%
which is the vertical equation of motion derived in \cite{BKMM}. The above
conservation law can be directly computed via equation $\left( \ref{Eq: G
abel paral transp}\right) $ since $\gamma _{1}^{1}=0$ (the underlying linear
connection in the $1-$dimensional bundle $\mathfrak{g}^{D}\longrightarrow
Q/G=S^{1}$ is flat, see sec. \ref{subsec: G abelian}). Consequently, $%
\overset{\cdot }{\theta }$ is constant. Finally, since we have solved for $J(%
\overset{\cdot }{c})$ using the eq. of motion and of constraints, we can
apply formula $\left( \ref{Eq: G abel solution}\right) $ obtaining%
\begin{equation*}
g(t)=\left( \overset{\cdot }{\theta }mR\left( \dint\limits_{t_{1}}^{t}ds\
cos\varphi (s)\right) ,\overset{\cdot }{\theta }mR\left(
\dint\limits_{t_{1}}^{t}ds\ cos\varphi (s)\right) ,I\overset{\cdot }{\theta }%
(t-t_{1})\right) .
\end{equation*}%
Note that $g_{Mech}(t)=(0,0,0)$ in this case. Finally, the full solution $%
c(t)\in Q$ is 
\begin{eqnarray*}
c(t) &=&g(t)\cdot d_{0}(t) \\
&=&\left( \overset{\cdot }{\theta }mR\left( \dint\limits_{t_{1}}^{t}ds\
cos\varphi (s)\right) ,\overset{\cdot }{\theta }mR\left(
\dint\limits_{t_{1}}^{t}ds\ cos\varphi (s)\right) ,I\overset{\cdot }{\theta }%
(t-t_{1})\right) \cdot (x_{0},y_{0},\theta _{0},\varphi (t)) \\
&=&\left( \overset{\cdot }{\theta }mR\left( \dint\limits_{t_{1}}^{t}ds\
cos\varphi (s)\right) +x_{0},\overset{\cdot }{\theta }mR\left(
\dint\limits_{t_{1}}^{t}ds\ cos\varphi (s)\right) +y_{0},I\overset{\cdot }{%
\theta }(t-t_{1})+\theta _{0},\varphi (t)\right)
\end{eqnarray*}%
from which we clearly see that motion is induced on the group variables from
the base controlled curve $\varphi (t)$ due to the presence of the
non-sliding nonholonomic ($D-$)constraints.

\subsection{Ball on a rotating turntable}

\label{Ex: ball on rotating table}We also recall the setting for describing
a ball on a rotating turntable\emph{\ }from \cite{BKMM}. This is an example
of the systems considered in sections \ref{subsec:affine constraints} and %
\ref{subsec: Trivial Q case}. The corresponding Lagrangian on $Q=\mathbb{R}%
^{2}\times SO(3)$ is%
\begin{equation*}
L=\frac{1}{2}m(\dot{x}^{2}+\dot{y}^{2})+\frac{1}{2}mk^{2}(\omega
_{x}^{2}+\omega _{y}^{2}+\omega _{z}^{2}),
\end{equation*}%
and the non-sliding \emph{affine }$D-$\emph{constraints} for the ball motion
are%
\begin{eqnarray*}
-\dot{x}+a\omega _{y} &=&\Omega y \\
\dot{y}+a\omega _{x} &=&\Omega x
\end{eqnarray*}%
where $(x,y)\in \mathbb{R}^{2}$ denote the ball%
\'{}%
s position and $\dot{g}=\omega _{x}\xi _{x}^{R}\left( g\right) +\omega
_{y}\xi _{y}^{R}\left( g\right) +\omega _{z}\xi _{z}^{R}\left( g\right) $
the angular velocity of $g(t)\in SO(3)$ representing the ball%
\'{}%
s rotation around its center. Here, $\xi _{i}^{R}(g)$ denotes the right
invariant vector in $T_{g}SO(3)\ $whose value at $e$ is $\xi _{i}\in 
\mathfrak{so(3)}$, the generator of rotations about the $i-$axis. Also
above, $a$ is the ball%
\'{}%
s radius, $mk^{2}$ its (any) principal\ moment of inertia and $\Omega $ the
given angular velocity of the rotating turntable. To take these eqs. to the
form of eq. $\left( \ref{Eq: aff constr}\right) $ we define%
\begin{eqnarray*}
\mathfrak{A}_{(x,y,g)}^{D}(\dot{x},\dot{y},\dot{g}) &=&\left( (\dot{x},\dot{y%
},\dot{g}),v_{q}^{4}\right) \frac{v_{q}^{4}}{\left\Vert v_{q}^{4}\right\Vert
^{2}}+\left( (\dot{x},\dot{y},\dot{g}),v_{q}^{5}\right) \frac{v_{q}^{5}}{%
\left\Vert v_{q}^{5}\right\Vert ^{2}} \\
\gamma (x,y,g) &=&\Omega y\frac{v_{q}^{4}}{\left\Vert v_{q}^{4}\right\Vert
^{2}}+\Omega x\frac{v_{q}^{5}}{\left\Vert v_{q}^{5}\right\Vert ^{2}}
\end{eqnarray*}%
where $\left( ,\right) =\left( ,\right) _{\mathbb{R}^{2}}+\left( ,\right) _{%
\mathfrak{so(3)}}$ denotes the kinetic energy inner product on $Q=\mathbb{R}%
^{2}\times G\ $with $G=SO(3)$ and $v_{q}^{4}=-\frac{\partial }{\partial x}%
+a\xi _{y}^{R}\left( g\right) $, $v^{5}=\frac{\partial }{\partial y}+a\xi
_{x}^{R}\left( g\right) $ in $TQ$. Note that both $D:=Ker(A^{D})=Span\{a%
\frac{\partial }{\partial x}+\xi _{y}^{R}\left( g\right) ;-a\frac{\partial }{%
\partial y}+\xi _{x}^{R}\left( g\right) ;\xi _{z}^{R}\left( g\right) \}$ and 
$\gamma $ are $G-$invariant for the natural \emph{right} action of $G$ on $Q$%
. Also notice that on the previous sections we considered a \emph{left} $G$
action on $Q$, so we turn the above natural right action into a left one by
defining $g\cdot (x,y,h)=(x,y,hg^{-1})$ in $\mathbb{R}^{2}\times G$.

In this case, since shape space $B\ $is $\mathbb{R}^{2}$, the controlled
curve $\tilde{c}(t)=(x(t),y(t))$ represents the position of the contact
point between the ball and the table as describing a given trajectory. So
the problem is to find out how the ball rotates (i.e. to find $g(t)$) due to
the presence of the non-sliding affine constraints and to the fact that the
contact point is \emph{moving in this known way} $(x(t),y(t))$. From section %
\ref{subsec:affine constraints}, we know that the corresponding equations
for the unknown $g(t)\in G$ are the eqs. of motion $\left( \ref{Eq. motion
for psi}\right) $ and the constraint eqs. $\left( \ref{Eq: aff constr
controlled}\right) $. Also from that section, we know that we can simplify
the constraint equation by considering an \emph{affine gauge }$d_{0}^{Aff}(t)
$ satisfying $\left( \ref{Eq: Aff gauge}\right) $. In the present example, $%
\mathfrak{g}^{(x,y,g)}=Span\{Ad_{g^{-1}}\xi _{z}\}\ $with $\xi _{z}\in 
\mathfrak{so(3)}$ the generator of rotations about the $z-$axis. Also, the\
momentum map for the above $G-$symmetric Lagrangian is $J(\dot{x},\dot{y},%
\dot{g})=-mk^{2}g^{-1}\dot{g}\in \mathfrak{so(3)}\simeq \mathfrak{so}^{\ast }%
\mathfrak{(3)}$. One possible affine gauge choice is%
\begin{equation*}
d_{0}^{Aff}(t)=(x(t),y(t),g_{Aff}(t))
\end{equation*}%
with $\dot{g}_{Aff}g_{Aff}^{-1}=\frac{1}{a}\left( -\dot{y}+\Omega x\right)
\xi _{x}+\frac{1}{a}\left( \dot{x}+\Omega y\right) \xi _{y}$, i.e., with no $%
z-$(spatial) angular velocity component. Consequently, the full solution $%
c(t)=((x(t),y(t),g_{tot}(t))$ is written as 
\begin{equation*}
c(t)=g(t)\cdot d_{0}^{Aff}(t)=(x(t),y(t),g_{Aff}(t)g^{-1}(t))
\end{equation*}%
with $g(t)$ satisfying:

\begin{enumerate}
\item \emph{(Constraints)} $g^{-1}\dot{g}\in \mathfrak{g}%
^{d_{0}^{Aff}(t)}=Span\{Ad_{g_{Aff}(t)^{-1}}\xi _{z}\}$

\item \emph{(Motion) }$\left( \frac{d}{dt}J(\dot{x},\dot{y},\dot{g}%
),Ad_{\left( g_{Aff}(t)g^{-1}(t)\right) ^{-1}}\xi _{z}\right) _{\mathfrak{%
so(3)}}=0$
\end{enumerate}

It is easy to see that, by calling $g_{tot}(t)=g_{Aff}(t)g^{-1}(t)$, eq. $%
(2) $ above reduces to $J_{z}^{S}(\dot{c}):=mk^{2}\left( \dot{g}%
_{tot}g_{tot}^{-1},\xi _{z}\right) _{\mathfrak{so(3)}}=const.$, i.e. the $z-$%
component of the (spatial) angular momentum is conserved, since%
\begin{eqnarray*}
\left( \frac{d}{dt}J(\dot{x},\dot{y},\dot{g}),Ad_{g_{tot}^{-1}}\xi
_{z}\right) _{\mathfrak{so(3)}} &=&\frac{d}{dt}\left( J(\dot{x},\dot{y},\dot{%
g}),Ad_{g_{tot}^{-1}}\xi _{z}\right) _{\mathfrak{so(3)}}+\left( J(\dot{x},%
\dot{y},\dot{g}),Ad_{g_{tot}^{-1}}ad_{\dot{g}_{tot}g_{tot}^{-1}}\xi
_{z}\right) _{\mathfrak{so(3)}} \\
&=&\frac{d}{dt}J_{z}^{S}(\dot{c})+mk^{2}\left( g_{tot}^{-1}\dot{g}%
_{tot},Ad_{g_{tot}^{-1}}ad_{\dot{g}_{tot}g_{tot}^{-1}}\xi _{z}\right) _{%
\mathfrak{so(3)}}
\end{eqnarray*}%
and the second term in the r.h.s. above vanishes. Notice that, although we
have a conservation law, it is a $1-$dimensional one and no non-trivial
reconstruction phase formulas for $g(t)$ follow from it.

\begin{remark}
\emph{(Conservation due to symmetry)\ }Using remark \ref{rmk: No D+TG}, we
can easily see, by considering $G$ as being only rotations about the $z$
axis and acting by \emph{left\ multiplication }on $Q$, that eq. $\left( \ref%
{Eq. motion c(t)}\right) $ becomes directly the above $z-$component
conservation of the corresponding (spatial)\ angular momentum. Nevertheless,
notice that this setting does not give any insight on the
constraint-base-induced motion $g(t)$.
\end{remark}

Now, from $(1)$ above, we get 
\begin{equation*}
g^{-1}\dot{g}=Ad_{g_{Aff}^{-1}(t)}\omega _{z}\xi _{z}
\end{equation*}%
and from $(2)\ $that 
\begin{equation*}
\omega _{z}=const.
\end{equation*}%
So, finally, the full base-induced group variable $g_{tot}(t)\ $in the full
system%
\'{}%
s motion $c(t)\ $is obtained as a product of the two simpler factors $%
g_{Aff}(t)g^{-1}(t)$ described above.

Note that, in this simple example, the factorization result we obtained
following our general considerations is the same as what we obtain by
proposing the solution $g_{tot}(t)=g_{Aff}(t)g^{-1}(t)$ for the constraints
plus conservation eqs. as expressed in ref. \cite{BKMM}:%
\begin{equation*}
\dot{g}_{tot}=\frac{1}{a}\left( -\dot{y}+\Omega x\right) \xi
_{x}^{R}(g_{tot})+\frac{1}{a}\left( \dot{x}+\Omega y\right) \xi
_{y}^{R}(g_{tot})+\left( const\right) \xi _{z}^{R}(g_{tot}).
\end{equation*}

\subsection{A non-holonomicaly constrained self-deforming body}

\label{Ex: 2 Balls}This is an example of a base controlled and $D-$%
constrained system presenting \emph{phase formulas} due to (non full)
horizontal symmetries (section \ref{subsubsec:Phases Horizontal symmetries}, 
\cite{BKMM}). The system consists of two rigid spheres as in Figure \ref%
{Fig: 1}. The small ball is attached to the inside of the big one (\emph{%
holonomic} constraint) which, in turn, can move freely. The key ingredient
is that the first rotates \emph{without sliding with respect to the second}.
This last requirement represents a non-holonomic $D-$constraint on the total
system and we further assume that no external forces are present. This gives
a simplified model for a small robot (the small ball) moving inside a
space-craft (the big ball). As we shall see below, this example generalizes
the treatment of \cite{C def body} by allowing \emph{non-holonomic
constraints} to induce total body motion from the arbitrarily controlled
(base) variables.

\begin{figure}
	\centering
		\includegraphics{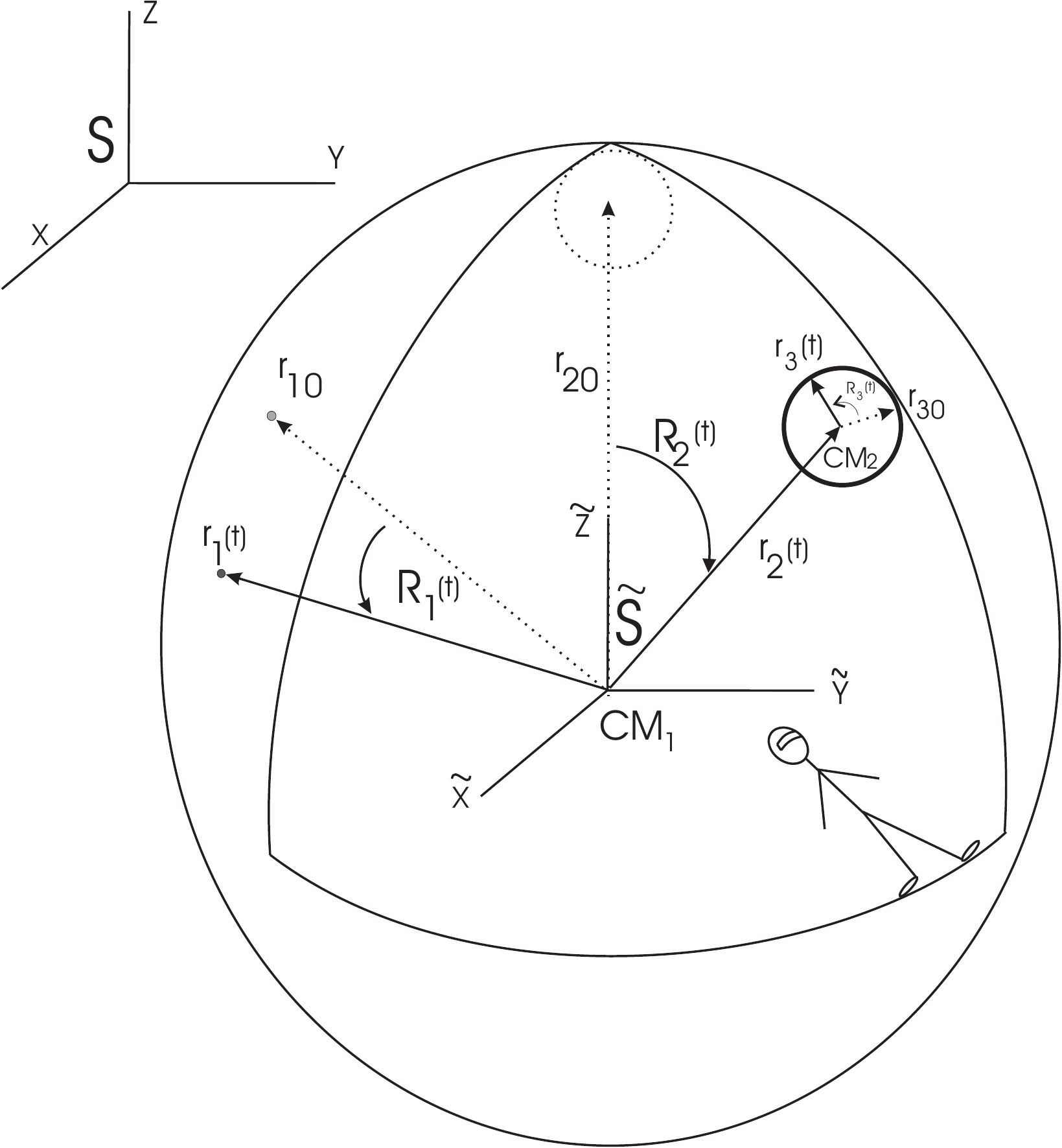}
	\caption{The big ball \'{} s rotation $R_{1}(t)$ and the position of the center $CM_{2}$ of the small
ball, both as seen from refeerence system $\tilde{S}$, are described by $%
r_{1}(t)=R_{1}(t)r_{10}$ and $r_{2}(t)=R_{2}(t)r_{20}$, respectivelly. $%
\tilde{S}$ has its origin at the center $CM_{1}$ of the big ball and axes
parallel to those of an inertial frame $S$. The rotation $R_{3}(t)\ $of the
small ball about its center $CM_{2}$ is described by the vector $%
r_{3}(t)=R_{3}(t)r_{30}$.}
	\label{Fig: 1}
\end{figure}

The configuration space is $Q=SO(3)\times S_{r}^{2}\times SO(3)\ni
(R_{1},r_{2},R_{3})$ defined by requiring $r_{i}(t)=R_{i}(t)r_{io}\in 
\mathbb{R}^{3}$ to be the position of the point $i$ with respect to a
reference system with axes parallel to those of a chosen inertial one and
with origin in the corresponding ball%
\'{}%
s center (see Figure \ref{Fig: 1}). We denoted by $S_{r}^{2}$ the $2-$sphere
of radius $r=\left\Vert r_{2}(t)\right\Vert =const$. In this coordinates,
the Lagrangian takes the simple kinetic energy form%
\begin{equation*}
L(\dot{R}_{i})=T(\dot{R}_{i})=\frac{1}{2}\left( R_{1}^{-1}\dot{R}%
_{1},I_{1}R_{1}^{-1}\dot{R}_{1}\right) _{\mathfrak{so(3)}}+\frac{1}{2}\mu
r^{2}\left( \dot{r}_{2}\cdot \dot{r}_{2}\right) +\frac{1}{2}\left( R_{3}^{-1}%
\dot{R}_{3},I_{3}R_{3}^{-1}\dot{R}_{3}\right) _{\mathfrak{so(3)}}
\end{equation*}%
and the $2$ non-sliding non-holonomic $D-$constraint eqs. (for $r_{20}=r\ 
\check{z})$ read%
\begin{eqnarray}
-\frac{a}{r}\left( Ad_{R_{2}^{-1}}\left( -\dot{R}_{1}R_{1}^{-1}+\dot{R}%
_{3}R_{3}^{-1}\right) ,\xi _{x}\right) _{\mathfrak{so(3)}} &=&\left(
Ad_{R_{2}^{-1}}\left( -\dot{R}_{1}R_{1}^{-1}+\dot{R}_{2}R_{2}^{-1}\right)
,\xi _{x}\right) _{\mathfrak{so(3)}}  \label{Eq: Ex 2balls constr} \\
-\frac{a}{r}\left( Ad_{R_{2}^{-1}}\left( -\dot{R}_{1}R_{1}^{-1}+\dot{R}%
_{3}R_{3}^{-1}\right) ,\xi _{y}\right) _{\mathfrak{so(3)}} &=&\left(
Ad_{R_{2}^{-1}}\left( -\dot{R}_{1}R_{1}^{-1}+\dot{R}_{2}R_{2}^{-1}\right)
,\xi _{y}\right) _{\mathfrak{so(3)}}.  \notag
\end{eqnarray}%
Above, $I_{1}=diag(\frac{2}{5}m_{1}(r+a)^{2})$, $I_{3}=diag(\frac{2}{5}%
m_{2}a^{2})$, with $a$ the small ball%
\'{}%
s radius, are the inertia tensors of the balls with respect to their
respective centers in the standard basis $\{\xi _{i}\}$ of $\mathfrak{so(3)}$
formed by the generators of $i-$axis rotations, $i=x,y,z$, and $\mu =\frac{%
m_{1}m_{2}}{m_{1}+m_{2}}$.

\begin{remark}
\emph{(}$R_{2}$ \emph{expressions)}. It can be easily seen that every
expression depending on $R_{2}$ given within this subsection, like the
constraints above, is invariant under $R_{2}(t)\rightsquigarrow
R_{2}(t)R_{z}(t)$ with $R_{z}(t)$ a rotation about the $z-$axis. This means
that they really depend on $r_{2}(t)=r\ R_{2}(t)\check{z}$, but we keep the
rotational dependence for simplicity. Given $r_{2}(t)\in S_{r}^{2}$, one
choice for $R_{2}(t)$ is given by the horizontal lift in the $U_{z}(1)$
bundle $SO(3)\longrightarrow S_{r}^{2}$ (see \cite{C def body}).
\end{remark}

The distribution $D\subset TQ$ of tangent vectors satisfying eqs. $\left( %
\ref{Eq: Ex 2balls constr}\right) $ has dimension $dimD=dimQ-2=6$.

Now, consider the group $G=SO(3)^{2}\ni (R,g_{3})$ (left)\ acting on $Q\ $via%
\begin{equation*}
(R,g_{3})\cdot (R_{1},r_{2},R_{3})=(RR_{1},Rr_{2},RR_{3}g_{3}^{-1}).
\end{equation*}%
It is easy to see that both $L$ and $D$ are $G-$invariant. Shape space $B=Q/G
$ can be parameterized by elements $r_{2,1}\in S_{r}^{2}$ and hypothesis $%
(H1)$ of sec. \ref{subsec:Kinematical setting} is satisfied. We also assume $%
(H2)$ to hold, which, in this case, means that the controlled part of the
motion is represented by a gauge curve $d_{0}(t)=(e,r_{2,1}(t),e)$. If $%
c(t)=(R_{1}(t),R_{2}(t),R_{3}(t))\in Q$ represents the full system%
\'{}%
s motion, then $r_{2,1}(t)=R_{1}^{-1}(t)r_{2}(t)$ represents the position of
the $CM_{2}$ as seen from a reference system with origin at $CM_{1}$ and
with axes rotating with $R_{1}$, i.e. a system rotating with the big ball.
Indeed, the full motion can be written as $%
c(t)=(R_{1}(t),R_{3}^{-1}(t)R_{1}(t))\cdot d_{0}(t)$ and note that \emph{no
constraints} remain on the \emph{controlled variable} $r_{2,1}$ (it can be
arbitrarily chosen within $B\simeq S_{r}^{2}$). Also, notice that from the $%
dimQ=8$ variables, as $2$ are being freely controlled, we are left with $4$ 
\emph{equations of motion }plus the $2$ \emph{constraints} to solve.

More physically, the problem is to find the \emph{total reorientation of the
system} $R_{1}(t)$ induced by the inside motion. This, in turn, is generated
by the inner translational motion $d_{0}(t)$ of the small ball and followed
by its $D-$induced rotational motion $R_{1}^{-1}(t)R_{3}(t)$, both as seen
from a system fixed to the big ball, due to the presence of the non-sliding
constraints.

\begin{remark}
\label{Rmk: Ex 2balls ref syst}\emph{(Measurement of }$r_{2,1}$\emph{)\ }The
curve $r_{2,1}(t)$ is the one that an astronaut standing in the space-craft,
modeled by the big ball, would see as the small ball%
\'{}%
s center moves (see Figure \ref{Fig: 1}). Consequently, it can also be
measured in lab conditions, when the space-craft is attached to the floor
(and cannot rotate), but when the small ball rehearses the same
translational motion $r_{2,1}$ that will occur in space.
\end{remark}

We now turn to the equations of motion. Consider the subgroup $H:=\{(R,e),\
R\in SO(3)\}\subset G$. It can be easily checked that $\mathfrak{h}%
_{Q}=(Lie(H))_{Q}\subset D$ and that, for $q=(R_{1},r_{2},R_{3})\in Q$,%
\begin{equation*}
\mathfrak{g}^{q}=\mathfrak{h}\oplus Span\{Ad_{R_{3}^{-1}}Ad_{R_{2}}\xi
_{z}^{3}\}
\end{equation*}%
with $\xi _{z}^{3}$ seen as an element of the second $\mathfrak{so(3)\ }$%
copy in $Lie(G)=\mathfrak{so(3)}\oplus \mathfrak{so(3)}$. The above means
that we are in the presence of \emph{non full} $\mathfrak{h}$-horizontal
symmetries (\cite{BKMM}). Consequently,%
\begin{equation*}
i_{\mathfrak{h}}^{\ast }J(\dot{c})=I_{1}\dot{R}_{1}R_{1}^{-1}+\mu
r^{2}\left( r_{2}\times \dot{r}_{2}\right) ^{\curlywedge }+I_{3}\dot{R}%
_{3}R_{3}^{-1}=I_{1}\dot{R}_{1}R_{1}^{-1}+Ad_{R_{2}}I_{20}R_{2}^{-1}\dot{R}%
_{2}+I_{3}\dot{R}_{3}R_{3}^{-1}
\end{equation*}%
in $\mathfrak{so(3)}\overset{metric}{\simeq }\mathfrak{so}^{\ast }\mathfrak{%
(3)}=Lie(H)^{\ast }$ is a \emph{conserved quantity}. Above, $^{\curlywedge }$
denotes the (Lie algebra) isomorphism $\mathbb{R}^{3}\longrightarrow 
\mathfrak{so(3)}$ and%
\begin{equation*}
I_{2,0}=\mu r^{2}\left( 
\begin{array}{ccc}
1 &  &  \\ 
& 1 &  \\ 
&  & 0%
\end{array}%
\right) .
\end{equation*}%
This horizontal momentum represents the \emph{total} \emph{angular momentum}
of the system \cite{Goldstein}.

\begin{remark}
\emph{(Relevance of the present approach due to constraints)\ }We would like
to remark that, if we considered only $H$ as symmetry group, as it is done
for non-constrained self deforming bodies (see \cite{C def body}), then the $%
D-$constraints are no longer vertical (remark \ref{rmk: Vertical D}). In
other words, the corresponding base variables become constrained and it
would make no sense to think of them as \emph{arbitrarily} controlled or
given. By considering the bigger $G$ instead, we restrict to the smaller
base variable space which are actually a priori arbitrarily controllable.
\end{remark}

Note that $dim\mathfrak{g}^{q}=4$, so the above conservation law represents
only $3$ of eqs. of motion $\left( \ref{Eq. motion c(t)}\right) $. The
remaining equation is%
\begin{equation}
\left( \frac{d}{dt}\left( R_{3}^{-1}\dot{R}_{3}\right)
,Ad_{R_{3}^{-1}}Ad_{R_{2}}\xi _{z}\right) _{\mathfrak{so(3)}}=0,
\label{Eq: Ex 2balls motion 3}
\end{equation}%
which tells that there is no angular acceleration of the smaller ball
rotation in the $CM_{1}-CM_{2}$ direction. This same effect is observed in
the ball on a rotating turn-table example (see \cite{BKMM} and the previous
section).

Finally, from section \ref{subsubsec:Phases Horizontal symmetries}, we know
that we can write (reconstruction) phase formulas for the system%
\'{}%
s motion due to the horizontal conservations. Below, we summarize the $Q-$%
reconstruction procedure for obtaining the solution $c(t)$ from the base
motion $\tilde{c}(t)$.

\begin{itemize}
\item We start with $d_{0}(t)=(e,r_{2,1}(t),e)$, and $c(t)=$ $%
(R_{1}(t),r_{2}(t),R_{3}(t))\in Q$ representing the desired solution.

\item To use the results of the previous sections, we choose a \emph{%
non-holonomic gauge }$d_{0}^{NH}$. We fix it by $%
d_{0}^{NH}(t)=(R_{1,NH},R_{3,NH}^{-1}R_{1,NH})(t)\cdot d_{0}(t)$ with 
\begin{gather*}
\left[ constraints+1eq.\right] \ -\frac{a}{r}\left( -\dot{R}%
_{1,NH}R_{1,NH}^{-1}+\dot{R}_{3,NH}R_{3,NH}^{-1}\right) =Ad_{R_{1,NH}}\dot{R}%
_{2,1}R_{2,1}^{-1} \\
\left[ i_{\mathfrak{h}}^{\ast }J(\dot{d}_{0}^{NH})=0\right] \ I_{3}\dot{R}%
_{3,NH}R_{3,NH}^{-1}+\left(
I_{1}+Ad_{R_{1,NH}R_{2,1}}I_{20}Ad_{R_{2,1}^{-1}}\right) \dot{R}%
_{1,NH}R_{1,NH}^{-1}+ \\
+Ad_{R_{1,NH}R_{2,1}}I_{20}R_{2,1}^{-1}\dot{R}_{2,1}=0
\end{gather*}%
with $r_{2,1}(t)=rR_{2,1}(t)\check{z}$ and trivial initial conditions for $%
R_{i,NH}$, $i=1,3$. Equivalently, we could have chosen the \emph{horizontal}
non-holonomic gauge $\left( \ref{Eq. nonholonomic gauge relation}\right) $
leading to the same $i_{\mathfrak{h}}^{\ast }J(\dot{d}_{0}^{NH})=0$ equation
plus constraint eqs. $\left( \ref{Eq: Ex 2balls constr}\right) $ plus one
more (involved) equation.

\item We now write $c(t)=(R,g_{3})(t)\cdot d_{0}^{NH}(t)$. Notice that,
since the horizontal symmetries are non-full, eq.$\left( \ref{Eq. D const
for psi in NHgauge}\right) $ for $g(t)\equiv (R,g_{3})(t)$ is non-trivial
and yields%
\begin{equation*}
g_{3}^{-1}\dot{g}_{3}=\lambda (t)\ Ad_{R_{3,NH}^{-1}}Ad_{R_{1,NH}R_{2,1}}\xi
_{z}
\end{equation*}%
with $\lambda (t)\in \mathbb{R}$ to be determined. The corresponding
vertical equations of motion for $g(t)$ read%
\begin{eqnarray}
\left[ \mathfrak{h-}conservation\right] \ i_{\mathfrak{h}}^{\ast }J(\dot{c})
&=&const=Ad_{R}\left( I_{d_{0}^{NH}}^{\mathfrak{h}}R^{-1}\dot{R}-\lambda
I_{3}Ad_{R_{1,NH}R_{2,1}}\xi _{z}\right) =:Ad_{R}\Pi ^{\mathfrak{h}}(t)
\label{Eq: Ex 2 balls motion} \\
\left[ Eq.\left( \ref{Eq: Ex 2balls motion 3}\right) \right] \ \overset{%
\cdot }{\lambda } &=&\left( \frac{d}{dt}\left[ Ad_{R_{3,NH}^{-1}}R^{-1}\dot{R%
}+R_{3,NH}^{-1}\dot{R}_{3,NH}\right] ,Ad_{R_{3,NH}^{-1}}Ad_{R_{1,NH}R_{2,1}}%
\xi _{z}\right) _{\mathfrak{so(3)}}  \notag
\end{eqnarray}%
with $I_{d_{0}^{NH}}^{\mathfrak{h}}=I_{1}+Ad_{R_{1,NH}R_{2,1}}I_{2,0}Ad_{%
\left( R_{1,NH}R_{2,1}\right) ^{-1}}+I_{3}:\mathfrak{h}\longrightarrow 
\mathfrak{h}\simeq \mathfrak{h}^{\ast }\simeq \mathfrak{so(3)}$ the
corresponding restricted inertia tensor.
\end{itemize}

Above, $g_{3}\ $is s.t. $R_{1,NH}^{-1}R_{3,NH}\
g_{3}^{-1}=R_{3,1}(t)=R_{1}^{-1}(t)R_{3}(t)$ represents the rotational
motion of the small ball as seen from a reference system with origin at $%
CM_{2}$ and axes rotating with the big ball, i.e., is what an astronaut
standing inside the big ball would see (see remark \ref{Rmk: Ex 2balls ref
syst} and Figure \ref{Fig: 1}).\ Also, $\lambda =\left( g_{3}^{-1}\dot{g}%
_{3},Ad_{R_{3,NH}^{-1}}Ad_{R_{1,NH}R_{2,1}}\xi _{z}\right) _{\mathfrak{so(3)}%
}=\left( \dot{g}_{3}g_{3}^{-1},Ad_{R_{3}^{-1}R_{2}}\xi _{z}\right) _{%
\mathfrak{so(3)}}\ $represents a dynamical correction to the (spatial)
angular velocity of the small ball in the direccion $CM_{1}-CM_{2}$ needed
for eq. $\left( \ref{Eq: Ex 2balls motion 3}\right) $ to be satisfyied from
an inertial reference frame.

Notice that the above equations of motion for $R$ and $\lambda $ are
coupled. Nevertheless, in the obtained factorization%
\begin{equation*}
c(t)=\left( R\ R_{1,NH},R\ R_{1,NH}\ R_{2,1},R\ R_{3,NH}\ g_{3}^{-1}\right)
\end{equation*}%
every element as defined above represents a simpler piece from which the
overall motion is constructively induced from the known one $R_{2,1}(t)$ on
the base. This shows how we can (geometrically)\ take advantage of the
kinematical structure of the system for writing the controlled solution.
Moreover, the global reorientation $R$ can be further factorized by
implementing the \emph{phase formulas} corresponding to the $\mathfrak{h-}$%
conservation reconstruction (sec. \ref{subsubsec:Phases Horizontal
symmetries}).

\paragraph{The phase formula for $R$.}

From sec. \ref{subsubsec:Phases Horizontal symmetries}, we know that $R(t)$
can be reconstructed from the \emph{body total angular momentum }$\Pi ^{%
\mathfrak{h}}(t)$ solution on $O_{i_{\mathfrak{h}}^{\ast }J(\dot{c})}\simeq
S_{radius=i_{\mathfrak{h}}^{\ast }J(\dot{c})}^{2}\subset Lie(H)=\mathfrak{%
so(3)}\simeq \mathbb{R}^{3}$ within the $U(1)-$bundle $SO(3)\longrightarrow
S_{radius=\left\Vert i_{\mathfrak{h}}^{\ast }J\right\Vert }^{2}$ (see
details in \cite{C def body}). In this case, $\Pi ^{\mathfrak{h}}(t)$ was
defined in eq. $\left( \ref{Eq: Ex 2 balls motion}\right) $ and, from $%
\left( \ref{Eq. total Pi HS}\right) $ via $\mathfrak{so(3)}\simeq \mathbb{R}%
^{3}$,%
\begin{equation*}
\dot{\Pi}^{\mathfrak{h}}(t)=\Pi ^{\mathfrak{h}}(t)\times \left(
(I_{d_{0}^{NH}(t)}^{\mathfrak{h}})^{-1}\left[ \Pi ^{\mathfrak{h}}(t)+\lambda
I_{3}R_{1,NH}\frac{r_{2,1}}{r}\right] \right)
\end{equation*}%
with $\times $ standing for the usual vector product in $\mathbb{R}^{3}$.
This equation coincides with the one generically presented in \cite{C def
body} but in a \emph{very precise non-holonomic gauge }$\tilde{d}%
_{0}^{NH}=(e,g_{3})\cdot d_{0}^{NH}$, which makes the whole procedure
compatible with the $D-$constraints. Also in this case, this equation
appears coupled another equation, i.e. that of $\lambda $, since the
horizontal symmetries are non-full.

The \emph{phase formula }corresponding to the reconstruction of sec. \ref%
{subsubsec:Phases Horizontal symmetries}, for $i_{\mathfrak{h}}^{\ast }J\neq
0$, reads%
\begin{equation*}
R(t)=exp\left( \theta ^{Dyn}(t)\frac{i_{\mathfrak{h}}^{\ast }J}{\left\Vert
i_{\mathfrak{h}}^{\ast }J\right\Vert }\right) R_{1}^{Geom}(t)
\end{equation*}%
with the constant $\ i_{\mathfrak{h}}^{\ast }J\in \mathfrak{so(3)}$. The 
\emph{geometric phase }$R^{Geom}(t)\ $is the \emph{horizontal lift} of the
body total angular momentum curve $\Pi ^{\mathfrak{h}}(t)$ in the $U(1)-$%
bundle $SO(3)\longrightarrow S_{radius=\left\Vert i_{\mathfrak{h}}^{\ast
}J\right\Vert }^{2}$ with respect to the connection $A_{g}(\dot{g})=\left( 
\dot{g}g^{-1},\frac{i_{\mathfrak{h}}^{\ast }J}{\left\Vert i_{\mathfrak{h}%
}^{\ast }J\right\Vert }\right) _{\mathfrak{so(3)}}$(for details, see \cite{C
def body}).

The \emph{dynamical phase }$\theta ^{Dyn}(t)\in U(1)=H_{i_{\mathfrak{h}%
}^{\ast }J}$ is defined by (recall sec. \ref{subsubsec:Phases Horizontal
symmetries}) 
\begin{gather*}
\theta ^{Dyn}(t)=\frac{1}{\left\Vert i_{\mathfrak{h}}^{\ast }J\right\Vert }%
\int_{t_{1}}^{t}ds\ [2K(\frac{d}{dt}c(s))-2K_{int}(s)+\lambda (s)\left( 
\frac{2}{5}m_{2}a^{2}\right) \left( \xi _{z},\left( I_{e}^{\mathfrak{h}%
}\right) ^{-1}Ad_{R_{2}^{-1}(s)}i_{\mathfrak{h}}^{\ast }J\right) _{\mathfrak{%
so(3)}}+ \\
+\frac{\lambda (s)^{2}\left( \frac{2}{5}m_{2}a^{2}\right) ^{2}}{\frac{2}{5}%
m_{1}(r+a)^{2}+\frac{2}{5}m_{2}a^{2}}]+\theta _{0}^{Dyn}
\end{gather*}%
where $K$ represents the kinetic energy of the whole $Q$ system given in
Appendix \ref{sec:App Kin En} and $I_{e}^{\mathfrak{h}}=I_{1}+I_{2,0}+I_{3}$%
. Rotation $R_{2}$ is defined by $r_{2}(t)=rR_{2}(t)\check{z}$, giving the
physical motion of $CM_{2}\ $in $c(t)$. Notice the unavoidable (dynamical) $%
\lambda $\ dependance of the dynamical phase formula due to the fact that
the horizontal symmetries are non-full (also compare to the non-$D-$%
constrained case of \cite{C def body}).

Finally, it is worth noting that, when the solution $\Pi ^{\mathfrak{h}}(t)$
is simple and closed for $t\in \lbrack t_{1},t_{2}]$, then 
\begin{equation*}
R(t_{2})=exp\left( \left( \theta ^{Dyn}(t_{2})+\theta ^{Geom}\right) \frac{%
i_{\mathfrak{h}}^{\ast }J}{\left\Vert i_{\mathfrak{h}}^{\ast }J\right\Vert }%
\right) R_{1}(t_{1})
\end{equation*}%
with $\theta ^{Dyn}(t_{2})$ as given above and $\theta ^{Geom}$ given (mod. $%
2\pi $) by minus the (signed)\ \emph{solid angle} enclosed by $\Pi ^{%
\mathfrak{h}}(t)$ in the $2-$sphere of radius $\left\Vert i_{\mathfrak{h}%
}^{\ast }J\right\Vert $ within $\mathbb{R}^{3}\simeq \mathfrak{so(3)}$. The
above is an example of a ($D-$)generalized self deforming body phase
formula, not encoded in \cite{C def body} .

\begin{remark}
\emph{(Control)\ }The above formulas can be useful for control purposes,
this is, when you want to find the suitable base curve $R_{2,1}(t)$ inducing
a certain desired global reorientation $R(t_{2})$.
\end{remark}

\begin{remark}
\emph{(The case }$i_{\mathfrak{h}}^{\ast }J=0$\emph{) }In this case, the
equation for $R$ is \emph{geometrical}, meaning that it is a horizontal lift
equation along $\tilde{d}_{0}^{NH}=(e,g_{3})\cdot d_{0}^{NH}$ with respect
to the $\mathfrak{h}-$mechanical connection. Nevertheless, this equation is
coupled to that of $g_{3}$ (equiv. $\lambda $) which is not of geometric
nature. Consequently, the complete motion induction from the initial \emph{%
controlled} base variables $\tilde{c}(t)=R_{2,1}(t)\in B$ is not entirely
geometrical. The cause is that horizontal symmetries are non-full (compare
with remark \ref{rmk: HS zero momentum}) and so they do not exhaust the
whole vertical dynamics (i.e. because of the additional dynamical eq. $%
\left( \ref{Eq: Ex 2balls motion 3}\right) $, see also \cite{BKMM} for
similar comments).
\end{remark}

\subsection{Deforming body with dipolar magnetic moment in an external
magnetic field}

\label{subsubsec:Example dipolar magnetic}Here we describe the motion of a
(deforming) body with magnetic moment $M\in \mathbb{R}^{3}$ in the presence
of an external magnetic field. This system is modeled as an affine $D-$%
constrained and controlled system for which momentum is not conserved
because of the magnetic applied forces and which is, thus, not covered by
the analysis of \cite{C def body}. We shall assume the following hypothesis
about the magnetic nature of the system to hold:

\begin{itemize}
\item the magnetic moment is proportional to the \emph{total angular momentum%
} $J$, i.e.%
\begin{equation*}
M=\gamma J
\end{equation*}%
where $\gamma $ is the \emph{giromagnetic ratio }(\cite{Goldstein}).

\item the interaction with an external magnetic field $B$ is of dipolar type
(\cite{Goldstein}), this is%
\begin{equation*}
\frac{d}{dt}J=M\times B
\end{equation*}%
where $M\times B$ is the external torque acting on the dipole and $\times $
denotes the standard vector product in $\mathbb{R}^{3}$.

\item the above holds even when the shape $c(t)\in Q$ (see \cite{Mont
gauge,SW})\ of the underlying body and the field $B(t)$ \bigskip are
changing with time.
\end{itemize}

From the above assumptions, the equation of motion for the angular momentum $%
J(\dot{c})$ of the body is%
\begin{equation*}
\frac{d}{dt}J(\dot{c})=\gamma J(\dot{c})\times B(t).
\end{equation*}%
If we define the corresponding \emph{Larmor frecuency vector }(\cite%
{Goldstein})\emph{\ }as $\omega _{l}(t):=-\gamma B(t)\in \mathbb{R}^{3}$,
then the above can be re-expressed as%
\begin{equation*}
J(\dot{c})=h_{M}(t)L_{0}\overset{\Psi }{\equiv }Ad_{h_{M}(t)}^{\ast }\hat{L}%
_{0}
\end{equation*}%
where $h_{M}(t)\in SO(3)$ is defined by%
\begin{eqnarray*}
\overset{\cdot }{h}_{M}h_{M}^{-1}(t) &=&\hat{\omega}_{l}(t) \\
h_{M}(t_{1}) &=&Id
\end{eqnarray*}%
and $\hat{\omega}_{l}=\Psi ^{-1}(\omega _{l})\in so(3)$ for the usual Lie
algebra isomorphism $\Psi :(so(3),[,])\longrightarrow (\mathbb{R}^{3},\times
)$. Also above, $\hat{L}_{0}$ denotes the initial value $J(\dot{c}(t_{1}))\ $%
seen as an element of $so(3)^{\ast }$ through the usual isomorphisms.

The equations for the motion of such a system can be derived from the
affine-constrained Lagragian system $(TQ,\mathcal{L},\mathcal{A}^{D},\Gamma
) $ where

\begin{itemize}
\item $Q\longrightarrow Q/G$ is the configuration space of the underlying
deforming body\ (\cite{Mont gauge,C def body}) with symmetry group $G=SO(3)$

\item the Lagrangian is given by the kinetic energy contribution $\mathcal{L}%
(\dot{q})=\frac{1}{2}k_{q}(\dot{q},\dot{q})$, where $k_{q}$ a $G-$invariant
metric on $TQ$ induced by the standard $\mathbb{R}^{3}-$metric (\cite{Mont
gauge}),

\item $\mathcal{A}^{D}$ is the \emph{mechanical} principal connection $1-$%
form on $Q\longrightarrow Q/G$ given by 
\begin{equation*}
\mathcal{A}^{D}(\dot{q})=I_{q}^{-1}J(\dot{q})
\end{equation*}%
where $I_{q}$ denotes the inertia tensor and $J:TQ\longrightarrow \mathfrak{g%
}^{\ast }$ the usual \emph{angular momentum} map,

\item $\Gamma :Q\longrightarrow \mathfrak{g}$ is the map given by 
\begin{equation*}
\Gamma (q)=I_{q}^{-1}(Ad_{h_{M}(t)}^{\ast }\hat{L}_{0}).
\end{equation*}
\end{itemize}

The \emph{affine constraints} for the physical curve $c(t)$ become%
\begin{equation}
\mathcal{A}^{D}(\dot{c}(t))=\Gamma (c(t)).  \label{Eq. aff const mag field}
\end{equation}

We now continue with the analysis in the \emph{controlled case}, i.e., we
add hypothesis $\left( H2\right) $ that the base curve $\tilde{c}(t)\in Q/G$%
, representing the changing body%
\'{}%
s shape, is given.

Note that the distribution $D$ corresponding to the \emph{mechanical
connection} $\mathcal{A}^{D}$ is transversal to the group orbit since it is
a principal connection (see details in \cite{Mont gauge}). Then, results
from section \ref{subsec:affine constraints} in this particular case, say
that eqs. of motion $\left( \ref{Eq. motion affine}\right) $ for $g(t)$ are
trivial (i.e. $0=0$ because $\mathfrak{g}^{q}=0\forall q)$. The only
remaining equations for $g(t)$ are the constraint ones $\left( \ref{Eq: aff
constr controlled}\right) $ which, in a \emph{mechanical gauge }$%
d_{0}^{Mec}(t)$ $\left( \ref{Eq: Mech gauge}\right) $ with $%
d_{0}^{Mec}(t_{1})=c(t_{1})$, read%
\begin{eqnarray*}
Ad_{g(t)}^{\ast }I_{d_{0}^{Mec}(t)}^{g}\left( g^{-1}\dot{g}\right)
&=&Ad_{h_{M}(t)}^{\ast }\hat{L}_{0} \\
g(0) &=&Id.
\end{eqnarray*}

Following section \ref{subsec: Phases for dipolar}, we call 
\begin{equation*}
R_{M}(t)=h_{M}^{-1}(t)g(t)\in SO(3)
\end{equation*}%
and note that 
\begin{equation*}
Ad_{R_{M}(t)}^{\ast }I_{d_{0}^{Mec}(t)}\left( g^{-1}\dot{g}\right) =\hat{L}%
_{0}
\end{equation*}%
is a \emph{conserved quantity}. The passage from $g$ to $R_{M}$ can be
understood as passing to describe the system from a new reference frame
which is rotating via $h_{M}(t)$ with respect to the original (inertial)
frame, i.e., with spatial angular velocity $\omega _{l}(t)$ (see \cite%
{Goldstein} pp. 231).

The above conservation equation can be turned into the form%
\begin{equation*}
L(R_{M}(t),\Pi (t))=\hat{L}_{0}\in \mathfrak{g}^{\ast }
\end{equation*}%
with $\Pi (t):=I_{d_{0}^{Mec}(t)}\left( g^{-1}\dot{g}\right) $ the \emph{%
body angular momentum }and $L:G\times \mathfrak{g}^{\ast }\longrightarrow 
\mathfrak{g}^{\ast }$, $L(R_{M},\Pi )=Ad_{R_{M}}^{\ast }\Pi $. We are in the
situation described in Appendix \ref{sec:App Reconstr}. The rotation $%
R_{M}(t)$ can be thus reconstructed (\cite{MMR}) from $\Pi (t)$ within the $%
U(1)-$bundle $L^{-1}(\hat{L}_{0})\simeq SO(3)\longrightarrow O_{\hat{L}_{0}}$%
. Note that $O_{\hat{L}_{0}}\simeq S^{2}$ for $\hat{L}_{0}\neq 0$. The above
defined $\Pi (t)$ must lie in the coadjoint orbit $O_{\hat{L}_{0}}$ and
satisfies 
\begin{equation*}
\frac{d}{dt}\Pi (t)=\Pi (t)\times \Psi \left( R_{M}^{-1}\overset{\cdot }{R}%
_{M}\right) =\Pi (t)\times \Psi \left( I_{d_{0}^{Mec}(t)}^{-1}\Pi
(t)-Ad_{g^{-1}}\hat{\omega}_{l}(t)\right) .
\end{equation*}%
The reconstruction procedure follows the lines of \cite{C def body} and sec. %
\ref{subsec: Phases for dipolar}.\ Suppose that the solution $\Pi (t)$
describes a closed simple curve on the sphere $S^{2}=O_{\hat{L}_{0}}$, $\Pi
(t_{1})=\Pi (t_{2})=\hat{L}_{0}$; reconstruction yields%
\begin{equation*}
R_{M}(t_{2})=exp\left( (\theta ^{Dyn}(t_{2})+\theta ^{Geom})\frac{\hat{L}_{0}%
}{\left\Vert \hat{L}_{0}\right\Vert }\right)
\end{equation*}%
where the \textbf{geometric phase} angle $\theta ^{Geom}$ can be shown to be
(mod $2\pi $) minus the signed solid angle determined by the closed path $%
\Pi (t)$ on the sphere (see \cite{C def body}), and the \emph{dynamical
phase }$\theta ^{Dyn}(t)$ is calculated by 
\begin{equation*}
\theta ^{Dyn}(t)=\frac{1}{\left\Vert \hat{L}_{0}\right\Vert }%
\dint\limits_{t_{1}}^{t}ds\left( \left\langle \Pi
(s),I_{d_{0}^{Mec}(s)}^{-1}\Pi (s)\right\rangle -\left\langle \hat{J}(\dot{c}%
),\hat{\omega}_{l}(s)\right\rangle \right) .
\end{equation*}%
In the above expression, the first term gives $2K-2K_{int}$, where $K$
represents the \emph{rotational kinetic energy }(see Appendix \ref{sec:App
Kin En})\ and the second term is the \emph{magnetic potential energy }of the
system (see \cite{Goldstein} pp. 230). Finally, we have a \textbf{phase
formula} for the physical curve $c(t)\in Q$%
\begin{equation*}
c(t_{2})=h_{M}(t_{2})\cdot exp(\theta ^{Dyn}(t_{2})\frac{\hat{L}_{0}}{%
\left\Vert \hat{L}_{0}\right\Vert })\cdot exp(\theta ^{Geom}\frac{\hat{L}_{0}%
}{\left\Vert \hat{L}_{0}\right\Vert })\cdot d_{0}^{Mec}(t_{2})
\end{equation*}%
which determines exactly the position of the system for the dynamically
defined time $t_{2}$ in which the \emph{body angular momentum }$\Pi (t)$
returns to its initial value. This is the \emph{affine-constrained }%
(magnetic) version of the result obtained in \cite{C def body}.

\section{Appendix:\ Kinetic Energy}

\label{sec:App Kin En}Here we derive an expression for the kinetic energy of
the mechanical system on $Q$, in terms of the \emph{controlled variables
curve} $d_{0}(t)$ and the \emph{group unknwon }$g(t)$.

We shall assume that $Q$ is a Riemannian manifold and that the kinetic
energy of the underlying simple mechanical system (with or without controls)
is given by the corresponding metric on $Q$. This means that, if $%
k:TQ\otimes TQ\longrightarrow TQ$ denotes this metric, then the \textbf{%
kinetic energy }reads:%
\begin{eqnarray*}
K &:&TQ\longrightarrow \mathbb{R} \\
K(v_{q}) &=&\frac{1}{2}k_{q}(v_{q},v_{q}).
\end{eqnarray*}%
Now, on our \emph{controlled system}, the physical curve $c(t)\in Q$ is of
form $\left( \ref{Eq constr def c(t)}\right) $ and then, the velocity $\dot{c%
}(t)$ is given by $\left( \ref{veloc c(t)}\right) $. Thus, the kinetic
energy on the controlled curve becomes%
\begin{equation*}
K(\dot{c}(t))=K_{int}(t)+\frac{1}{2}\left\langle I_{0}(t)(\xi (t)),\xi
(t)\right\rangle +\left\langle J_{0}(t),\xi (t)\right\rangle
\end{equation*}%
where 
\begin{equation*}
K_{int}(t)=\frac{1}{2}k_{d_{0(t)}}\left( \dot{d}_{0}(t),\dot{d}_{0}(t)\right)
\end{equation*}%
shall be called the \emph{internal }(or \emph{gauge})\emph{\ kinetic energy}
and \emph{\ }$\xi (t)=g^{-1}\frac{d}{dt}g(t)\in \mathfrak{g}$, $%
I_{0}(t)=I_{d_{0}(t)}$, $J_{0}(t):=J(\dot{d}_{0}(t))$ as in section \ref%
{subsubsec:Eqs g(t)}.

In terms of the \emph{body momentum} $\Pi (t)$ defined by equation $\left( %
\ref{eq. def Pi}\right) $, the expression takes the form:%
\begin{equation*}
K(\frac{d}{dt}c(t))=K_{int}(t)+\frac{1}{2}\left\langle I_{0}^{-1}(t)(\Pi
(t)),\Pi (t)\right\rangle -\frac{1}{2}\left\langle
I_{0}^{-1}(t)(J_{0}(t)),J_{0}(t)\right\rangle
\end{equation*}%
where the last term can be interpreted as a \emph{gauge dependent energy
contribution }which appears because of the use of the \emph{%
\"{}%
moving\ reference\ system%
\"{}
}represented by $d_{0}(t)$ in $Q$.

\begin{remark}
\emph{(Mechanical energy) }If there are also potential forces present in the
mechanical system on $Q$, represented by a \emph{potential} $%
V:Q\longrightarrow \mathbb{R}$, then the total \emph{mechanical energy }is $%
E=K(\frac{d}{dt}c(t))+V(g\cdot d_{0}(t))$. If $V$ is $G-$invariant, then $%
E=K(\frac{d}{dt}c(t))+V(d_{0}(t))$. Notice that as, in general, the \emph{%
control forces }are non potential and time-dependent, they do \emph{work }on
the system. So the above mechanical energy is \emph{not conserved} during
the controlled motion.
\end{remark}

\begin{remark}
\emph{(Gauge potential interaction) }In terms of (1-d)\ gauge field
theories, the term $\left\langle J_{0}(t),\xi (t)\right\rangle $ can be seen
as a coupling between the \emph{gauge field} $J_{0}$ and the \emph{gauge
variables} $\xi $ (see also remark \ref{rmk: 1d fields}).
\end{remark}

Recall the \textbf{mechanical connection} on $Q\overset{\pi }{%
\longrightarrow }Q/G$ (see \cite{Mont gauge}). The gauge curve $d_{0}(t)$ is 
\emph{horizontal }with respect to this connection iff $J_{0}(t)=0$ for all $%
t $. In this \emph{mechanical gauge}, the kinetic energy is given by two 
\emph{uncoupled} contributions:%
\begin{equation*}
K(\frac{d}{dt}c(t))=K_{int}(t)+\frac{1}{2}\left\langle I_{0}(t)(\xi (t)),\xi
(t)\right\rangle =K_{int}(t)+\frac{1}{2}\left\langle I_{0}^{-1}(t)(\Pi
(t)),\Pi (t)\right\rangle .
\end{equation*}

\section{Appendix:\ Reconstruction on $G\longrightarrow O_{\protect\mu }$}

\label{sec:App Reconstr}Consider the two maps (\cite{MR}) $T^{\ast }G\overset%
{Body\ coord.}{\leftrightarrow }G\times \mathfrak{g}^{\ast }\overset{L}{%
\underset{\pi }{\rightrightarrows }}\mathfrak{g}^{\ast }$ given by $L(g,\Pi
)=Ad_{g}^{\ast }\Pi $ and $\pi (g,\Pi )=\Pi $, and suppose that we have a
curve $(g(t),\Pi (t))\in G\times \mathfrak{g}^{\ast }$ satisfying $%
L(g(t),\Pi (t))=\mu =const$. The idea of this appendix is to reconstruct $%
g(t)$ from $\Pi (t)$ by means of the fact that $\Pi (t)=Ad_{g^{-1}(t)}^{\ast
}\mu $. Note that $\Pi (t)$ lies in the coadjoint orbit $O_{\mu }\subset 
\mathfrak{g}^{\ast }$ through $\mu $. For reconstruction (\cite{MMR}), we
need to consider a \emph{principal connection} on the $G_{\mu }-$principal
bundle $G\overset{\pi }{\longrightarrow }O_{\mu }$, where $G_{\mu }:=\{g\in
G;\ Ad_{g}^{\ast }\mu =\mu \}$ denotes the stabilizer subgroup. Recall that
this bundle corresponds to the reduction $L^{-1}(\mu )\approx G\overset{\pi }%
{\longrightarrow }O_{\mu }$, where the $G_{\mu }$ action on $L^{-1}(\mu
)\subset G\times \mathfrak{g}^{\ast }$ is the one induced by usual left
action (in body coordinates) of $G$ on $T^{\ast }G$. Using the principal
bundle isomorphism 
\begin{eqnarray*}
&:&L^{-1}(\mu )\overset{\approx }{\longrightarrow }G \\
&:&(g,Ad_{g^{-1}}^{\ast }\mu )\longmapsto g
\end{eqnarray*}%
we see that a principal connection on $G\overset{\pi }{\longrightarrow }%
O_{\mu }$ can be defined by a choice of a complement $HOR_{e}\subset 
\mathfrak{g}$ to the \emph{isotropy Lie algebra }$\mathfrak{g}_{\mu
}=Lie(G_{\mu })$, i.e., $\mathfrak{g}=HOR_{e}\oplus \mathfrak{g}_{\mu }$,
and by then \emph{right translating} this complement to any point $g\in G$.
There is no canonical way of choosing $HOR_{e}$ in general. So, let $P:%
\mathfrak{g}\longrightarrow \mathfrak{g}_{\mu }$ be a linear projector onto $%
\mathfrak{g}_{\mu }$ such that%
\begin{equation}
Ad_{h}\circ P=P\circ Ad_{h}  \label{Eq: App compat P}
\end{equation}%
for all $h\in G_{\mu }$ and define $HOR_{e}=Ker(P)$. The corresponding \emph{%
connection }$\emph{1-}$\emph{form }$A_{P}:TG\longrightarrow \mathfrak{g}%
_{\mu }$ induced by $P$ is then given by%
\begin{equation*}
A_{P}(v_{g})_{g}:=P(v_{g}g^{-1})
\end{equation*}%
for $v_{g}\in T_{g}G$ and $v_{g}g^{-1}$ denoting the derivative at $g$ of
the right translation by $g^{-1}$ in $G$.

\begin{example}
\label{ex:Ad invar metric}\emph{(}$\emph{Ad-}$\emph{invariant metrics) }If
the Lie algebra $\mathfrak{g}$ is equipped with an $Ad-$invariant scalar
product $(,)$, then let $P$ to be the \emph{orthogonal projector} with
respect to $(,)$ onto $\mathfrak{g}_{\mu }$. It can easily be seen that this
projector $P$ satisfies $\left( \ref{Eq: App compat P}\right) $, inducing a
principal connection on $G\overset{\pi }{\longrightarrow }O_{\mu } $.
\end{example}

Now, we shall make use of this connection to \textbf{reconstruct} $g(t)$
from a solution $\Pi (t)$ on the coadjoint orbit $O_{\mu }$. Following \cite%
{MMR}:

\begin{itemize}
\item consider the \emph{horizontal lift }$g_{G}(t)\in G$ from $%
g_{G}(t_{1})=g(t_{1})$ of the base curve $\Pi (t)\in O_{\mu }$ with respect
to the connection $A_{P}$,

\item find $h_{D}(t)$ as the curve in $G_{\mu }$ fixed by requiring that 
\begin{equation*}
g(t)=h_{D}(t)\cdot g_{G}(t)
\end{equation*}%
be a solution of the reconstruction equation $\Pi (t)=Ad_{g^{-1}(t)}^{\ast
}\mu $, for the initial value $g(t_{1})$.
\end{itemize}

The group elements in the above decomposition of $g(t)$ at time $t$, $%
h_{D}(t)$ and $g_{G}(t)$, are usually called the \textbf{dynamic phase }and
the \textbf{geometric phase}, respectively. The curve $h_{D}(t)$ must be a
solution of%
\begin{equation}
\frac{d}{dt}h_{D}h_{D}^{-1}(t)=A_{P}(\frac{d}{dt}g)_{g}
\end{equation}%
with $h_{D}(t_{1})=e$.

Suppose now that $\mathfrak{g}$ has an $Ad-$invariant scalar product $(,)$
as in example \ref{ex:Ad invar metric}. This bilinear form induces a vector
space isomorphism $\Psi :\mathfrak{g}^{\ast }\longrightarrow \mathfrak{g}$
which transforms the coadjoint action into the adjoint action of $G$. Let $%
u_{1}=\frac{\Psi (\mu )}{\left\Vert \Psi (\mu )\right\Vert }$ and $%
\{u_{i}\}_{i=1}^{dim\mathfrak{g}_{\mu }}$ denote an orthonormal basis with
respect to $(,)$ of the vector subspace $\mathfrak{g}_{\mu }\subset 
\mathfrak{g}$. Note that this can always be done since $\Psi (\mu )\in 
\mathfrak{g}_{\mu }$. The orthogonal projector, in this case, can be written
as 
\begin{equation*}
P(v)=\Sigma _{i=1}^{dim\mathfrak{g}_{\mu }}(u_{i},v)\ u_{i}
\end{equation*}%
and\bigskip , thus, 
\begin{equation}
\frac{d}{dt}h_{D}h_{D}^{-1}(t)=\Sigma _{i=1}^{dim\mathfrak{g}_{\mu }}(u_{i},%
\frac{d}{dt}gg^{-1})\ u_{i}.
\end{equation}

\end{document}